\documentclass[12pt]{article}
\usepackage{amsmath}
\usepackage{graphicx}
\usepackage{enumerate}
\usepackage{natbib}
\usepackage{url} 

\newcommand{\blind}{1}

\newcommand{\E}{\mathbbm{E}}     
\newcommand{\PR}{\mathbbm{P}}    

\newcommand{\var}{\mbox{\rm var}}

\newcommand{\sd}{\mbox{\rm sd}}       

\newcommand{\bX}{\mathbf{X}} 
\newcommand{\bZ}{\mathbf{Z}} 
\newcommand{\bbe}{\boldsymbol{\beta}} 
\newcommand{\blam}{\boldsymbol{\lambda}}
\newcommand{\bkap}{\boldsymbol{\kappa}}

\newcommand{\pspRsq}{R^2_{x|z}} 
\newcommand{\phixz}{\ensuremath{\phi_{x|z}}} 
\newcommand{\fsq}{f^2} 
\newcommand{\fsqphi}{\ensuremath{f^2_{\phi}}} 
\newcommand{\fsqR}{f^2_{R}} 

\newcommand{\EndPf}{\hfill$\square$ \medskip} 

\usepackage{xcolor}

\usepackage{amsthm, amsmath, amssymb, bbm, booktabs, graphicx, tikz}
\usepackage{multirow}
\usepackage{float}
\usepackage{diagbox}

\newtheorem{assumption}{Assumption}
\newtheorem{lemma}{Lemma}
\newtheorem{theorem}{Theorem}
\newtheorem{corollary}{Corollary}

\usepackage{xr-hyper} 
\usepackage{hyperref}  

\begin{document}

\def\spacingset#1{\renewcommand{\baselinestretch}%
{#1}\small\normalsize} \spacingset{1}


\if1\blind
{
  \title{\bf General measures of effect size to calculate power and sample size for Wald tests with generalized linear models}
  \author{Amy L Cochran\\
    Department of Mathematics \\
    Department of Population Health Sciences\\
    University of Wisconsin Madison\\
    and \\
    Shijie Yuan \\
    Department of Statistics and Data Sciences \\
    University of Texas Austin \\
    and \\
    Paul J Rathouz\thanks{
    The authors gratefully acknowledge R21MH133371}\hspace{.2cm} \\
    Department of Population Health \\
    University of Texas Austin
    }
  \maketitle
} \fi

\if0\blind
{
  \bigskip
  \bigskip
  \bigskip
  \begin{center}
    {\large\bf General measures of effect size to calculate power and sample size for Wald tests with generalized linear models}
\end{center}
  \medskip
} \fi

\bigskip
\begin{abstract}
Power and sample size calculations for Wald tests in generalized linear models (GLMs) are often limited to specific cases like logistic regression. More general methods typically require detailed study parameters that are difficult to obtain during planning. We introduce two new effect size measures for estimating power and sample size in studies using Wald tests across any GLM. These measures accommodate any number of predictors or adjusters and require only basic study information. We provide practical guidance for interpreting and applying these measures to approximate a key parameter in power calculations. We also derive asymptotic bounds on the relative error of these approximations, showing that accuracy depends on features of the GLM such as the nonlinearity of the link function. To complement this analysis, we conduct simulation studies across common model specifications, identifying best use cases and opportunities for improvement. Finally, we test the methods in finite samples to confirm their practical utility, using a case study on the relationship between education and receipt of mental health treatment.
\end{abstract}

\noindent%
\vfill

\newpage
\spacingset{1.05} 
\section{Introduction}
Accurate power and sample size (PSS) calculations are essential for study planning. Many studies are designed to conduct Wald tests within the framework of generalized linear models (GLMs). For instance, logistic regression might be used to determine if smoking predicts disease risk. Similarly, Poisson regression could analyse how air pollution influences the rate of depression episodes. Typically, these analyses adjust for variables like age or gender. They become more complex when multiple predictors are tested jointly, such as air pollution \textit{and} pollen levels, or treatment plans with three options. In these cases, researchers could plan to fit a GLM with multiple adjustors and predictors, and use a Wald test to assess whether the combined influence of these predictors on the mean outcome differs from zero.

Over the years, various methods have been developed for performing PSS calculations in the context of GLMs. Early work by \citet{whittemore1981sample} on logistic regression provided foundational methods by fully specifying the moment generating function of predictors and adjustors. Later, several authors expanded on these ideas for more general GLM contexts, introducing PSS calculations using Wald \citep{wilson1986calculating,shieh2005power}, score \citep{self1988power}, and likelihood ratio tests \citep{self1992power,shieh2000power}. Despite these advancements, these methods often require extensive and specific assumptions about the data, such as the joint distribution of predictors and adjustors and the exact relationships between predictors and outcomes, which can be cumbersome to specify in real-world applications. For instance, \citet{self1988power} required categorical predictors with defined category frequencies and log-odds values, making their approach less flexible when dealing with continuous or more complex predictor structures.

Recent methods, like those proposed by \citet{lyles2007practical}, use a design matrix to define a discrete distribution for predictors and adjustors, but still demand details about how predictors relate to outcomes, information that can be difficult to obtain in practice. Many methods apply specifically to logistic or Poisson regression
\citep{whittemore1981sample,signorini1991sample,hsieh1998simple,shieh2001sample,schoenfeld2005calculating,demidenko2007sample,demidenko2008sample,novikov2010modified, bush2015sample}.  Many also only consider single predictors \citep{whittemore1981sample,wilson1986calculating,hsieh1998simple,signorini1991sample,shieh2001sample,demidenko2008sample,novikov2010modified,demidenko2007sample},
with some exceptions
\citep{self1988power,self1992power,shieh2000power,shieh2005power,lyles2007practical, bush2015sample}. While these methods work in specific cases, a general method is still needed for GLMs with multiple predictors and adjustors that avoids full specification of predictor distributions and outcome relationships.

Alongside theoretical advancements, software tools have been created to support PSS calculations for GLMs. Programs like \textit{proc power} in SAS, \textit{G*Power}, and \textit{PASS} offer user-friendly interfaces that accept key inputs such as effect sizes, predictor distributions, and expected outcome probabilities. However, these tools are typically limited to specific models like logistic and Poisson regression, and often do not support testing multiple predictors at once. Some programs, like SAS \textit{proc power}, can handle multiple predictors, but only if researchers use detailed procedures like those in \citet{lyles2007practical} or \citet{shieh2005power}, which require the full distributions of all variables involved. To accommodate multiple adjustors, 
they typically apply heuristic methods, such as inflating the sample size based on the multiple correlation coefficient between the predictor and adjustors \citep{hsieh1998simple}.  Additionally, these tools often require detailed knowledge of the full distribution of predictors and adjustors, which can be difficult to specify accurately. As a result, while these software tools are valuable for many PSS calculations, they may not be sufficient for more complex scenarios, limiting their effectiveness in diverse research settings.

By comparison, \citet{gatsonis1989multiple} introduced a seminal framework for linear regression that bypasses many challenges of GLMs. It remains remains widely effective due to four key benefits. First, it uses the multiple $R^2$ to measure effect size, giving a uniform and interpretable way to quantify the variation in outcomes explained by predictors, whether predictors are continuous, categorical, or a mix of both. Second, it supports joint testing of multiple predictors, including interactions and categorical variables with multiple levels. Third, it incorporates multiple adjustors via partial $R^2$, which (like $R^2$) can be readily interpreted and communicated, even to non-statisticians. Finally, the framework requires only the first two moments of the predictor and adjustor distributions.

In this paper, we introduce two new effect size measures for PSS calculations involving Wald tests in GLMs, recapturing the benefits of partial $R^2$ from linear regression. The first, $\phixz^2$, quantifies the added variance in the linear predictor due to the predictors beyond what is accounted for by adjustors. The second, $\pspRsq$, reflects the portion of mean square error on the outcome scale attributable to predictors, beyond the adjustors. These measures apply to any GLM, accommodate arbitrary predictors or adjustors, and require only first and second moments. We show how to interpret and use these measures to approximate the non-centrality parameter needed for PSS calculations. We assess the error in this approximation across varying conditions. Lastly, we evaluate finite sample performance.

\section{Novel measures of effect size}

\subsection{Motivation and intuition}

Let $\bX$ be our $p$-dimensional predictor and $Y$ the outcome. To adjust for other factors, we include a $k$-dimensional vector $\bZ$, which also contains a constant term for the intercept. We assume a distribution for $\bX$ and $\bZ$ arises by simple random sampling, though results can extend to planned designs.

Cohen's $d$ is a common measure of effect size that captures the difference in means between two groups, scaled by residual standard deviation (SD). In the special case of a linear regression model,
\begin{align*}
    Y  = \bX'\bbe + \bZ'\blam  + \epsilon
\end{align*}
for independent normal $\epsilon$ with mean zero and variance $\sigma^2$, we can connect $\bbe$ to Cohen's $d$. Suppose $\bX$ is one-dimensional and binary with mean .5, and is uncorrelated with $\bZ$. Then, within levels of $\bZ$, the coefficient $\bbe$ equals the mean difference in $Y$ comparing units with $\bX = 1$ and $\bX = 0$:
\begin{align*}
 \bbe = \E[Y \mid \bZ, \bX = 1] - \E[Y \mid \bZ, \bX = 0].
\end{align*}
Because such binary $\bX$ has a SD of .5, this difference corresponds to a two SD contrast in $\bX$. For general $\bX$, an analogous effect size is $2\,\sd(\bX'\bbe)$, measuring the mean difference in Y comparing two units whose $\bX'\bbe$ values differ by two SDs, holding $\bZ$ fixed. When $\sigma^2 = 1$, this effect size becomes Cohen's $d$ within levels of $\bZ$. 

We will define an effect size measure, $\phixz$, that coincides with Cohen’s $d$ in this setting, providing a familiar reference point. However, it also admits connections to other standard quantities. For example, take logistic regression for binary $Y$, with binary $\bX$ (mean $.5$) and $\bZ=1$:
\begin{align*}
  \textrm{logit} \, \PR\left(Y=1 \mid \bZ, \bX\right) = \bX'\bbe + \bZ'\blam.
\end{align*}
Here, $\bbe$ is the change in the log odds of the outcome between $\bX=1$ and $\bX=0$, corresponding to a two SD contrast in $\bX$.  For general $\bX$, one might instead consider $2\,\sd(\bX'\bbe)$, reflecting the change in log odds between two units whose $\bX'\bbe$ values differ by two SD. Our measure, $\phixz$, agrees with this two SD contrast in log odds.

While $\phixz$ quantifies effect size on the linear‐predictor scale (e.g., log odds), it can be useful to work on the outcome scale, in a form more familiar to those who use $R^2$–type measures. For this reason, we introduce another effect size measure $\pspRsq$ that reduces exactly to partial $R^2$ from linear regression, which captures the incremental explanatory power of $\bX$ beyond what is already accounted for by $\bZ$. 

To clarify this connection, we review partial $R^2$. The full mean model, $\bX'\bbe + \bZ'\blam,$ minimizes mean squared error (MSE), 
$$\E\left[(Y - \bX'\bbe - \bZ'\blam)^2\right],$$ 
over all linear predictors in $\bX$ and $\bZ$. To evaluate the added value of $\bX$, we compare this to the MSE when predicting $Y$ using only $\bZ$, 
$$\E\left[(Y - \bZ'\bkap)^2\right],$$ 
where $\bZ'\bkap$ is the best linear predictor in $\bZ$. Asymptotically, partial $R^2$ is then the proportion of MSE (from the $\bZ$-only model) reduced by including $\bX$:
\begin{align*}
\text{Partial } R^2 = \frac{\E\left[(Y - \bZ'\bkap)^2\right] - \E\left[(Y - \bX'\bbe - \bZ'\blam)^2\right]}{\E\left[(Y - \bZ'\bkap)^2\right]}.
\end{align*}

In the simple cases we discussed, our effect size measures correspond directly to familiar quantities such as Cohen's $d$, partial $R^2$, or two SD contrasts in log odds. While this alignment helps with intuition, it relies on these specific conditions. When we move to more complex adjustors, variances, or link functions, this correspondence breaks down, and a weighting term emerges, complicating effect size measurement. 

\subsection{Definition and interpretation}

A GLM describes $Y$ conditional on $\bX$ and $\bZ$ using a mean model with link function~$g$,
\begin{equation}
\label{glmmean}
g(\mu) = g\left( \E\left[ Y | \bX, \bZ \right] \right) =  \bX'\bbe + \bZ'\blam 
:= \eta\ ,
\end{equation}
where $\bbe$ and $\blam$ are unknown parameters and $\mu$ is defined implicitly as $\E\left[ Y | \bX, \bZ \right]$. Expectations and conditional expectations are always defined with respect to the joint distribution of $(Y,\bX,\bZ)$. We suppress the dependence of $\mu$ and $\eta$ on $\bX$, $\bZ$, $\bbe$, and $\blam.$ We assume $g$ and its inverse $g^{-1}$ are continuously differentiable on their respective domains.

The conditional distribution of $Y$ given $\bX$ and $\bZ$ belongs to the exponential family with density (or probability mass function for discrete $Y$),
\begin{equation}
\label{glmdistr}
\exp( y \theta - b(\theta) + c(y) ),
\end{equation}
where $\theta$ is defined implicitly through the relation $b'(\theta) = \mu = g^{-1}\left( \eta \right).$ Additionally, $b''(\theta)=\var\left[ Y | \bX, \bZ\right]:= v(\mu).$ We write $v$ for simplicity when the meaning is clear.

A GLM defined by \eqref{glmmean} and \eqref{glmdistr} is fitted via maximum likelihood estimation (MLE). Estimates $\hat{\blam}$ and $\hat{\bbe}$ are obtained from $n$ independent observations of $(Y, \bX, \bZ)$ using an iteratively re-weighted least squares procedure \citep{mccullagh2019generalized}. Each iteration evaluates a linearized version of $g(Y)$ around $\mu$ and a weight term:
\begin{align*} g(Y) \approx Y_l := \eta + \frac{\partial \eta}{\partial \mu}(Y - \mu), \qquad \text{and} \qquad w := \frac{1}{v} \left( \frac{\partial \mu}{\partial \eta} \right)^2, \end{align*}
at our current estimates. Each iteration regresses $Y_l$ onto $\bX$ and $\bZ$ with weights $w$ to recover new estimates, or equivalently,
$w^{1/2}Y_l$ onto $w^{1/2}\bX$ and $ w^{1/2} \bZ$. This process is repeated until convergence. Supplementary Text~\ref{app:weight} lists $w$ for common GLMs.

From this perspective, $\eta$ is the best linear predictor of $Y_l$ in $\bX$ and $\bZ$ on a \textit{weighted scale}. That is, it minimizes \textit{weighted} MSE:
\begin{align*}
    \E\left[ w \right( Y_l - \eta \left)^2 \right] := \text{WMSE}.
\end{align*}
The added value of $\bX$ can be evaluated by comparing predictions from the full model, $\eta$, to those from a model using only $\bZ$. Specifically, let $\eta_z$ denote the best linear predictor based solely on $\bZ$, minimizing weighted MSE:
\begin{align*}
\E\left[ w \right( Y_l - \eta_z \left)^2 \right] =   \min_{\bkap \in \mathbb{R}^q } \E\left[ w \right( Y_l - \bZ'\bkap \left)^2 \right] := \text{WMSE}_0,
\end{align*}
where the weight $w$ is evaluated at the full model $\eta$ and where we assume a unique minimizer exists.
Accordingly, we can write
\begin{align*}
    \eta = \eta_z + (\eta - \eta_z),
\end{align*}
with $\eta_z$ summarizing the information from $\bZ$ alone, and $(\eta - \eta_z)$ capturing the incremental contribution of $\bX$.

Define our first measure of effect size as
\begin{equation}
\label{effect-phi}
    \phixz:=2\sd(\eta-\eta_z),
\end{equation}
where SD is defined over the distribution of~$\bX$ and $\bZ$. It quantifies the variation in $\eta$ attributable to adding $\bX$. We expect $\phixz$ to appeal to the applied investigator who uses odds ratios and similar contrasts when planning studies. 

For logistic regression, $\phixz$ represents a log odds ratio comparing two units that differ by 2 SDs in $\eta-\eta_z$. For log-linear regression with Poisson $Y$, $\phixz$ is a log rate ratio comparing two units that differ by 2 SDs in the linear predictor $\eta-\eta_z$. For linear regression with unit residual variance, $\phixz$ is Cohen's $d$ comparing two units that differ by 2 SDs in the linear predictor $\eta-\eta_z$. And, for the cases we considered earlier (i.e. linear regression with uncorrelated $\bX$ and $\bZ$, or logistic regression with constant $\bZ$), $\phixz$ is simply $2\sd(\bX'\bbe)$, which motivates our choice of the multiplier 2 in \eqref{effect-phi} since $\phixz$ equals $|\bbe|$ for binary $\bX$ with mean $.5.$ 


For our second effect measure, we use an equivalent expression for $\text{WMSE}$:
\begin{align*}
\text{WMSE} = \E\left[ (Y - \mu)^2 / v \right],
\end{align*}
representing a MSE for predicting $Y$ by its mean $\mu$, standardized by the SD $\sqrt{v}$. Building on this idea, we can define a standardized MSE for predicting $Y$ using only the adjustors $\bZ$. Let $\mu_z$ be our predictor based on $\bZ$ alone, which we define as the linear predictor $\eta_z$ transformed back to the original scale: $\mu_z := g^{-1}(\eta_z)$. Using the same SD scale as before, we write:
\begin{align*}
\text{SMSE}_0 = \E\left[ (Y - \mu_z)^2 / v \right],
\end{align*}
Define a partial pseudo-$R^2$:
\begin{align}
\pspRsq := \frac{\text{SMSE}_0 - \text{WMSE}}{\text{SMSE}_0} = \frac{\E\left[ (Y - \mu_z)^2 / v \right] - \E\left[ (Y - \mu)^2 / v \right]}{\E\left[ (Y - \mu_z)^2 / v \right]},
\end{align}
representing the fraction of the MSE on the SD scale that is explained by letting the conditional mean for $Y$ depend on $\bX$ in addition to $\bZ$.


Comparing $\phixz$ and $\pspRsq$, differences emerge in interpretation. The measure $\phixz$ is on the scale of the linear predictor (e.g., a log odds ratio). It ranges from 0 to infinity, with higher values indicating a stronger effect of $\bX$ on $Y$ given $\bZ$. In contrast, $\pspRsq$ is a relative measure on the outcome scale. It falls between 0 and 1, where values closer to 1 suggest a stronger relationship between $\bX$ and $Y$ after adjusting for $\bZ$. Both measures equal 0 when $\bbe = 0$. 

\section{Power and sample size}

We introduced two measures designed to work with any predictor $\bX$ and any adjustor $\bZ$ that includes a constant term. They do not require full knowledge of the distributions of $\bX$ or $\bZ$, or the parameters $\blam$ or $\bbe$. We now use them to approximate power under the alternative hypothesis $\bbe\neq 0$ for the Wald test.

\subsection{A review of Wald tests}

Under regularity conditions, the MLE estimate $\left[ \hat \blam' \quad \hat\bbe' \right]'$ is asymptotically normal with mean $\left[ \blam' \quad \bbe' \right]'$. Its variance is $(n \mathcal{I})^{-1}$ where $\mathcal{I}$ is the expected Fisher information for $\blam$ and $\bbe$ from a single observation of $(Y, \bX, \bZ)$:
\begin{align*}
\mathcal{I} :=
\begin{bmatrix}
\E\left[w \bZ \bZ'\right] & \E\left[w \bZ \bX'\right] \\
\E\left[w \bX \bZ'\right] & \E\left[w \bX \bX'\right]
\end{bmatrix},
\end{align*}
obtained by squaring and averaging the score function of a single observation:
\begin{align*}
   w (Y-\mu) \frac{\partial \eta}{ \partial \mu} \left[ \bZ' \quad \bX' \right]'.
\end{align*}
For $n$ independent observations of $(Y, \bX, \bZ)$, the expected Fisher information is $n \mathcal{I}$.

Focusing on the predictor, $\hat{\bbe}$ is asymptotically normal with mean $\bbe$ and variance given by the lower $p \times p$ block of $(n\mathcal{I})^{-1}$. It is computed as $(n\mathcal{I}_{\bbe|\blam})^{-1}$ where
\begin{align*}
{\mathcal{I}}_{\bbe | \blam} := \E\left[ w \bX \bX' \right] - \E\left[ w \bX \bZ' \right] \E\left[ w \bZ \bZ' \right]^{-1} \E\left[ w \bZ \bX' \right]. \end{align*}
Thus, $\sqrt{n}{\mathcal{I}}_{\bbe | \blam}^{1/2} \hat{\bbe}$ is asymptotically normal with mean $\sqrt{n}{\mathcal{I}}_{\bbe | \blam}^{1/2} \bbe$ and variance given by the identity matrix. We view $\mathcal{I}_{\bbe | \blam}$ as the expected Fisher information for $\bbe$ from a single observation  after adjusting for $\blam$, and $n\mathcal{I}_{\bbe | \blam}$ as the expected Fisher information for $\bbe$ from $n$ independent observations after adjusting for $\blam$.  

A Wald test for the null hypothesis, $H_0:\bbe=0$, uses the test statistic
\begin{equation*}
    n \hat \fsq := n \hat{\bbe}' \hat{\mathcal{I}}_{\bbe | \blam} \hat{\bbe},
\end{equation*}
where $\hat{\mathcal{I}}_{\bbe | \blam}$ is a consistent estimator of ${\mathcal{I}}_{\bbe | \blam}$. Because $\hat{\mathcal{I}}_{\bbe | \blam}$ is consistent for ${\mathcal{I}}_{\bbe | \blam}$, then  $\sqrt{n} \hat{\mathcal{I}}_{\bbe | \blam}^{1/2} \hat{\bbe}$ is asymptotically normal with mean 
$\sqrt{n}{\mathcal{I}}_{\bbe | \blam}^{1/2} \bbe$ and identity variance matrix. Thus, $n\hat \fsq$ follows, asymptotically, a non-central $\chi^2$ distribution with $p$ degrees of freedom (df) and non-centrality parameter $n \fsq$, where
\begin{equation*}
\fsq :=\bbe'{\mathcal{I}}_{\bbe | \blam}\bbe.
\end{equation*}
Equivalently, $\fsq = \E[ w(\eta-\eta_z)^2]$ (see Supplementary Text~\ref{app:derivations} for key derivations). Under the null hypothesis, $\fsq$ is zero. A Wald test rejects the null at significance level $\alpha$ if $n\hat \fsq$ exceeds the $(1-\alpha)$ quantile of a $\chi^2$ distribution with $p$ df and non-centrality parameter $0$. This quantile is $F_{\chi^2_p(0)}^{-1}(1-\alpha)$, where $F_{\chi^2_p(u)}$ denotes the cumulative distribution function of a non-central $\chi^2$ distribution with $p$ df and non-centrality $u$. 

\subsection{Proposed power and sample size calculations}

The power of the Wald test is the probability that $n \hat\fsq$ exceeds $F_{\chi^2_p(0)}^{-1}(1-\alpha)$. If we knew $n \fsq$, we can approximate this probability using:
\begin{align*}
q := 1 - F_{\chi^2_p(n\fsq)}\left(F_{\chi^2_p(0)}^{-1}(1-\alpha)\right).
\end{align*}
For fixed $n$, $q$ approaches 1 as $\fsq$ increases and approaches $\alpha$ as $\fsq$ approaches zero. However, $\fsq$ depends on the joint distribution of $(\bX,\bZ)$ and the parameters $\blam,\bbe$. 


We propose the approximation:
\begin{align*}
\fsq \approx  w_1 \phixz^2 / 4 := \fsqphi,
\end{align*}
where $w_1\neq 0$ is a constant chosen to be close to the weight $w$. To justify this form, we can express $\fsq$ in the following form:
\begin{align*}
    \fsq = w_1 \phixz^2/4 + \E[(w-w_1)(\eta-\eta_z)^2] + \E[(w-w_1)(\eta-\eta_z)]^2 / w_1.
\end{align*}
When $w_1$ is close to $w$, the remaining two terms become small, leaving $w_1 \phixz^2/4$ as a natural approximation. In practice, we let $w_1$ be $w$ evaluated at $g(\E[Y])$, since the applied investigator can typically provide a reasonable value for $\E[Y]$.

Additionally, it can be shown that
\begin{align*}
    \frac{\fsq}{1+\fsq} = \frac{\E\left[ w \right( Y_l - \eta_z \left)^2 \right]-\E\left[ w \right( Y_l - \eta \left)^2 \right]}{\E\left[ w \right( Y_l - \eta_z \left)^2 \right]} := R_W^2,
\end{align*}
capturing the portion that $\bX$ explains in the linearized outcome $Y_l$ above $\bZ$ on the weighted scale. With $\fsq = R_W^2/(1-R_W^2)$, we derive an alternative approximation:
\begin{align*}
   \fsq \approx \frac{\pspRsq}{1-\pspRsq} :=\fsqR.
\end{align*}

These approximations allow power or sample size to be computed during study design:

\begin{enumerate}
    \item \textbf{Solicit information}: Work with investigator to define the GLM, including the distribution, outcome, predictors, adjustors, and link function. Select an effect size measure, $\phixz$ or $\pspRsq$, and specify the anticipated mean, $\E[Y]$ (\textit{if needed}).

    \item \textbf{Set statistical criteria}: Specify the significance level $\alpha$. Define the desired target: either the sample size $n$ or the target power $q_*$.

    \item \textbf{Approximate $\fsq$}: Use the solicited information to approximate $\fsq$. From $\phixz$, calculate $\fsqphi = w_1 \phixz^2/4$, where $w_1$ is the value of the weight function $w$ evaluated at $g(\E[Y])$. Alternatively, use $\fsqR = \pspRsq/(1 - \pspRsq)$.

    \item \textbf{Perform calculations}: For sample size, solve for $n$ in the equation  
    \[
    q_* = 1 - F_{\chi^2_p(n\tilde \fsq)}\left(F_{\chi^2_p(0)}^{-1}(1-\alpha)\right),
    \]  
    where $q_*$ is the target power and $\tilde \fsq$ is either $\fsqR$ or $\fsqphi$.  For power, use the same equation to compute $q_*$ for a given sample size $n$.
\end{enumerate}

\subsection{Connection to linear regression}

For linear regression, the identity link and constant variance simplify the setup: $Y_l=Y$, $w$ is constant, and $\eta,\eta_z$ reduce to $\mu,\mu_z$. This yields
\begin{align*}
\fsq = \E[w(\eta-\eta_z)^2]
= \E\left[(\mu-\mu_z)^2  / \sigma^2 \right],
\end{align*}
and both $\fsqphi$ and $\fsqR$ collapse to this same expression. In addition, $\fsq$ coincides with the familiar partial $f^2$ from PSS calculations in linear regression and corresponds directly to partial $R^2$:
\begin{align*}
\text{Partial } R^2
= \frac{ \fsq}{1+ \fsq} = R_W^2 = \pspRsq.
\end{align*}
Thus, in linear regression $\fsq$, $\fsqphi$, $\fsqR$, partial $f^2$ coincide, as do $R_W^2$, $\pspRsq$, and partial $R^2$. Our framework therefore generalizes these classical quantities to all GLMs.

\section{Approximation error}\label{s:simulation}

We investigate the accuracy of the approximations ($\fsqphi$ and $\fsqR$) when used to determine sample size or predict power. A key concern is how much powers $q_\phi$ and $q_R$ computed using $\fsqphi$ or $\fsqR$ differ from power $q$ computed using $\fsq$, especially when they are overly optimistic (i.e., inflated).

\subsection{The role of relative error}

Our starting point is a design that targets a specific power $q_*$ (e.g., 80\%) for fixed $\alpha$ and $p$ df. We choose the sample size $n_\phi$ so that the power computed with $\fsqphi$ equals the target, $q_\phi = q_*$. Likewise, $n_R$ is chosen so that $q_R = q_*$. Equivalently,
\begin{align*}
    n_\phi \fsqphi = n_R \fsqR := \nu_*,
\end{align*}
where $\nu_*$ is the non-centrality parameter required to achieve a power of $q_*$:
\begin{align*}
    q_* = 1 - F_{\chi^2_p(\nu_*)}\!\left(F_{\chi^2_p(0)}^{-1}(1-\alpha)\right).
\end{align*}

The difference in power depends on the relative error in our approximations:
\begin{align*}
    (\fsq - \fsqphi)/\fsqphi := \text{re}_\phi, \qquad \text{and} \qquad ( \fsq-\fsqR)/\fsqR := \text{re}_R, 
\end{align*}
measured relative to the approximated value, not the true value. To see this, note:
\begin{align*}
    q  - q_\phi =  &1 - F_{\chi^2_p(n_\phi \fsq)}\left(F_{\chi^2_p(0)}^{-1}(1-\alpha)\right) - q_* \\
    = &1 - F_{\chi^2_p(\nu_* (1+\text{re}_{\phi}))}\left(F_{\chi^2_p(0)}^{-1}(1-\alpha)\right) - q_*.
\end{align*}
Above, the difference in power is written as a function of the target power $q_*$, its corresponding non-centrality parameter $\nu_*$, and relative error $\text{re}_\phi$. 
Similarly, 
\begin{align*}
    q  - q_R =  1 - F_{\chi^2_p(\nu_* (1+\text{re}_R))}\left(F_{\chi^2_p(0)}^{-1}(1-\alpha)\right) - q_*.
\end{align*}
That is, once the design is fixed to achieve a chosen target power, the actual power is determined entirely by the relative error. Table~\ref{tab:power_error} shows this relationship for $\alpha=0.05$ and $p=1$. For instance, if the target power is 80\% but $\fsq$ is 15\% smaller than $\fsqphi$, the actual asymptotic power falls by 6.7 percentage points to 73.3\%.

\begin{table}[!ht]
\centering
\caption{Differences in power as a function of relative error ($\text{re}_\phi$ or $\text{re}_R$). Target power is the power computed using an approximated effect size ($\fsqphi$ or $\fsqR$). All other cells show the difference between the true asymptotic power computed using the actual effect size ($\fsq$) and the target power. The significance level is $\alpha = 0.05$, and degrees of freedom are $p = 1$.}

\smallskip

{
\begin{tabular}{cllllll}
\hline
& \multicolumn{6}{c}{\textbf{Relative error ($\text{re}_\phi$ or $\text{re}_R$)}} \\
\textbf{Target power} & \textbf{-15\%} & \textbf{-10\%} & \textbf{-5\%}  & \textbf{5\%} & \textbf{10\%} & \textbf{15\%} \\
\hline
\textbf{60} & -6.8 & -4.4 & -2.2 & 2.1 & 4.1 & 6.0 \\
\textbf{64} & -7.0 & -4.5 & -2.2 & 2.1 & 4.1 & 6.1 \\
\textbf{68} & -7.0 & -4.6 & -2.2 & 2.1 & 4.1 & 6.0 \\
\textbf{72} & -7.0 & -4.6 & -2.2 & 2.1 & 4.0 & 5.8 \\
\textbf{76} & -6.9 & -4.5 & -2.1 & 2.0 & 3.9 & 5.6 \\
\textbf{80} & -6.7 & -4.3 & -2.0 & 1.9 & 3.6 & 5.2 \\
\textbf{84} & -6.2 & -4.0 & -1.9 & 1.7 & 3.3 & 4.7 \\
\textbf{88} & -5.6 & -3.5 & -1.7 & 1.5 & 2.8 & 4.0 \\
\hline
\end{tabular}
}
\label{tab:power_error}
\end{table}

\subsection{A sequence of local alternatives}

These calculations motivate an asymptotic regime for evaluating our approximations. We fix a target power $q_*$ and consider a sequence of effect sizes shrinking toward zero, bringing the alternative closer to the null ($\bbe=0$). As the effect shrinks, the required sample size diverges to keep power fixed. In this regime, the approximation error in power is governed entirely by $\text{re}_\phi$ and $\text{re}_R$. This mirrors the theory of \textit{local power} 
where parameters drift toward the null as sample size grows. If alternatives shrink too quickly, power collapses to the significance level; too slowly, and power converges to one. Staying in the intermediate regime yields meaningful limits in which approximation accuracy can be assessed. 

Concretely, we fix a target power $q_*$, a target mean $\mu_*$, reference coefficients $\bbe_*$ and $\bkap_*$, and the joint distribution for $(\bX,\bZ)$. Next, we choose scalars $\delta_{\beta} \geq 0$ and $\delta_{\kappa} \geq 0$ so that
\begin{align*}
    \bbe = \delta_{\beta} \beta_*, \qquad \text{ and } \qquad 
    \blam = [\iota \,\,\, \delta_{\kappa} \bkap_*' ].'
\end{align*}
For each $\delta_{\beta}$ and $\delta_{\kappa}$, the intercept $\iota$ is chosen to keep $\E[Y]$ fixed at $\mu_*$. Once $\blam$ and $\bbe$ are defined, so are $\eta$ and effect measures $\fsq$, $\fsqphi$, and $\fsqR$. Finally, sample sizes are selected so that $\fsqphi$ and $\fsqR$ achieve power $q_*$. The resulting power error is then controlled by the corresponding relative error, $\text{re}_\phi$ or $\text{re}_R$. 

We study limits as $\delta_\kappa \to 0$ and $\delta_\beta \to 0$, which drive the effect measures ($\fsq$, $\fsqphi$, $\fsqR$) to zero and the required sample sizes to infinity. Our main interest is how the relative error scales with $\delta_\kappa$ and $\delta_\beta$. This analysis requires several assumptions (detailed in Supplementary Text~\ref{app:proofs}). The key one controls how close $w$ remains to $w_*$, the weight obtained when $\eta = g(\mu_*)$. We assume the mean square error in $w$ is bounded by a term proportional to $(\delta_\kappa + \delta_\beta)^2$:

\begin{assumption} \label{assm:smooth_w}
Fix $\mu_*$, $\bkap_*$, and $\bbe_*$, and let $w_*$ denote $w$ evaluated at $\eta=g(\mu_*)$. There exists constants $M$ and $\delta_*>0$ so that
\begin{align*}
    \E[ (w-w_*)^2 ] \leq (\delta_{\kappa} + \delta_{\beta})^2 M < \infty
\end{align*}
whenever $0 \leq \delta_{\kappa}, \delta_{\beta} < \delta_*$ with $\E[Y] = \mu_{*}$.
\end{assumption}

The constant M measures how far $w$ is from being constant and is central to analysing $\mathrm{re}_\phi$. The condition can hold in a few ways. If $w$ is constant, then $w=w_*$ and the assumption is trivial with $M=0$. More generally, since $\eta=g(\mu_*)$ and $w=w_*$ when $\delta_\kappa=\delta_\beta=0$ and $\E[Y]=\mu_*$, smoothness of $w$ in $\delta_\kappa$ and $\delta_\beta$ (while keeping $\E[Y]=\mu_*$ fixed) allows us to bound its deviation near $\delta_\kappa=\delta_\beta=0$:
$$
|w - w_*| \leq (\delta_{\kappa} + \delta_{\beta}) \text{Rem}_w
$$
for some random variable $\text{Rem}_w$. The assumption could be satisfied with a bound on $\E[\text{Rem}_w^2]$, which might involve bounds on moments of $\bX$ and $\bZ$, and on derivatives of $w$ and $\eta$ with respect to $\delta_{\kappa}$ and $\delta_{\beta}$.

With these assumptions, we state our main result on relative error for $\fsqphi$:

\begin{theorem} \label{thm:error_phi}
Fix $\mu_{*}$, $\bkap_*$, and $\bbe_*$. Under Assumptions \ref{assm:smooth_w}, \ref{assm:inverse_fcn}, and \ref{assm:regular}  with $M$ defined therein,
\[
\left|\rm{re}_{\phi}\right| = \left\vert \fsq - \fsqphi \right\vert / \fsqphi  = {\mathcal{O}}\left( \sqrt{M} \left\{\delta_{\kappa} + \delta_{\beta} \right\} \right).
\]
\end{theorem}

This theorem has several implications. As the regression coefficients for both $\bZ$ and $\bX$ (excluding the intercept) shrink to zero, the approximation error also vanishes, at a rate linear in the size of the coefficients and governed by how much $w$ varies around $\eta = g(\mu_*)$ (through the constant $M$). When these coefficients are small, the relative error is near zero, so $\fsqphi$ closely matches $\fsq$ and $q_\phi$ closely matches $q$. In effect, both predictors and adjustors must have only modest influence on the outcome to ensure the power computed using $\phixz$ is accurate.

Plus, if $w$ is constant, as in the case of a GLM with a Gamma distribution and a log link, then $M = 0$, eliminating any relative error. This is stated as a corollary:

\begin{corollary} \label{corr:error_phi}
Under the conditions of Theorem~\ref{thm:error_phi}, $\rm{re}_{\phi} = 0$ if $w$ is a constant function.
\end{corollary}

To analyse the relative error for our second effect measure, we introduce one more assumption. As defined earlier, we use $\mu_z$ to denote $\mu$ evaluated at $\eta=\eta_z$. 

\begin{assumption} \label{assm:smooth_mu}
Fix $\mu_{*}$, $\bkap_*$, and $\bbe_*$. There exist constants $K$ and $\delta_*>0$ so that 
\begin{align*}
    \E\left[ \text{Rem}_{\mu}^2 / v \right] \leq  K \delta_{\beta}^4 < \infty
\end{align*}
whenever $0 \leq \delta_{\kappa}, \delta_{\beta} < \delta_*$ with $\E[Y] = \mu_{*}$, where $\text{Rem}_{\mu}$ is the remainder from a linear approximation of $\mu_z$ around $\eta_z = \eta$:
\begin{align*}
   \mu_z = \mu + \frac{\partial \mu}{\partial \eta} (\eta_z - \eta) + \text{Rem}_{\mu}.
\end{align*}
\end{assumption}

Like our assumption on $w$, this assumption is to ensure that $\mu$ is sufficiently smooth so that $\mu_z$ gets close to $\mu$ relative to $v$ in expectation when $\delta_{\beta}$ is small. Here, the constant $K$ captures how much the link $g(\mu)$ deviates from a linear function. With this assumption, along with the same assumptions for the last theorem, we can state our main result on the relative error from using $\fsqR$ in place of $\fsq$:

\begin{theorem} \label{thm:error_R}
Fix $\mu_{*}$, $\bkap_*$, and $\bbe_*$. Under Assumptions \ref{assm:smooth_w}, \ref{assm:smooth_mu},  \ref{assm:inverse_fcn}, and \ref{assm:regular} with $K$ defined therein,
\[
\left|\rm{re}_{R}\right| = \left\vert \fsq - \fsqR \right\vert/ \fsqR = {\mathcal{O}}\left(\sqrt{K} \delta_{\beta}\right).
\]
\end{theorem}

This theorem says as the regression coefficients for $\bZ$ and $\bX$ go to zero, the approximation error also vanishes, at a rate linear in the size of the predictor coefficients and governed by how much $\mu$ varies. When the predictor effect is small, relative error is near zero, so $\fsqR$ is close to $\fsq$ and $q_R$ close to $q$. Under an identity link ($\mu=\eta$), $K=0$, eliminating relative error entirely. This yields the corollary:
\begin{corollary} \label{corr:error_R}
Under the conditions of Theorem~\ref{thm:error_R}, $\rm{re}_R = 0$
for an identity link.
\end{corollary}

Given what we learned about relative error in the asymptotic regime, the choice between $\phixz$ and $\pspRsq$ depends on the GLM. The measure $\phixz$ is preferable when $w$ is roughly constant, whereas $\pspRsq$ is preferable when the link function $g(\mu)$ is nearly linear. Our analysis also suggests using $\pspRsq$ when only predictor coefficients are small, and either measure when both predictor and adjustor coefficients are small. Proofs appear in Supplementary Text~\ref{app:proofs}.

\section{Simulation}

Outside the sequence of local alternatives, relative error depends on the chosen GLM and on the joint distribution of $\eta_z,\eta$, itself induced by $\bX,\bZ$ and the parameters $\blam,\bbe$. To explore broader settings, we simulate different GLMs by modelling $\eta_z$ and $\eta$ directly. For $\eta$, we use
\begin{align*}
    \eta = c_0 + c_1 B_z + c_2 B_x,
\end{align*}
where $B_x$ and $B_z$ are Beta distributed, allowing control over skewness and kurtosis. Correlation between them is introduced via a Gaussian copula with parameter $\rho$. Constants $c_0,c_1,c_2$ are chosen so that $\E[\eta]=\iota$, $\var(c_1 B_z)=s_z$, and $\var(c_2 B_x)=s_x$. Shape parameters $a_z,b_z$ (and similarly $a_x,b_x$) adjust skewness and modality.

To obtain $\eta_z$, we assume $\blam'\bZ = c_0 + c_1 B_z$ for two-dimensional $\bZ$ (one dimension being the constant $1$). Then, $\eta_z = \E[w\eta \bZ']\E[w \bZ\bZ']^{-1} \bZ$ can be computed by forming the weight $w$ from $\eta$ and then regressing $\eta$ onto $1$ and $B_z$ with weights $w$. This avoids assumptions about $\blam$ or $\bX$ beyond $\eta=\blam'\bZ+\bbe'\bX$. The resulting distributions of $\eta$, $\eta_z$, and $\eta - \eta_z$ vary with these parameters, producing changes in location, spread, and shape. Representative examples appear in Figure~\ref{fig:dist_logistic} and in Supplementary Text~\ref{app:distribution_plots}.

\begin{figure}[ht!]
\centering
\includegraphics[width=.9\textwidth]{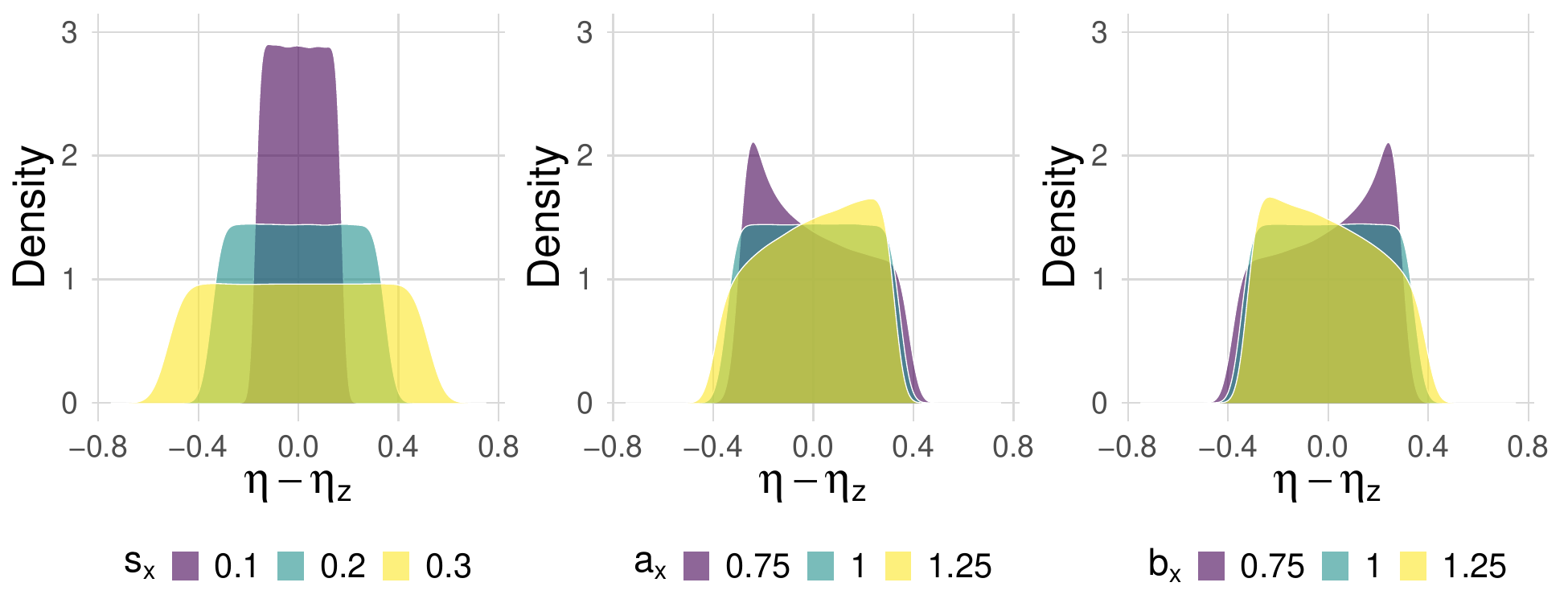}
\caption{Simulated distributions of $\eta - \eta_z$ for logistic regression, as we vary parameters affecting $\bbe'\bX$ ($=c_2 B_x$): standard deviation $s_x$ and shape parameters $a_x$ and $b_x$. Unless noted, we fix $a_x = b_x = a_z = b_z = 1$, $s_x = s_z = 0.2$, $\rho = 0$, and $g^{-1}(\iota) = 0.25$.}
\label{fig:dist_logistic}
\end{figure}

Because relative error is defined through expectations over $(Y, \bX, \bZ)$, it lacks a closed-form expression. We therefore approximate these expectations using Monte Carlo integration with 50,000 samples per setting. All simulations were performed in R (version 4.4.3) using code available at \href{https://github.com/cochran4/glm_pss}{https://github.com/cochran4/glm\_pss}. 

\subsection{Logistic regression}

\begin{figure}[ht!]
\centering
\includegraphics[width=.9\textwidth]{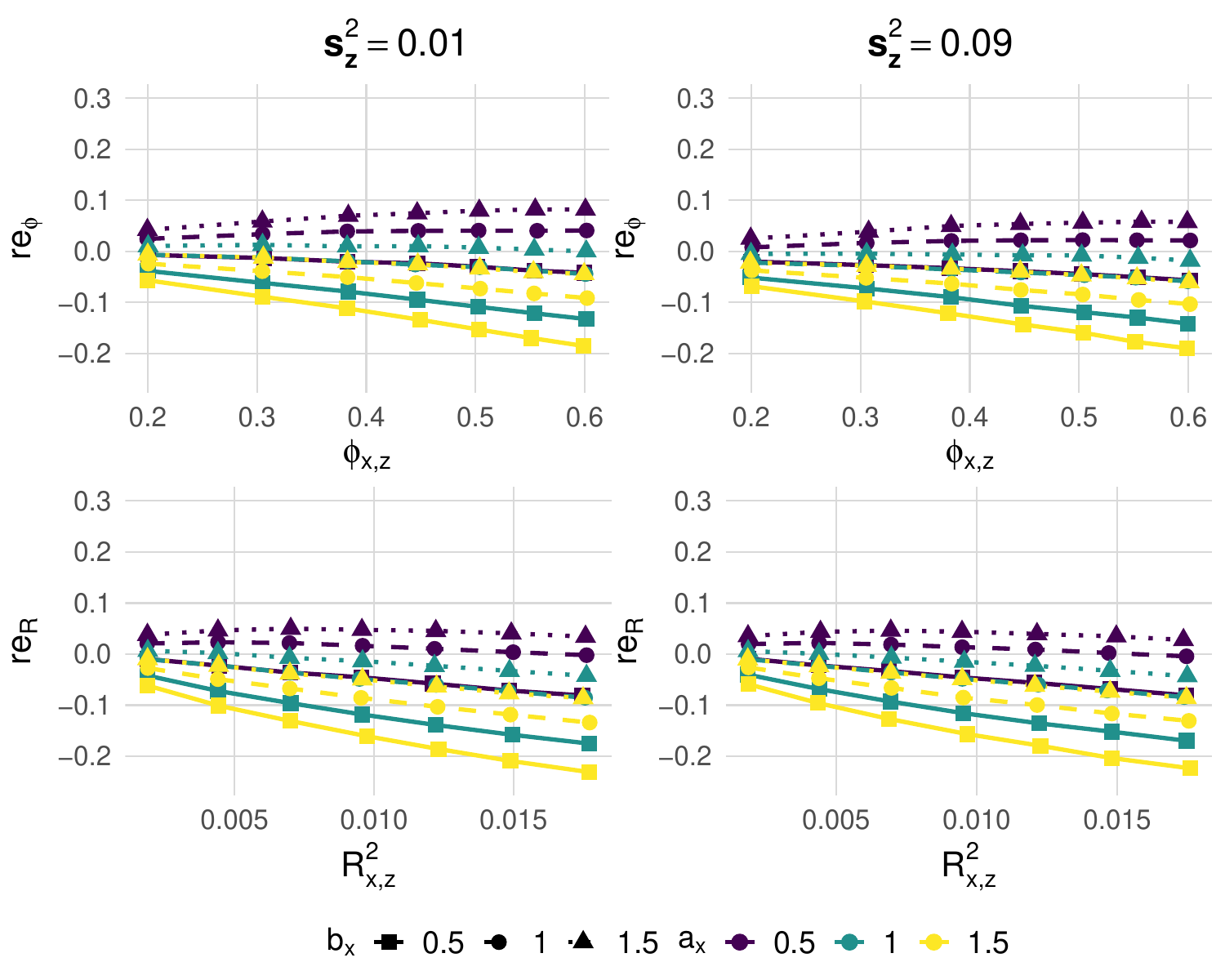}
\caption{Relative error for logistic regression, plotted against $\phixz$ (top panels) and $\pspRsq$ (bottom panels). Left and right panels correspond to two levels of $s_z^2$. Within each panel, we vary $a_x$ and $b_x$ over all combinations of values in $\{0.5, 1, 1.5\}$. Each point reflects a value of $s_x^2$, evenly spaced from $0.01$ to $0.09$. Other parameters are fixed: $a_z = b_z = 1$, $\rho = 0$, and $g^{-1}(\iota) = .25$.}
\label{fig:re_logistic}
\end{figure}

Figure~\ref{fig:re_logistic} shows relative error for $\fsqphi$ and $\fsqR$ in logistic regression, fixing $a_z=b_z=1$, $\rho=0$, and $g^{-1}(\iota)=.25$. Relative error increases as $\phixz$ and $\pspRsq$ grow, with an approximately linear relationship. Errors can reach about -20\%, meaning the approximations overestimate $\fsq$ and thus overestimate power. Recall from Table~\ref{tab:power_error} that a relative error of -15\% corresponds, at worst, to a 7.1 percentage point overestimate of power for $\alpha=.05$ and $p=1$.

These results align with our theoretical findings: $\text{re}_R$ is near zero for small $s_x^2$, and $\text{re}_\phi$ is near zero when both $s_x^2$ and $s_z^2$ are small. The relative error $\text{re}_R$ changes little as $s_z^2$ increases from $0.01$ to $0.09$, whereas $\text{re}_\phi$ shifts slightly downward, becoming more negative at larger $s_z^2$. The shape parameters of $B_x$ ($a_x$, $b_x$) also strongly affect relative error. Negatively skewed $B_x$ ($a_x>b_x$) produces negative errors and overestimates power, while positively skewed $B_x$ ($b_x>a_x$) produces positive errors and underestimates power; these directions would likely reverse if $g^{-1}(\iota)=.75$.


Supplementary Text~\ref{app:extra_simulations} examines relative error while varying $a_z$, $b_z$, $\rho$, and $g^{-1}(\iota)$, showing these have far less influence than the parameters governing $B_x$. A formal sensitivity analysis confirms that $s_x^2$ and the shape parameters of $B_x$ are the primary drivers of relative error (Supplementary Text~\ref{app:sensitivity}).

\subsection{Bernoulli distribution with identity link}

\begin{figure}[ht!]
\centering
\includegraphics[width=.9\textwidth]{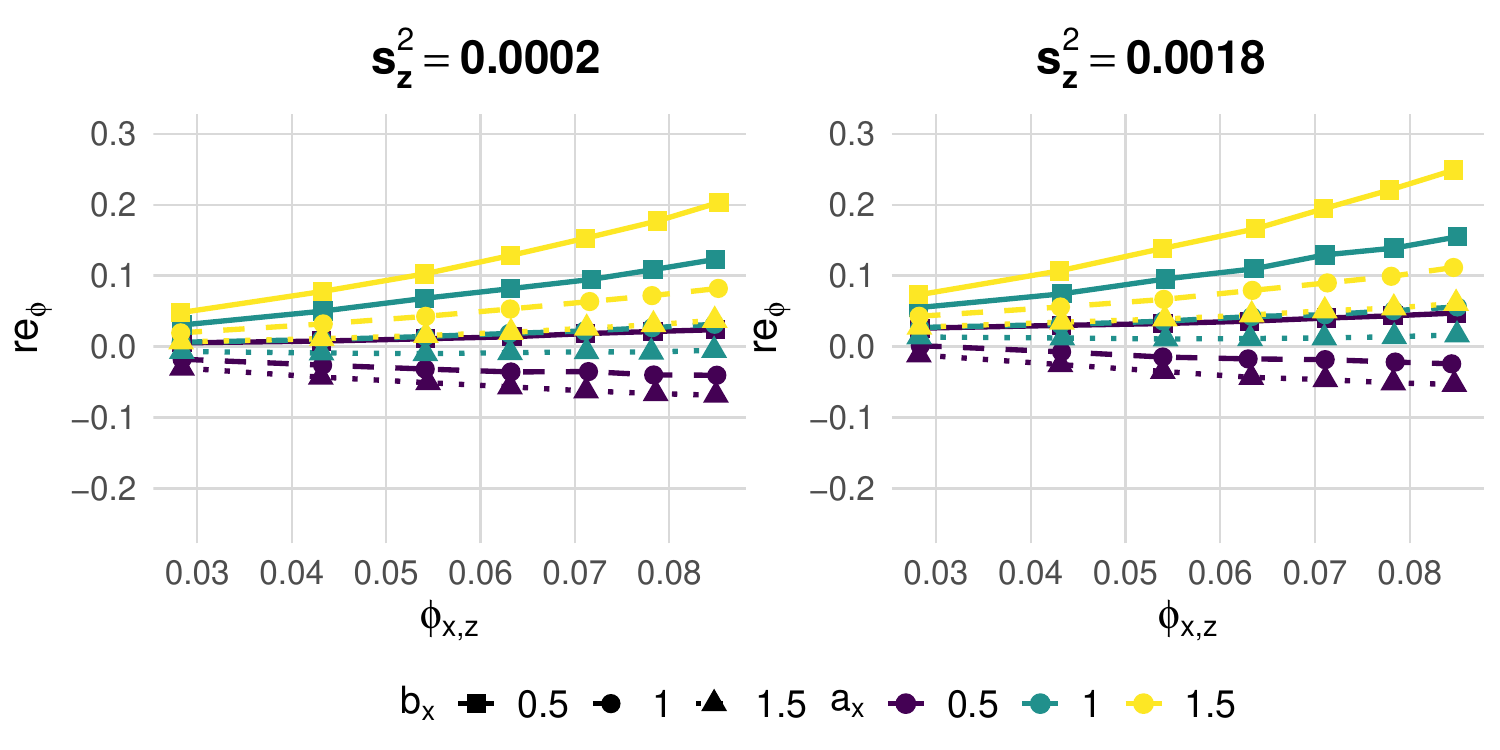}
\caption{Relative error $\text{re}_\phi$ for a Bernoulli distribution and identity link, plotted against $\phixz$; $\text{re}_R$ is not shown because it is zero. Left and right panels correspond to two levels of $s_z^2$. Within each panel, we vary $a_x$ and $b_x$ over all combinations of values in $\{0.5, 1, 1.5\}$. Each point reflects a value of $s_x^2$, evenly spaced from $0.0002$ to $0.0018$. Other parameters are fixed: $a_z = b_z = 1$, $\rho = 0$, and $g^{-1}(\iota) = .25$.}
\label{fig:re_logistic_ident}
\end{figure}

Figure~\ref{fig:re_logistic_ident} shows relative error for $\fsqphi$ under a Bernoulli model with an identity link (a linear probability model), fixing $a_z=b_z=1$, $\rho=0$, and $g^{-1}(\iota)=.25$. For $\fsqR$, the relative error is zero because of the identity link (Corollary~\ref{corr:error_R}). The relative error $\text{re}_\phi$ increases roughly linearly with $\phixz$, with slope determined by the shape parameters $a_x$ and $b_x$: it is most positive when $B_x$ is negatively skewed ($a_x=1.5$, $b_x=0.5$) and most negative when $B_x$ is positively skewed ($a_x=0.5$, $b_x=1.5$). Relative error varies little with the shape of $B_z$ or the correlation $\rho$, but does vary with $g^{-1}(\iota)$, with larger errors when this parameter is small (Supplementary Text~\ref{app:extra_simulations}). In such cases, however, an identity link is less appropriate for modelling a binary outcome than when the mean is close to~0.5.

\subsection{Poisson distribution with log link}

\begin{figure}[ht!]
\centering
\includegraphics[width=.9\textwidth]{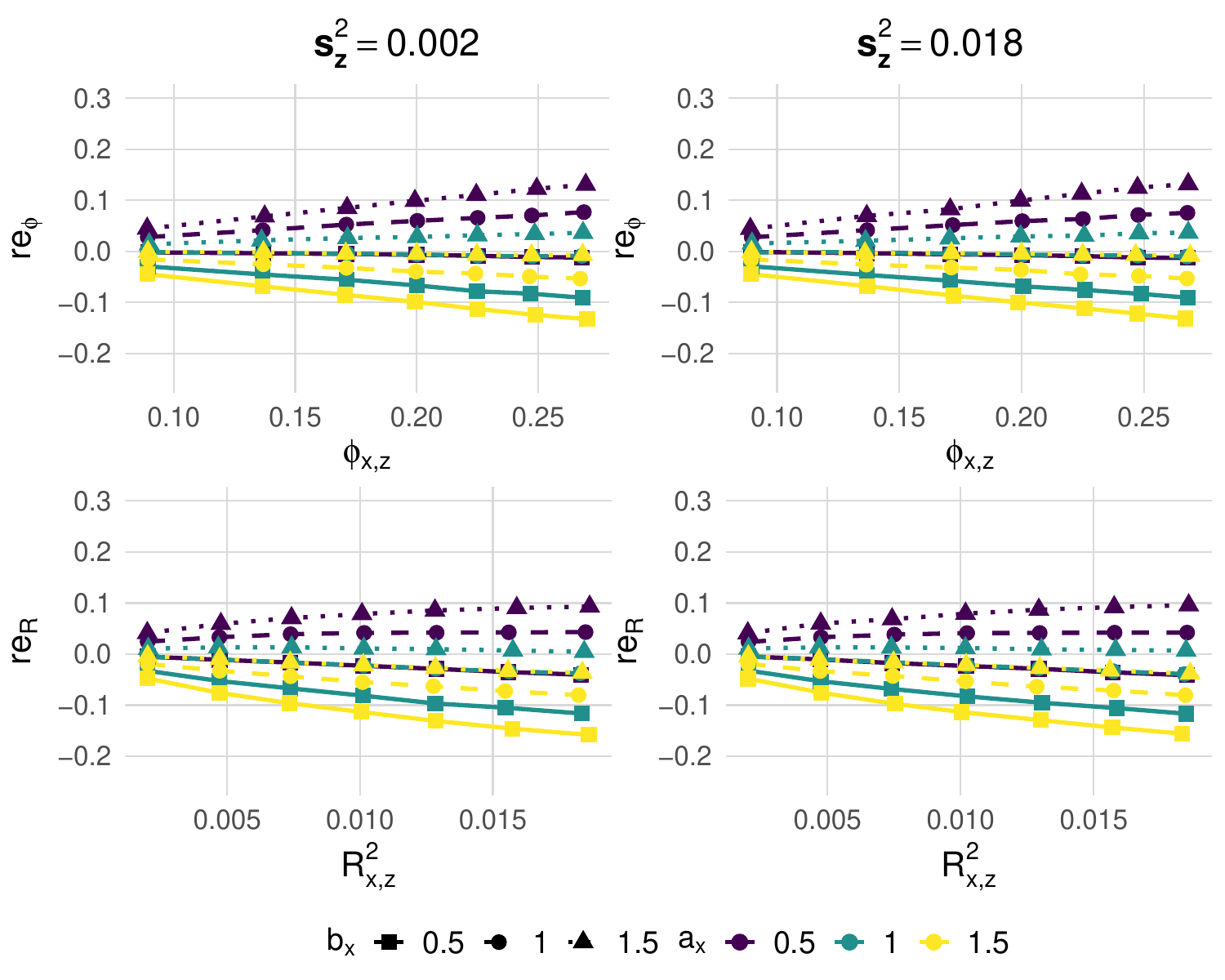}
\caption{Relative error for a Poisson distribution with a log link, plotted against $\phixz$ (top panels) and $\pspRsq$ (bottom panels). Left and right panels correspond to two levels of $s_z^2$. Within each panel, we vary $a_x$ and $b_x$ over all combinations of values in $\{0.5, 1, 1.5\}$. Each point reflects a value of $s_x^2$, evenly spaced from $0.002$ to $0.018$. Other parameters are fixed: $a_z = b_z = 1$, $\rho = 0$, and $g^{-1}(\iota) = 1$.}
\label{fig:re_poisson}
\end{figure}

Our next example is Poisson regression with a log link. Figure~\ref{fig:re_poisson} shows relative errors, fixing $a_z = b_z = 1$, $\rho = 0$, and $g^{-1}(\iota)=1$. As in earlier GLMs, both $\text{re}_\phi$ and $\text{re}_R$ are roughly linear in $\phixz$ and $\pspRsq$, with slopes determined by the shape parameters $a_x$ and $b_x$. Slopes are most positive when $B_x$ is positively skewed ($a_x=0.5$, $b_x=1.5$) and most negative when it is negatively skewed ($a_x=1.5$, $b_x=0.5$). As in logistic regression, relative errors are less sensitive to $B_z$'s shape parameters ($a_z$, $b_z$), to $\rho$, and to $g^{-1}(\iota)$ than to the parameters governing $B_x$ (Supplementary Text~\ref{app:extra_simulations}).

\subsection{Gamma distribution with log link}

\begin{figure}[ht!]
\centering
\includegraphics[width=.9\textwidth]{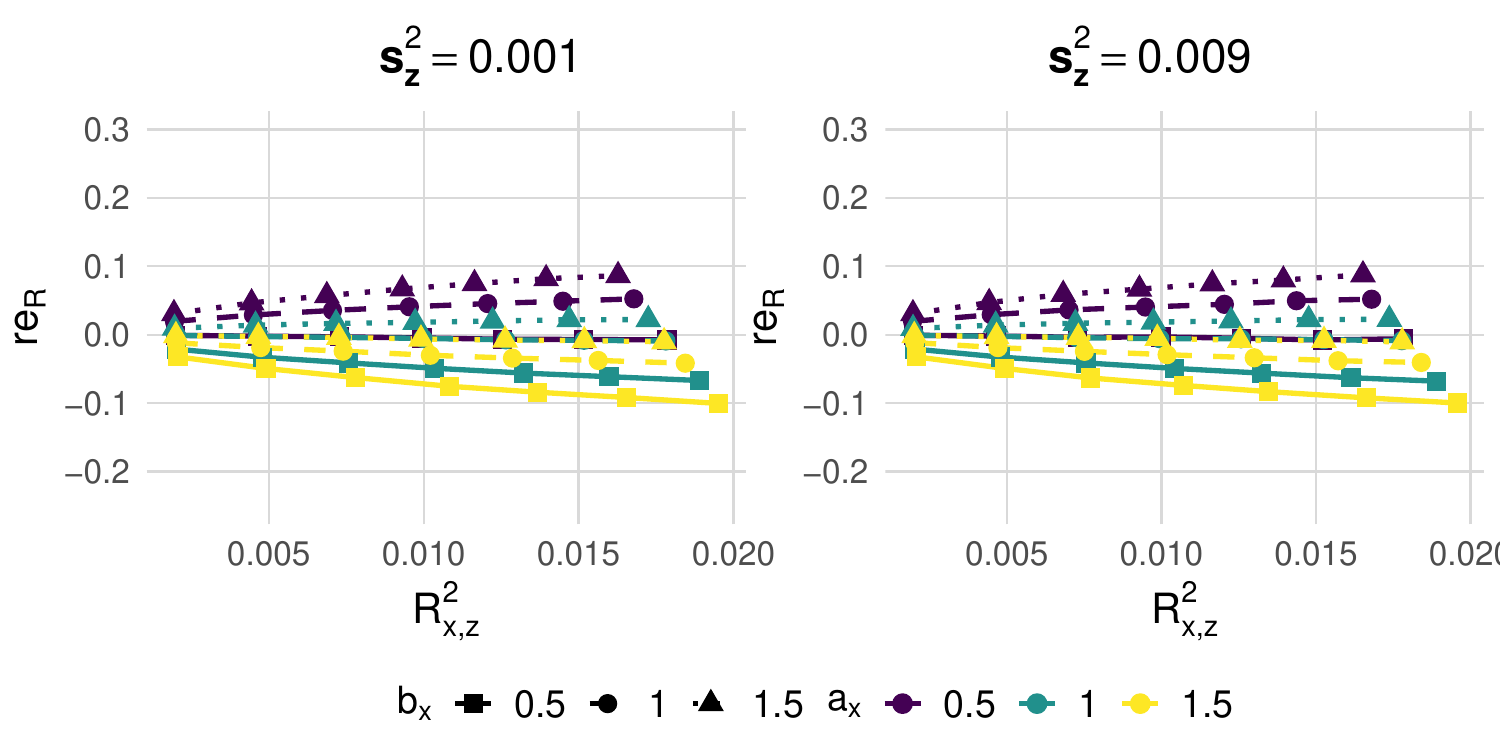}
\caption{Relative error $\text{re}_R$ for a Gamma model with a log link, plotted against $\pspRsq$; $\text{re}_\phi$ is not shown because it is zero. Left and right panels correspond to two levels of $s_z^2$. Within each panel, we vary $a_x$ and $b_x$ over all combinations of values in $\{0.5, 1, 1.5\}$. Each point reflects a value of $s_x^2$, evenly spaced from $0.001$ to $0.009$. Other parameters are fixed: $a_z = b_z = 1$, $\rho = 0$, $g^{-1}(\iota) = 4$; shape parameter is $2$.}
\label{fig:re_gamma}
\end{figure}

Our final example is Gamma regression with a log link. Figure~\ref{fig:re_gamma} shows relative error for $\fsqR$, fixing $a_z=b_z=1$, $\rho=0$, and $g^{-1}(\iota)=1$. For $\fsqphi$, the relative error is zero because the weight term $w$ is constant (Corollary~\ref{corr:error_phi}). As with the other GLMs, $\text{re}_R$ is roughly linear in $\pspRsq$, with slope determined by the shape parameters $a_x$ and $b_x$: most positive when $B_x$ is positively skewed ($a_x=0.5$, $b_x=1.5$) and most negative when negatively skewed ($a_x=1.5$, $b_x=0.5$). Relative error is less sensitive to $a_z$, $b_z$, $\rho$, and $g^{-1}(\iota)$ than to the parameters governing $B_x$ (Supplementary Text~\ref{app:extra_simulations}).

\section{Case study}

We demonstrate how our approach can be applied while also assessing the accuracy of our approximations in finite samples. Using data from the 2023 National Surveys on Drug Use and Health \citep{usdhhs2018nsduh}, we focus on adults who experienced major depression in the past year. Table~\ref{tab:case_study_summary} in the Supplement summarizes key dataset characteristics. Outcomes include receipt of any mental health treatment (yes/no), the number of treatment types used (0--4), and functional impairment via the Sheehan Disability Scale (SDS; 0--40). These are analysed using GLMs with Bernoulli, Poisson, and Gamma distributions and logit, identity, and log links, with sociodemographic variables included.

Each example assumes the empirical distribution for $(\bX,\bZ)$ and treats the fitted GLM coefficients $\bbe$ and $\blam$ as the true values. For the Gamma model, we also estimate a shape parameter $k$. Results are summarized in Supplementary Text~\ref{app:case_study}.

\subsection{Effect sizes for logistic regression}

Studies with a binary outcome and a multi-level predictor, such as race or education, are often analysed with logistic regression. Here, we use education (Less than High School, High School, Some College/Associate Degree, College) to examine whether differences in education explain variability in mental health treatment.

To illustrate, let $Y$ indicate whether an individual received any mental health treatment in the past year. The predictor $\bX$ includes three binary indicators for education, using `College' as the reference. We model $Y$ with logistic regression:
\begin{align*}
    \text{logit}(\E[Y|\bX]) = \blam + \bbe'\bX.
\end{align*}
The null hypothesis is that treatment rates do not differ across groups, i.e., $\bbe = 0$. 

Based on our assumptions, under the alternative hypothesis, we have
\begin{align*}
    \begin{array}{c c c c c c}
        \bbe' = & \big[ & -0.195 &  -0.620 & -0.612 & \big], \\
    && \shortstack{\textit{Some} \\ \textit{college}} & \shortstack{ \textit{HS} \\ {\color{white} fill} } & \shortstack{$<$ \textit{HS} \\ {\color{white} fill} }
    \end{array}
\end{align*}
showing that those with at least some college education are more likely to receive mental health treatment than those with only a high school education. The effect size measures are $\phixz = 0.51$ ($e^{\phixz}=1.67$) and $\pspRsq = 0.015$. For a binary $\bX$ with mean 0.5, this $\phixz$ corresponds to an odds ratio of 1.67. We also find $\fsq = 0.0149$, compared to $\fsqphi = 0.0149$ ($\text{re}_\phi = 0.2\%$) and $\fsqR = 0.0151$ ($\text{re}_R = 1.7\%$), indicating that both approximations slightly overestimate $\fsq$.

Going further, we may ask whether education affects $Y$ beyond what is explained by other variables such as age and sex at birth. This requires updating the model to include $\bZ$, consisting of a constant, age (3 levels), and sex (binary):
\begin{align*}
    \text{logit}(\E[Y|\bX,\bZ]) = \blam'\bZ + \bbe'\bX.
\end{align*}
As expected, education's effect on treatment receipt has diminished:
\begin{align*}
    \begin{array}{c c c c c c}
        \bbe' = & \big[ & -0.121 &  -0.475 & -0.468 & \big]. \\
    && \shortstack{\textit{Some} \\ \textit{college}} & \shortstack{ \textit{HS} \\ {\color{white} fill} } & \shortstack{$<$ \textit{HS} \\ {\color{white} fill} }
    \end{array}
\end{align*}
Our effect size measures also diminished: $\phixz = 0.38$ ($e^{\phixz} = 1.47$) and $\pspRsq=.008$. This yields $\fsq = 0.0083$, which is well-approximated by $\fsqphi = 0.0084$ ($\text{re}_\phi = 0.9\%$) and $\fsqR = 0.0083$ ($\text{re}_R = -0.7\%$).

Another common question is whether a factor and its interaction with another variable add explanatory value. For example, we might ask whether education and its interactions with sex help explain receipt of mental health treatment beyond sex and age. In our prior model, this corresponds to expanding $\bX$ to include sex–education interactions and testing whether these terms jointly contribute to the model.

We observe moderate effects of education for males and small interaction effects:
\begin{align*}
    \begin{array}{c c c c c c c c c}
        \bbe' = & \big[ & -0.078 &  -0.438 & -0.449 & 0.193 & 0.192 &  0.074 & \big], \\
    && \shortstack{\textit{Some} \\ \textit{college}} & \shortstack{ \textit{HS} \\ {\color{white} fill} } & \shortstack{$<$ \textit{HS} \\ {\color{white} fill} } & \shortstack{\textit{Some college:} \\ \textit{female}} & \shortstack{\textit{HS:} \\\textit{female}} & \shortstack{$<$ \textit{HS:} \\ \textit{female}}
    \end{array}
\end{align*}
showing that education has a stronger effect on treatment receipt for males than females. The overall effect size measures for education and its interaction with sex are $\phixz = 0.40$ ($e^{\phixz}=1.50$) and $\pspRsq = 0.009$. This gives $\fsq = 0.0090$, close to its approximations: $\fsqphi = 0.0093$ ($\text{re}_\phi = 3.6\%$) and $\fsqR = 0.0091$ ($\text{re}_R = 0.7\%$). These effect sizes are similar to those for education alone after adjustment, suggesting that including a sex-by-education interaction adds little.

\subsection{Other examples of effect sizes}

We repeated the examples above using other GLMs and outcomes. Focusing on the effect size measures for education adjusted for age and sex, the next example uses a linear probability model (Bernoulli with an identity link), keeping the outcome and distribution unchanged. The regression terms are now
\begin{align*}
    \begin{array}{c c c c c c}
        \bbe' = & \big[ & -0.025 &  -0.106 & -0.105 & \big], \\
    && \shortstack{\textit{Some} \\ \textit{college}} & \shortstack{ \textit{HS} \\ {\color{white} fill} } & \shortstack{$<$ \textit{HS} \\ {\color{white} fill} }
    \end{array}
\end{align*} 
confirming that education has a positive effect on treatment receipt. The change in link function places these terms on a probability scale rather than a log-odds scale. Accordingly, the first effect-size measure is also on this scale: $\phixz = 0.09$, equivalent to a 9 percentage point change in treatment probability for a binary predictor with mean 0.5. The scale of the second measure does not change: $\pspRsq = 0.008$, similar to its logistic-regression value. With the identity link, $\pspRsq$ perfectly recovers $\fsq$: $\fsq = \fsqR = 0.0082$, compared to $\fsqphi = 0.0082$ ($\text{re}_\phi = -0.6\%$). 


Next, we change the outcome to the number of mental health treatment types received. As a count variable, it is modeled using a Poisson distribution with a log link. The regression terms, now on the log scale, are
\begin{align*}
    \begin{array}{c c c c c c}
        \bbe' = & \big[ & -0.042 &  -0.208 & -0.175 & \big], \\
    && \shortstack{\textit{Some} \\ \textit{college}} & \shortstack{ \textit{HS} \\ {\color{white} fill} } & \shortstack{$<$ \textit{HS} \\ {\color{white} fill} }
    \end{array}
\end{align*} 
demonstrating that education increases the number of treatment types. The first effect size measure is $\phixz = 0.117$ ($e^{\phixz}=1.18$), equivalent to a 1.18-fold increase in the average number of treatment types for a binary predictor with mean 0.5. The second measure is $\pspRsq = 0.010$. We also find $\fsq = 0.0097$, which is slightly smaller than its approximations: $\fsqphi = 0.0103$ ($\text{re}_\phi = 5.8\%$) and $\fsqR = 0.0103$ ($\text{re}_R = 6.3\%$).


In our final example, we use the SDS total score (positively skewed; range 0-40), modelling it with a Gamma distribution and a log link. To avoid taking the log of zero, we shift scores by 0.5. The regression terms, on a log scale, are
\begin{align*}
    \begin{array}{c c c c c c}
        \bbe' = & \big[ & 0.072 &  0.079 & 0.076 & \big]. \\
    && \shortstack{\textit{Some} \\ \textit{college}} & \shortstack{ \textit{HS} \\ {\color{white} fill} } & \shortstack{$<$ \textit{HS} \\ {\color{white} fill} }
    \end{array}
\end{align*} 
demonstrating college education is associated with lower functional impairment, as measured by SDS scores. Our first measure is $\phixz = 0.06$ ($e^{\phixz} = 1.07$), equivalent to a 1.07-fold increase in mean SDS score for a binary predictor with mean 0.5. Our second measure is $\pspRsq = 0.006$. Because $w$ is constant for this GLM, $\phixz$ exactly recovers $\fsq$: $\fsq = \fsqphi = 0.0064$. By contrast, $\fsqR = 0.0066$ is larger ($\text{re}_R = 2.5\%$).

\subsection{Power for finite samples}

To close the case study, we evaluate how well our approximations support PSS calculations in finite samples. We return to the 12 model configurations and rescale $\bbe$ to fix $\fsq = 0.02$ for comparability. For each configuration, we draw $n$ observations of $(\bX,\bZ)$ from the empirical distribution with replacement, compute the linear predictor $\eta$, and generate $Y$ from the GLM. This produces a simulated dataset of size $n$, to which we fit the GLM and perform a Wald test of $\bbe=0$. Repeating this 2,000 times, we compute actual power as the rejection proportion and compare it to the asymptotic power predicted by $\fsq$ and its approximations.

\begin{figure}[ht!]
    \centering
    \includegraphics[width=.9\textwidth]{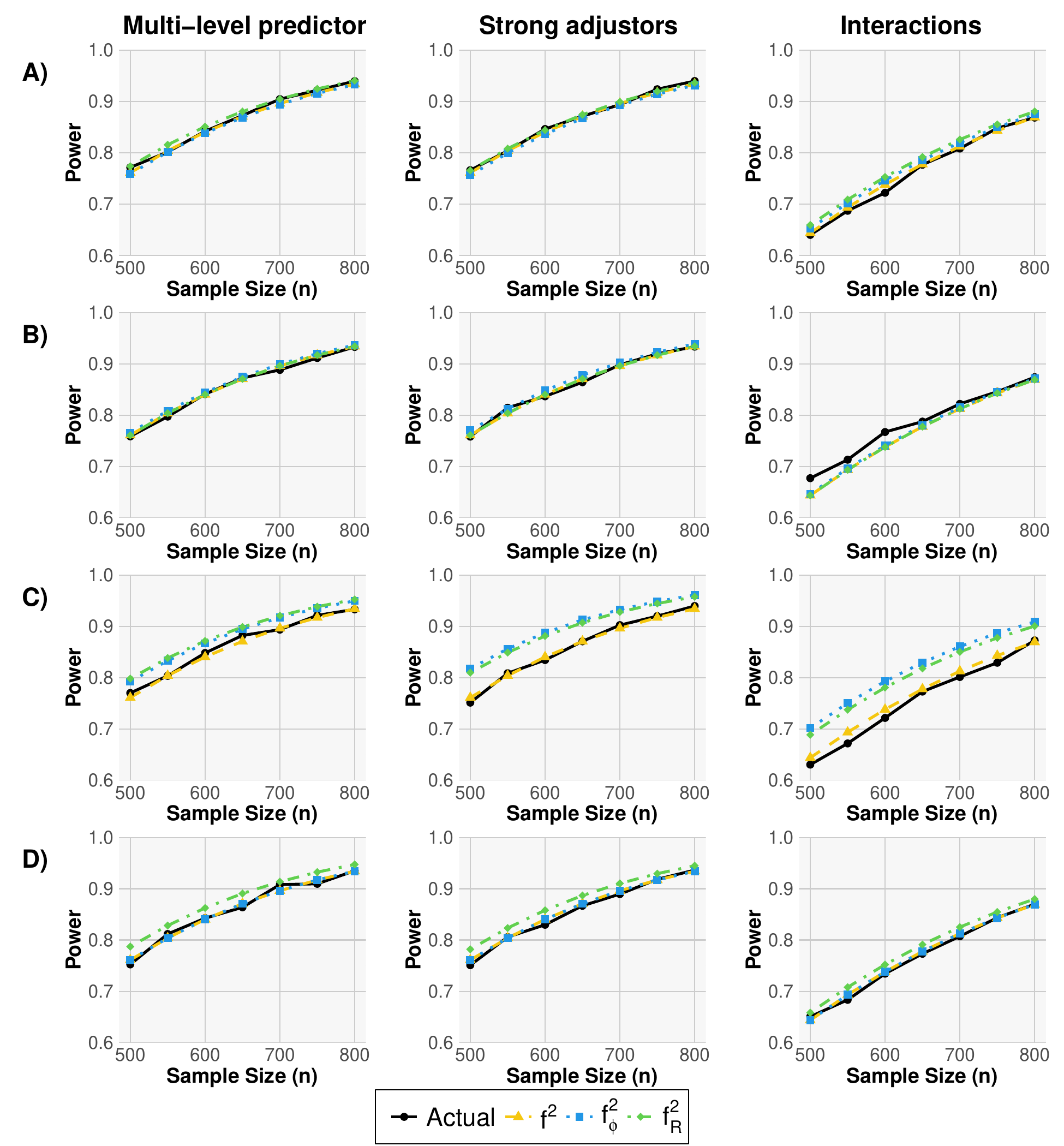}

    \caption{Comparison of actual power in a finite sample against power predicted using $\fsq$ and its approximations ($\fsqphi$ and $\fsqR$). Outcomes include: receipt of mental health treatment using  \textbf{A)} logistic regression and  \textbf{B)} a Bernoulli distribution and identity link;  \textbf{C)}  number of mental health treatments using a Poisson distribution and log link, and  \textbf{D)}  functional impairment, via Sheehan Disability Scale total score, using a Gamma distribution and log link. Panels correspond to modelling scenarios with multi-level predictors, strong adjustors, or interactions.}
    \label{fig:case_study}
\end{figure}

Figure~\ref{fig:case_study}A compares finite-sample power with predicted power for logistic regression, using receipt of mental health treatment as the outcome. Sample sizes range from $n=500$–$800$, yielding power in the typical 70\%–90\% range. The panels show models with education as a multi-level predictor: unadjusted, adjusted for age and sex, and further including interactions with sex. Finite-sample power aligns closely with predictions from $\fsq$, $\fsqphi$, and $\fsqR$, with $\fsq$ showing the best agreement.

Similar conclusions hold in Figure~\ref{fig:case_study}B, which uses a Bernoulli model with an identity link. Predicted powers again closely track actual finite-sample power across all three scenarios. For the Poisson model with a log link (Figure~\ref{fig:case_study}C), both approximations overestimate power, though the asymptotic calculation using $\fsq$ remains reasonably accurate. In the Gamma model with a log link (Figure~\ref{fig:case_study}D), $\pspRsq$ approximation overestimates power, while both the asymptotic calculation and the $\fsqphi$ approximation perform well, with the latter exactly matching $\fsq$ for this GLM.

\section{Conclusion}
\label{sec:conc}

We introduced two new effect size measures to simplify PSS calculations for Wald tests in GLMs. The first, $\phixz$, is defined on the linear predictor scale and reduces to familiar quantities, like the log odds ratio in logistic regression or the log rate ratio in Poisson regression for binary predictors. The second, $\pspRsq$, captures the proportion of MSE explained by predictors beyond adjustors, scaled by the modelled SD of the outcome. Both provide approximations to the noncentrality parameter required for power under the alternative. Unlike methods that rely on strong distributional assumptions or GLM-specific formulas, our approach adapts the strengths of linear regression (interpretable effect sizes, minimal assumptions, and flexibility with multiple predictors and adjustors) \citep{gatsonis1989multiple} to any GLM.

In the asymptotic regime where coefficients shrink toward zero and sample size increases to maintain fixed power, $\phixz$ approximates $f^2$ well when both predictor and adjustor effects are small, whereas $\pspRsq$ requires only predictor effects to be small. The approximations become exact when the weight term $w$ is constant for $\phixz$ or when the link is linear for $\pspRsq$. In linear regression, they reduce to the familiar partial $f^2$ and partial $R^2$, making them natural generalizations. Simulations show that accuracy is driven largely by the variance and skewness of $\bbe'\bX$, a limitation that cannot be overcome without higher-order distributional information.

In a case study predicting treatment receipt from education, $\pspRsq$ produced the same value of $0.015$ under both logistic and identity links. This stability reflects its anchoring to the response scale, giving it a consistent interpretation across link functions. In contrast, $\phixz$ varied considerably (e.g., $0.51$ for logit vs.\ $0.12$ for identity), since it depends on the linear predictor scale and therefore on the chosen link. Both measures behaved as expected: strong adjustors like age and sex reduced their values, while additional predictors increased them. Their performance was consistent across GLMs and insensitive to sample size, indicating that the main challenge lies in approximating $\fsq$, not issues related to asymptotic assumptions.

The astute reader may notice that we did not include benchmarks, even though power and sample size is a long-standing area. No general method can accommodate the combination of multiple predictors, adjustors, and GLMs we consider, aside from approaches requiring full specification of the joint distribution. But full specification yields $\fsq$ directly, precisely the quantity we aim to approximate. We therefore compare our measures to $\fsq$ itself, using relative error to evaluate $\fsqphi$ and $\fsqR$.

Our approach has limitations. Although it applies to standard GLMs, extending it to other settings is important. We focused on Wald tests, neglecting potentially more robust alternatives such as score or likelihood ratio tests. We also assumed fully specified GLMs, excluding quasi-likelihood models and unknown dispersions. Using an $F$ distribution could improve robustness by accounting for variance estimation. Moreover, our approximations can be compromised by skewness in $\bbe'\bX$, which cannot be resolved without additional higher-order moment information.

Lastly, although our measures are designed to be interpretable, their practical value depends on whether statisticians can work with study teams to extract the information needed to compute them. Doing so may demand more effort than in linear regression, such as explaining how to residualize adjustors on a \textit{weighted} rather than \textit{unweighted} scale. Developing these practices is not the main focus of this paper; our goal has been to introduce the measures and study their behaviour in simulations and asymptotically. We cannot yet guarantee that these practical steps will always succeed, only that they require deliberate effort and form a key direction for future work, including clearer guidelines, software, and applied examples.


\bigskip
\begin{center}
{\large\bf SUPPLEMENTARY MATERIAL}
\end{center}

\begin{description}

\item[Supplementary text:] This text includes detailed derivations of key equations, proofs of theorems, additional simulation results including a parameter sensitivity analysis, and case study models.  (.pdf file)

\end{description}

\section*{Acknowledgements}

This work is supported by NIH grant R21MH133371 to author Paul J Rathouz.

\newpage

\begin{center}
\large \bf Supplementary Text for ``General measures of effect size to calculate power and sample size for Wald tests with generalized linear models"
\end{center}

\setcounter{equation}{0}
\setcounter{figure}{0}
\setcounter{table}{0}
\setcounter{assumption}{0}
\setcounter{lemma}{0}
\setcounter{page}{1}
\makeatletter
\renewcommand{\theassumption}{S\arabic{assumption}}
\renewcommand{\thelemma}{S\arabic{lemma}}
\renewcommand{\theequation}{S\arabic{equation}}
\renewcommand{\thefigure}{S\arabic{figure}}
\renewcommand{\thetable}{S\arabic{table}}
\renewcommand{\bibnumfmt}[1]{[S#1]}
\renewcommand{\citenumfont}[1]{S#1}

\spacingset{1.05} 

\appendix

\section{Weight terms for specific GLMs} \label{app:weight}

A key variable in our paper is the weight term $w$, defined as:
\begin{align*}
 w := \frac{1}{v} \left( \frac{\partial \mu}{\partial \eta} \right)^2.
\end{align*}
Here, $v$ is the conditional variance of $Y$ given $\bX$ and $\bZ$, which depends on the distribution. For instance, in the binomial case, $v = \mu(1-\mu)$. The term $\partial \mu/\partial \eta$ represents the derivative of the inverse link function $g^{-1}(\eta)$ with respect to the linear predictor $\eta$, which depends on the chosen link function. Table~\ref{tab:weights} presents $w$ in terms of $\mu$ and relevant parameters for common GLMs, along with corresponding values of $v$ and $\partial \mu/\partial \eta$.
\begin{table}[!ht]
 \caption{Comparison of weight terms $w$ for common GLM distributions and link functions. Since $w$ depends on the variance $v$ and the derivative $\partial \mu / \partial \eta$, both are included for clarity.}
    \label{tab:weights}
\begin{tabular}{lllll}
\toprule
\textbf{Distribution} & \textbf{Link} & $\boldsymbol{v}$ 
& $\boldsymbol{\frac{\partial\mu}{\partial\eta}}$
& $\boldsymbol{w}$ \\
\midrule
\textbf{Normal (variance $\boldsymbol{\sigma^2}$)} 
& Identity 
& $\sigma^2$ 
& $1$ 
& $\frac{1}{\sigma^2}$ \\

\midrule

\textbf{Binomial} 
& Identity 
& $\mu(1-\mu)$ 
& $1$ 
& $\frac{1}{\mu(1-\mu)}$ \\

& Logit 
& $\mu(1-\mu)$ 
& $\mu(1-\mu)$ 
& $\mu(1-\mu)$ \\

& Log 
& $\mu(1-\mu)$ 
& $\mu$ 
& $\frac{\mu}{1-\mu}$ \\

\midrule

\textbf{Poisson} 
& Identity 
& $\mu$ 
& $1$ 
& $\frac{1}{\mu}$ \\

& Log 
& $\mu$ 
& $\mu$ 
& \(\mu\) \\

& Inverse 
& $\mu$ 
& $\mu^2$ 
& $\mu^3$ \\
\midrule

 \textbf{Gamma (shape $\boldsymbol{k}$)}
& Identity 
& $\mu^2/k$ 
& $1$ 
& $\frac{k}{\mu^2}$ \\

& Log 
& $\mu^2/k$ 
& $\mu$ 
& $k$ \\

& Inverse 
& $\mu^2/k$ 
& $\mu^2$ 
& $k\mu^2$ \\
\midrule

\textbf{Inverse Gaussian (shape $\boldsymbol{k}$)}
& Identity 
& $\mu^3/k$ 
& $1$ 
& $\frac{k}{\mu^3}$ \\

& Log 
& $\mu^3/k$ 
& $\mu$ 
& $\frac{k}{\mu}$ \\

& Inverse 
& $\mu^3/k$ 
& $\mu^2$ 
& $k\mu$ \\
\bottomrule
\end{tabular}
\end{table}

\section{Detailed derivations} \label{app:derivations}

This section provides details for several key equations presented in the main text.

\vspace{1em}
\noindent \textbf{Derivation of $\text{WMSE} = 1.$}

Recall the definition of WMSE:
\begin{align*}
    \text{WMSE} = \E[w(Y_l - \eta)^2].
\end{align*}
Substituting $Y_l = \eta + (\partial \eta / \partial \mu) (Y-\mu)$, we obtain
\begin{align*}
    \text{WMSE} = \E\left[w(Y-\mu)^2 \left( \frac{\partial \eta}{\partial \mu} \right)^2\right].
\end{align*}
Since $w = \left( \frac{\partial \mu}{\partial \eta} \right)^2 / v$, this simplifies to
\begin{align*}
    \text{WMSE} = \E[(Y-\mu)^2 / v].
\end{align*}
Applying iterated expectations:
\begin{align*}
    \E[(Y-\mu)^2 / v] 
    &=  \E\left[ \E[(Y-\mu)^2 / v \mid \bX, \bZ] \right]  \\
    &= \E\left[ (1/v) \E[(Y-\mu)^2  \mid \bX, \bZ] \right].
\end{align*}
By definition of $v$, we have $\E[(Y-\mu)^2  \mid \bX, \bZ] = v$, yielding
\begin{align*}
    \text{WMSE} = \E[(1/v) v] = 1.
\end{align*}

\vspace{1em}
\noindent \textbf{Derivation of $\eta_z = \E[w \eta \bZ'] \E[w \bZ \bZ']^{-1} \bZ$}.

The variable $\eta_z$ is defined as the best linear predictor of $Y_l$ given $\bZ$ under the weighted mean squared error (WMSE) criterion:
\begin{align*}
    \E \left[ w (Y_l - \eta_z)^2 \right] = \min_{\kappa \in \mathbb{R}^q} \E \left[ w (Y_l - \kappa' \bZ)^2 \right].
\end{align*}
To find the minimizer, we differentiate and set the gradient to zero:
\begin{align*}
    \frac{\partial}{\partial \kappa} \E \left[ w (Y_l - \kappa' \bZ)^2 \right] = 0,
\end{align*}
which simplifies to the orthogonality condition:
\begin{align*}
    \E \left[ w (Y_l - \kappa' \bZ) \bZ \right] = 0.
\end{align*}
Rearranging gives:
\begin{align*}
    \E \left[ w Y_l \bZ \right] = \E \left[ w \bZ \bZ' \right] \kappa.
\end{align*}
Substituting $Y_l = \eta + (\partial \eta / \partial \mu) (Y - \mu)$, we obtain:
\begin{align*}
    \E \left[ w Y_l \bZ \right] = \E \left[ w \eta \bZ \right] + \E \left[ w (\partial \eta / \partial \mu) (Y - \mu) \bZ \right].
\end{align*}
Applying iterated expectations:
\begin{align*}
    \E \left[ w (\partial \eta / \partial \mu) (Y - \mu) \bZ \right] 
    &= \E \left[ \E \left[ w (\partial \eta / \partial \mu) (Y - \mu) \bZ \mid \bX, \bZ \right] \right] \\
    &= \E \left[ w (\partial \eta / \partial \mu) \bZ \, \E \left[ Y - \mu \mid \bX, \bZ \right] \right].
\end{align*}
Since $\E \left[ Y - \mu \mid \bX, \bZ \right] = 0$ by definition of $\mu$, the last expression is zero, leaving:
\begin{align*}
    \E \left[ w \eta \bZ \right] = \E \left[ w \bZ \bZ' \right] \kappa.
\end{align*}
Solving for $\kappa$:
\begin{align*}
    \kappa = \E \left[ w \bZ \bZ' \right]^{-1} \E \left[ w \eta \bZ \right].
\end{align*}
Thus, the best linear predictor is:
\begin{align*}
    \eta_z = \kappa' \bZ = \E \left[ w \eta \bZ' \right] \E \left[ w \bZ \bZ' \right]^{-1} \bZ.
\end{align*}

\vspace{1em}
\noindent \textbf{Derivation of $\text{WMSE}_0 - \text{WMSE} = \E[w(\eta-\eta_z)^2] = \fsq$.}

We start with the definition: 
\begin{align*} \text{WMSE}_0 = \E\left[ w (Y_l - \eta_z)^2 \right]. 
\end{align*} 
Expanding using $\eta$: 
\begin{align*} \text{WMSE}_0 &= \E\left[ w \left( Y_l - \eta + \eta - \eta_z \right)^2 \right] \\
&= \E\left[ w ( \eta - \eta_z)^2 \right] + 2 \E\left[ w (Y_l - \eta)(\eta - \eta_z) \right] + \E\left[ w (Y_l - \eta)^2 \right]. 
\end{align*}
Applying iterated expectations to the middle term: 
\begin{align*}
    \E\left[ w (Y_l - \eta )(\eta - \eta_z) \right] &= \E\left[ \E\left[ w (Y_l - \eta )(\eta - \eta_z) | \bX, \bZ \right] \right] \\
    &=\E\left[ w (\eta-\eta_z)(\partial \eta/\partial \mu) \E\left[ Y-\mu | \bX, \bZ\right] \right].
\end{align*}
Since $\E \left[ Y - \mu \mid \bX, \bZ \right] = 0$, this term vanishes, leaving:
\begin{align*}
\text{WMSE}_0 = \E\left[ w ( \eta - \eta_z)^2 \right] + \E\left[ w (Y_l - \eta )^2 \right].
\end{align*}
With 
\begin{align*}
\text{WMSE} = \E\left[ w (Y_l - \eta )^2 \right],
\end{align*}
we obtain
\begin{align*}
    \text{WMSE}_0 - \text{WMSE} = \E[w(\eta - \eta_z)^2].
\end{align*}
Now, we verify that $\E[w(\eta - \eta_z)^2]=\fsq$. By definition,
\begin{align*}
    \fsq &= \bbe'\mathcal{I}_{\bbe | \blam} \bbe \\
    &= \bbe'\left( \E[w \bX \bX'] - \E[w \bX \bZ'] \E[w \bZ \bZ']^{-1} \E[w \bZ \bX'] \right) \bbe.
\end{align*}
Using the identity:
\begin{align*}
    &\E\left[w\left(\bX-\E[w \bX \bZ'] \E[w \bZ \bZ']^{-1}\bZ \right)\left(\bX-\E[w \bX \bZ'] \E[w \bZ \bZ']^{-1}\bZ \right)'\right] \\
    &\quad = \E\left[w \bX \bX'\right]-\E[w \bX \bZ'] \E[w \bZ \bZ']^{-1}\E[w \bZ \bX'] - \E[w \bX \bZ'] \E[w \bZ \bZ']^{-1}\E[w \bZ \bX'] + \\
    &\quad \quad\,\, \E[w \bX \bZ'] \E[w \bZ \bZ']^{-1}\E[w \bZ \bZ'] \E[w \bZ \bZ']^{-1} \E[w \bZ \bX'] \\
    &\quad = \E\left[w \bX \bX'\right]-\E[w \bX \bZ'] \E[w \bZ \bZ']^{-1}\E[w \bZ \bX'].
\end{align*}
and the identity:
\begin{align*}
    \eta - \eta_z &= \bbe'\bX + \blam'\bZ - \E[w \bbe'\bX \bZ'] \E[w \bZ \bZ']^{-1} \bZ - \E[w \blam'\bZ \bZ'] \E[w \bZ \bZ']^{-1} \bZ \\
    &= \bbe'\bX - \E[w\bbe'\bX \bZ']\E[w \bZ \bZ']^{-1}\bZ + \blam'\bZ - \blam'\bZ \\
    &= \bbe'\left(\bX - \E[w \bX \bZ']\E[w \bZ \bZ']^{-1}\bZ\right), 
\end{align*}
we rewrite
\begin{align*}
    \fsq &= \bbe'\E\left[w\left(\bX-\E[w \bX \bZ'] \E[w \bZ \bZ']^{-1}\bZ \right)\left(\bX-\E[w \bX \bZ'] \E[w \bZ \bZ']^{-1}\bZ \right)'\right] \bbe' \\
    &=\E\left[w \bbe'\left(\bX-\E[w \bX \bZ'] \E[w \bZ \bZ']^{-1}\bZ \right)\left(\bX-\E[w \bX \bZ'] \E[w \bZ \bZ']^{-1}\bZ \right)'\bbe'\right] \\
    &=\E\left[ w (\eta-\eta_z)^2 \right],
\end{align*}
completing the derivation.

\vspace{1em}
\noindent \textbf{Derivation of $\pspRsq = \frac{\E [ (\mu - \mu_z)^2 / v ]}{1 + \E[(\mu-\mu_z)^2)/v]}$}

Expanding using $\mu$:
\begin{align*}
 \E\left[ (Y - \mu_z)^2 / v \right] &= \E\left[ (Y - \mu + \mu - \mu_z)^2 / v \right]  \\
 &= \E\left[ (Y - \mu)^2 / v \right] + 2\E\left[ (Y - \mu)(\mu-\mu_z) / v \right]  + \E\left[ (\mu - \mu_z)^2 / v \right].  
\end{align*}
An earlier derivation gave:
\begin{align*}
     \E\left[ (Y - \mu)^2 / v \right] = \text{WMSE} = 1.
\end{align*}
Further, iterative expectations gives
\begin{align*}
    \E\left[ (Y - \mu)(\mu-\mu_z) / v \right] &= \E\left[ \E[(Y - \mu)(\mu-\mu_z) / v |\bX,\bZ] \right] \\
    &= \E\left[ (\mu-\mu_z)/v \E[Y - \mu |\bX,\bZ] \right].
\end{align*}
With $\E[Y - \mu |\bX,\bZ]=0$, we arrive at:
\begin{align*}
    \E\left[ (Y - \mu_z)^2 / v \right] = \E\left[ (Y - \mu)^2 / v \right] +  \E\left[ (\mu - \mu_z)^2 / v \right] = 1 +  \E\left[ (\mu - \mu_z)^2 / v \right]. 
\end{align*}
Plugging these expressions into the definition of $\pspRsq$:
\begin{align*}
\pspRsq &= \frac{ \E\left[ (Y - \mu_z)^2 / v \right] -  \E\left[ (Y - \mu)^2 / v \right]}{\E\left[ (Y - \mu_z)^2 / v \right]} \\
&= \frac{ \E\left[ (Y - \mu)^2 / v \right] +  \E\left[ (\mu - \mu_z)^2 / v \right] -  \E\left[ (Y - \mu)^2 / v \right]}{1 +  \E\left[ (\mu - \mu_z)^2 / v \right]} \\
&= \frac{ \E\left[ (\mu - \mu_z)^2 / v \right]}{1 +  \E\left[ (\mu - \mu_z)^2 / v \right]}.
\end{align*}

\vspace{1em}
\noindent \textbf{Derivation of $\fsq = \frac{w_1 \phi^2}{4} + \E[(w-w_1) (\eta-\eta_z)^2] +\E[(w-w_1)(\eta-\eta_z)]^2/w_1$ for any constant $w_1 \neq 0$.}

From a prior derivation, we found
\begin{align*}
    \fsq = \E[w(\eta-\eta_z)^2].
\end{align*}
Therefore, 
\begin{align*}
    \fsq &=  \E[w_1(\eta-\eta_z)^2] +  \E[(w-w_1)(\eta-\eta_z)^2] \\
    &= w_1 \left\{ \var(\eta-\eta_z) + \E[\eta-\eta_z]^2 \right\} +  \E[(w-w_1)(\eta-\eta_z)^2] \\
    &= w_1 \var(\eta-\eta_z) + \E[w_1(\eta-\eta_z)]^2/w_1 +  \E[(w-w_1)(\eta-\eta_z)^2],
\end{align*}
for any constant $w_1 \neq 0$. Observe that
$\var(\eta-\eta_z) = \phixz^2/4$, and, since $\eta_z = \E[w\eta \bZ']\E[w \bZ \bZ']^{-1} \bZ$, 
\begin{align*}
    \E[w(\eta-\eta_z) \bZ ] &= \E[w \eta \bZ] - \E[w \eta_z \bZ] \\
    &= \E[w \eta \bZ] - \E[w \eta \bZ] \E[w \bZ \bZ']^{-1} \E[ w \bZ \bZ'] \\
    &= 0.
\end{align*}
As the first element of $\bZ$ is the constant one, it follows that
\begin{align*}
    \E[w(\eta-\eta_z)] = 0,
\end{align*}
so that we can freely subtract it from the term $\E[w_1(\eta-\eta_z)]$. Hence,
\begin{align*}
    \fsq = w_1 \phixz^2/4 + \E[(w-w_1)(\eta-\eta_z)]^2/w_1 +  \E[(w-w_1)(\eta-\eta_z)^2].
\end{align*}

\section{Proof of theorems} \label{app:proofs}

Our goal is to analyze the relative error in our approximations,
\[
\frac{\fsq - \fsqphi}{\fsqphi} \qquad \text{ and } \qquad \frac{\fsq - \fsqR}{\fsqR},
\]
as the influence of the predictors and adjustors on $Y$ diminish. To be precise, we treat $\blam$ and $\bbe$ as functions of $(\iota, \delta_{\kappa}, \delta_{\bbe})$ given by:
\begin{align*}
    \bbe = \delta_{\bbe} \bbe_*, \\
    \blam = [\iota \,\,\, \delta_{\kappa} \kappa_*' ]'
\end{align*}
for fixed $\bbe_* \in \mathbbm{R}^p$ and $\kappa_* \in \mathbbm{R}^{q-1}$ and a fixed joint distribution for $\bX$ and $\bZ$. Consequently, each choice in $\iota$, $\delta_{\kappa}$, and $\delta_{\bbe}$ determines $\blam$ and $\bbe$, which in turn determines $\eta$ and then subsequent quantities like $\eta_z$, $w$, and the effect measures $\fsq$, $\fsqphi$, and $\fsqR$. We will investigate limits as $\delta_{\kappa} \rightarrow 0$ and $\delta_{\bbe} \rightarrow 0$, while keeping $\kappa_*$, $\bbe_*$, and $\E[Y]=\E[\mu]:=\mu_*$ fixed. 

To analyze this limit, we make several assumptions. Our first assumption ensures we can find a value of $\iota$ that, for any sufficiently small $\delta_{\kappa}$ and $\delta_{\bbe}$, keeps $\E[Y]$ fixed at $\mu_*$. We use the inverse function theorem and make our assumptions accordingly:

\begin{assumption} \label{assm:inverse_fcn}
Fix $\mu_{*}$, $\kappa_*$, and $\bbe_*$. The function
\begin{align*}
    (\iota, \delta_{\kappa}, \delta_{\bbe}) \mapsto \left(\E[\mu], \delta_{\kappa}, \delta_{\bbe}\right)
\end{align*}
is continuously differentiable in a neighborhood around $(g(\mu_{*}),0,0)$ and its derivative at $(g(\mu_{*}),0, 0)$ is invertible.
\end{assumption}

This assumption lets us apply the inverse function theorem, ensuring the function is bijective near $(g(\mu_{*}),0,0)$. Therefore, we can find $(\iota, \delta_{\kappa}, \delta_{\bbe})$ mapping to $(g(\mu_{*}), \delta_{\kappa}, \delta_{\bbe})$ for points close to $(g(\mu_{*}), 0, 0)$. This yields the following lemma:

\begin{lemma} \label{lem:inverse_fcn}
Fix $\mu_{*}$, $\kappa_*$, and $\bbe_*$, and suppose Assumption~\ref{assm:inverse_fcn} holds. Then, there exists $\delta_* > 0$ so that we can find $\iota$ whenever $0 \leq \delta_{\kappa}, \delta_{\bbe} < \delta_*$ so that
$\E[Y] = \mu_{*}$.
\end{lemma}

We apply Lemma~\ref{lem:inverse_fcn} to select $\iota$ such that $\E[Y] = \E[\mu]= \mu_{*}$ for sufficiently small $\delta_{\kappa}$ and $\delta_{\bbe}$. As a result, by choosing $\delta_{\kappa}$ and $\delta_{\bbe}$, we determine $\iota$, which in turn dictates $\eta$, followed by the other quantities like $w$ and the effect measures ($\fsq$, $\fsqphi$, and $\fsqR$). Hence, we can view these quantities as functions of only $\delta_{\kappa}$ and $\delta_{\bbe}$, with $\iota$ being implicitly determined by the values of $\delta_{\kappa}$ and $\delta_{\bbe}$. 

This lets us introduce the following asymptotic notation. For arbitrary functions $f$ and $g$ of $\delta_{\bbe}$ and $\delta_{\kappa}$, we use
$f(\delta_{\bbe}, \delta_{\kappa}) = \mathcal{O}\left(g(\delta_{\bbe}, \delta_{\kappa})\right)$
to mean there exist constants $\delta_*$ and $C$ so that
\begin{align*}
    f(\delta_{\bbe}, \delta_{\kappa}) \leq  C g(\delta_{\bbe}, \delta_{\kappa})
\end{align*}
whenever $0 \leq \delta_{\kappa}, \delta_{\bbe} < \delta_*$. We use $f(\delta_{\bbe}, \delta_{\kappa}) = \Omega\left(g(\delta_{\bbe}, \delta_{\kappa})\right)$
to mean there exist constants $\delta_*$ and $C$ so that
\begin{align*}
    f(\delta_{\bbe}, \delta_{\kappa}) \geq  C g(\delta_{\bbe}, \delta_{\kappa})
\end{align*}
whenever $0 \leq  \delta_{\kappa}, \delta_{\bbe} < \delta_*$. Further, we use $f(\delta_{\bbe}, \delta_{\kappa}) = \Theta\left(g(\delta_{\bbe}, \delta_{\kappa})\right)$ to mean both $f(\delta_{\bbe}, \delta_{\kappa}) = \mathcal{O}\left(g(\delta_{\bbe}, \delta_{\kappa})\right)$ and $f(\delta_{\bbe}, \delta_{\kappa}) = \Omega\left(g(\delta_{\bbe}, \delta_{\kappa})\right)$. 

Our next assumption puts a bound on the moments of $\bX$ and $\bZ$. Here, $\Vert x \Vert$ denotes the 2-norm on any vector $x$ (i.e. $x'x$ or $x \cdot x$) and $\Vert \textbf{B} \Vert = \sup_{x: \Vert x \Vert = 1} \Vert B x \Vert$ denotes the associated operator norm for any matrix $B$:

\begin{assumption} \label{assm:bounded_moments}
Random variables $\bX$ and $\bZ$ have bounded moments up to fourth order:
\begin{align*}
    \E\left[ \Vert \bX \Vert^i \Vert \bZ \Vert^j \right] < \infty
\end{align*}
for any $1 \leq i + j \leq 4$.
\end{assumption}

Our third assumption requires that the effect measures are well-behaved when $\delta_{\bbe}=\delta_{\kappa}=0$. To formalize this, we introduce the matrix $A$, defined as  
\begin{align*}
A := \E[ w \bX \bZ' ] \E[ w \bZ \bZ' ]^{-1},
\end{align*}
which simplifies the expression for $\eta - \eta_z$. Specifically, we have:
\begin{align*}
\eta - \eta_z = \eta - \E[ w \eta \bZ] \E[w \bZ \bZ']^{-1} \bZ.
\end{align*}
Expanding $\eta$, this becomes:
\begin{align*}
\blam'\bZ + \bbe'\bX - \blam' \E[ w \bZ \bZ] \E[w \bZ \bZ']^{-1} \bZ - \bbe' \E[w \bX \bZ] \E[w \bZ \bZ']^{-1} \bZ.
\end{align*}
Using the definition of $A$, we simplify to:
\begin{align*}
\blam' \bZ + \bbe'\bX - \blam' \bZ - \bbe'A \bZ = \bbe' (\bX - A \bZ).
\end{align*}
This form highlights how $A$ helps capture $\eta-\eta_z$.

\begin{assumption} \label{assm:regular}
Fix $\mu_*$, $\kappa_*$, and $\bbe_*$, and let $w_*$ and $A_*$ denote $w$ and $A$ evaluated at $\eta = g(\mu_*)$. Matrix $\E[w_* \bZ \bZ']$ is defined and invertible, and $\E[ w_* \left\{ \bbe_*'(\bX-A_* \bZ) \right\}^2 ] > 0$.
\end{assumption}

Our final assumption has already been stated in the main text (Assumption~\ref{assm:smooth_w}). It  requires that the mean square error in $w$ remains bounded by some scaling of $(\delta_{\kappa} + \delta_{\bbe})^2$.

With these assumptions, we are ready to provide the results we need to establish our theorems. We start with two lemmas. The first tells us that $A$ and $A_*$ are reasonably close:

\begin{lemma} \label{lem:A}
   Under the conditions of Theorem~\ref{thm:error_phi}, $\Vert A - A_* \Vert = \mathcal{O}(\delta_{\bbe}+\delta_{\kappa}).$
\end{lemma}

\textit{\textbf{Proof of Lemma~\ref{lem:A}}}. First, observe that 
\begin{align*}
    A_* = \E[w_* \bX \bZ'] \E[w_* \bZ \bZ']^{-1}.
\end{align*}
is defined and finite. That is, matrix $\E[w_* \bZ \bZ']^{-1}$ exists by Assumption~\ref{assm:regular}, and since $w_*$ is constant and the moments of $\bX$ and $\bZ$ up to the fourth order are finite (Assumption~\ref{assm:bounded_moments}), $\E[w_* \bX \bZ']$ must also be defined and finite. 

Second, observe that
\begin{align*}
    A = \E[w \bX \bZ'] \E[w \bZ \bZ]^{-1}
\end{align*}
is defined for $(\iota, \delta_{\kappa}, \delta_{\bbe})$ in a neighborhood of $(g(\mu_*), 0, 0)$. In this case, 
\begin{align*}
\E[w \bX \bZ'] = \E[w_* \bX \bZ'] + \E[(w-w_*) \bX \bZ'].
\end{align*}
Again, we have $\E[w_* \bX \bZ']$ defined and finite. In addition, $\E[(w-w_*) \bX \bZ']$ is defined and finite in a neighborhood of $(\eta_*, 0, 0)$, since each entry in $\E[|(w-w_*) \bX \bZ'|]$ is bounded above by
\begin{align*}
\E\left[|w-w_*| \Vert \bX \Vert \Vert \bZ \Vert\right] \leq \sqrt{ \E\left[(w-w_*)^2\right] } \sqrt{\E\left[ \Vert \bX \Vert^2 \Vert \bZ \Vert^2 \right] },
\end{align*}
where the first term is bounded above by $\sqrt{K}(\delta_{\kappa} + \delta_{\bbe})$ in a neighborhood of $(g(\mu_*), 0, 0)$ (Assumption~\ref{assm:smooth_w}), and the second term is finite (Assumption~\ref{assm:bounded_moments}). Similarly, 
\begin{align*}
\E[w \bZ \bZ'] = \E[w_* \bZ \bZ']\left( I + \E[w_* \bZ \bZ']^{-1} \E[(w-w_*) \bZ \bZ'] \right).
\end{align*}
In this case, $\E[w_* \bZ \bZ']^{-1}$ exists by Assumption~\ref{assm:regular}. Further, each entry in $\E[|(w-w_*) \bZ \bZ'|]$ is bounded above in a neighborhood of $(g(\mu_*), 0, 0)$ by 
\begin{align*}
 \E\left[|w-w_*| \lVert \bZ \rVert^2 \right] \leq \sqrt{ \E\left[(w-w_*)^2\right] } \sqrt{\E\left[ \Vert \bZ \Vert^4 \right] },
\end{align*}
with the first term bounded by $\sqrt{K}(\delta_{\kappa} + \delta_{\bbe})$  (Assumption~\ref{assm:smooth_w}) and the second term finite (Assumption~\ref{assm:bounded_moments}). In fact, we can always choose $\delta_{\kappa}$ and $\delta_{\bbe}$ small enough so that
\begin{align*}
    \Vert \E[w_* \bZ \bZ']^{-1} \E[(w-w_*) \bZ \bZ'] \Vert < 1/2.
\end{align*}
By Neumann's Lemma, which states that $I-B$ is invertible when $\Vert B \Vert < 1$, it follows that $I + \E[w_* \bZ \bZ']^{-1} \E[(w-w_*) \bZ \bZ']$ is not only defined and finite, but also invertible in a neighborhood of $(g(\mu_*), 0, 0)$.  Moreover, its inverse is also bounded in a neighborhood:
\begin{align*}
    \lVert \left( I + \E[w_* \bZ \bZ']^{-1} \E[(w-w_*) \bZ \bZ'] \right)^{-1} \Vert \leq \frac{1}{1- \lVert \E[w_* \bZ \bZ']^{-1} \E[(w-w_*) \bZ \bZ'] \rVert } \leq 2.
\end{align*}
So, $\E[w \bZ \bZ']$ is the product of two matrices---$\E[w_* \bZ \bZ']$ and $I + \E[w_* \bZ \bZ']^{-1} \E[(w-w_*) \bZ \bZ']$---each locally defined, finite, and invertible with bounded inverses. Hence, we can conclude that $\E[w \bZ \bZ']^{-1}$, and consequently $A$, is defined and finite in a neighborhood of  $(g(\mu_*), 0, 0)$.

Last, we observe that
\begin{align*}
A - A_* &= \E[w \bX \bZ'] \E[w \bZ \bZ']^{-1} - \E[w_* \bX \bZ'] \E[w_* \bZ \bZ']^{-1} \\
&= \left( \E[w \bX \bZ'] - \E[w_* \bX \bZ'] \E[w_* \bZ \bZ']^{-1} \E[w \bZ \bZ]  \right) \E[w \bZ \bZ']^{-1} \\
&= \left( \E[(w-w_*) \bX \bZ'] - \E[w_* \bX \bZ'] \E[w_* \bZ \bZ']^{-1} \E[(w-w_*) \bZ \bZ] \right) \E[w \bZ \bZ']^{-1}
\end{align*}
and hence $\Vert A - A_* \Vert$ is bounded above by
\begin{align*}
     \left( \lVert \E[(w-w_*) \bX \bZ'] \rVert + \lVert \E[w_* \bX \bZ'] \rVert \lVert \E[w_* \bZ \bZ']^{-1} \lVert \rVert \E[(w-w_*) \bZ \bZ'] \rVert \right) \lVert \E[w \bZ \bZ']^{-1} \rVert
\end{align*}
in a neighborhood of $(g(\mu_*), 0, 0)$. Our work above shows that $\lVert \E[w_* \bX \bZ'] \rVert$, $\lVert \E[w_* \bZ \bZ']^{-1} \lVert$, and $\lVert \E[w \bZ \bZ']^{-1} \rVert$ are also bounded around $(g(\mu_*), 0, 0)$, leaving
\begin{align*}
\lVert \E[(w-w_*) \bX \bZ'] \rVert \leq \E[|w-w_*| \lVert \bX \rVert \lVert \bZ \rVert ] \leq \sqrt{K}(\delta_{\kappa} + \delta_{\bbe}) \sqrt{ \E[\lVert \bX \rVert^2 \lVert \bZ \rVert^2 ] }
\end{align*}
and similarly,
\begin{align*}
\lVert \E[(w-w_*) \bZ \bZ'] \rVert \leq \E[|w-w_*| \lVert \bZ \rVert^2 ] \leq \sqrt{K}(\delta_{\kappa} + \delta_{\bbe}) \sqrt{ \E[\lVert \bZ \rVert^4 ] }
\end{align*}
with the moments of $\bZ$ and $\bX$ bounded. In the end, we have a constant $C$ so that
\begin{align*}
    \lVert A - A_* \rVert \leq C(\delta_{\kappa} + \delta_{\bbe} )
\end{align*}
for $(\iota, \delta_{\kappa}, \delta_{\bbe})$ in a neighborhood of $(g(\mu_*), 0, 0)$. We conclude $\lVert A - A_* \rVert$ is $\mathcal{O}(\delta_{\kappa} + \delta_{\bbe})$. 

\EndPf

Our second lemma tells us that $\fsq$ behaves like $\delta_{\bbe}^2$ for small $\delta_{\bbe}$ and $\delta_{\kappa}$ with $\E[Y]$ fixed.

\begin{lemma} \label{lem:zeta}
   Under the conditions of Theorem~\ref{thm:error_phi}, $\fsq = \Theta\left(\delta_{\bbe}^2\right)$.
\end{lemma}

\textit{\textbf{Proof of Lemma~\ref{lem:zeta}}}. We have
\begin{align*}
    \fsq &= \E[w (\eta-\eta_z)^2 ] = \delta_{\bbe}^2 \E\left[(w_* + (w-w_*))\left\{ \bbe_*' (\bX - A_* \bZ - (A-A_*) \bZ ) \right\}^2 \right].
\end{align*}
Expanding, 
\begin{align*}
&\E\left[(w_* + (w-w_*))\left\{ \bbe_*' (\bX - A_* \bZ - (A-A_*) \bZ ) \right\}^2 \right] \\
& = \E\left[w_* \left\{ \bbe_*'(\bX-A_* \bZ) \right\}^2 \right]  + \E\left[w_* \left\{ \bbe_*' (A-A_*) \bZ \right\}^2 \right] + \\
&\quad \E\left[(w-w_*) \left\{ \bbe_*'(\bX-A_* \bZ) \right\}^2 \right]  + \E\left[(w-w_*) \left\{ \bbe_*' (A-A_*) \bZ \right\}^2 \right] -  \\
&\quad 2 \E\left[w_* \bbe_*'(\bX-A_* \bZ) \bbe_*' (A-A_*) \bZ \right]  - 2 \E\left[(w-w_*) \bbe_*'(\bX-A_* \bZ) \bbe_*' (A-A_*) \bZ \right].
\end{align*}
Note that
$$\E\left[w_* \left\{ \bbe_*' (A-A_*) \bZ \right\}^2 \right] = \mathcal{O}(\delta_{\bbe} + \delta_{\kappa} ).$$
This follows from $\Vert A - A_* \Vert = \mathcal{O}( \delta_{\bbe} + \delta_{\kappa} )$ (Lemma~\ref{lem:A}), the bounded moments on $\bX$ and $\bZ$ (Assumption~\ref{assm:bounded_moments}), and the relation:
\begin{align*}
\left| \E\left[w_* \left\{ \bbe_*' (A-A_*) \bZ \right\}^2 \right] \right| \leq w_*  \lVert \bbe_* \rVert^2 \lVert A - A_* \Vert^2 \E\left[ \lVert \bZ \rVert^2 \right]. 
\end{align*}
We also have that
$$\E\left[(w-w_*) \left\{ \bbe_*'(\bX-A_* \bZ) \right\}^2 \right] = \mathcal{O}( \delta_{\kappa} + \delta_{\bbe}),$$
which follows from the smoothness of $w$ (Assumption~\ref{assm:smooth_w}), the bounded moments (Assumption~\ref{assm:bounded_moments}), the finiteness of $A_*$ (Assumption~\ref{assm:regular}), and the relation
\begin{align*}
\left| \E\left[(w-w_*) \left\{ \bbe_*'(\bX-A_* \bZ) \right\}^2 \right] \right| \leq \sqrt{\E[(w-w_*)^2]} \Vert \bbe_* \Vert^2 \sqrt{ \E\left[ \lVert \bX - A_* \bZ \Vert^4 \right] }. 
\end{align*}
Applying the same arguments, we find that every term in the above expansion is $\mathcal{O}(\delta_{\bbe} + \delta_{\kappa}),$ except for the first term:
\begin{align*}
\E\left[w_* \left\{ \bbe_*'(\bX-A_* \bZ) \right\}^2 \right].
\end{align*}
However, Assumption~\ref{assm:bounded_moments} (bounded moments) and Assumption~\ref{assm:regular} ensure that 
$$0< \E\left[w_* \left\{ \bbe_*'(\bX-A_* \bZ) \right\}^2 \right] = \Theta(1).$$ 
This means that
\begin{align*}
    \fsq = \Theta(\delta_{\bbe}^2) + \mathcal{O}\left(\delta_{\bbe}^2 (\delta_{\kappa} + \delta_{\bbe}) \right).
\end{align*}
Since we can make the last term arbitrary small relative to the first term, we have that $\fsq = \Theta(\delta_{\bbe}^2)$, as desired.

\EndPf

We are now ready to prove our theorem:

\textbf{Proof of Theorem~\ref{thm:error_phi}}. We first work with the numerator, $\fsq - \fsqphi$. Observe that $\fsqphi$ is
\begin{align*}
    w_1 \var\left(\eta-\eta_z \right) = w_* \E[(\eta-\eta_z)^2] - w_* \E[\eta-\eta_z]^2,
\end{align*}
where $w_* = w_1$, since $w_*$ is $w$ when $\eta$ is set to $g(\mu_*)$ and $w_1$ is $w$ when $\eta$ is set to $g(\E[Y])$,  which we keep fixed at $g(\mu_*)$. We also notice
\begin{align*}
   \E[w (\bX-A \bZ) \bZ'] = \E[w \bX \bZ'] - \E[w \bX \bZ'] \E[w \bZ \bZ']^{-1} \E[w \bZ \bZ] = 0,
\end{align*}
which, since the first entry in $\bZ$ is the constant $1$, means
\begin{align*}
    0 = \bbe'\E[w(\bX-A \bZ)] = \E[w (\eta-\eta_z)].
\end{align*}
Therefore,
\begin{align*}
    \fsq- \fsqphi &= \E[ (w-w_*) (\eta-\eta_z)^2 ] - \E[(w-w_*) (\eta-\eta_z)]^2 / w_*.
\end{align*}
Note we are not dividing by zero, since Assumption~\ref{assm:regular} requires that $w_* \neq 0$. Working with the first term, we apply Cauchy-Schwartz:
\begin{align*}
    \left\vert \E[ (w-w_*) (\eta-\eta_z)^2 ] \right\vert &= \left\vert \E\left[ (w-w_*) \left\{\bbe'(\bX-A \bZ)\right\}^2 \right] \right\vert \\
    &\leq \delta_{\bbe}^2 \sqrt{ \E\left[(w-w_*)^2\right] \E\left[ \left\{\bbe_*'(\bX-A \bZ)\right\}^4 \right] }.
\end{align*}
Cauchy-Schwartz and triangle inequalities, and the definition of the operator norm, also imply
\begin{align*}
    \left\{\bbe_*'(\bX-A \bZ)\right\}^4 &\leq \lVert \bbe_* \Vert^4 \lVert \bX - A \bZ \rVert^4 \\
    &\leq \lVert \bbe_* \Vert^4 \left( \lVert \bX \rVert + \lVert A_* \rVert \lVert \bZ \rVert + \lVert A - A_* \rVert \lVert \bZ \lVert \right)^4
\end{align*}
Importantly, the expression on the right hand side of the inequality is bounded in expectation for sufficiently small $\delta_{\kappa}$ and $\delta_{\bbe}$, which follows from $\Vert \bbe_* \Vert$ and $\Vert A_* \Vert$ being fixed and finite, $\bX$ and $\bZ$ having bounded moments up to fourth order (Assumption~\ref{assm:bounded_moments}), and $\lVert A - A_* \rVert$ being $\mathcal{O}(\delta_{\kappa} + \delta_{\bbe})$ (Lemma~\ref{lem:A}). Further, Assumption~\ref{assm:smooth_w} tells us
\begin{align*}
    \sqrt{ \E\left[(w-w_*)^2\right] } = \mathcal{O}( \sqrt{M} \left\{\delta_{\bbe} + \delta_{\kappa}\right\} ).
\end{align*}
This means that 
\begin{align*}
    \left\vert \E[ (w-w_*) (\eta-\eta_z)^2 ] \right\vert &= \mathcal{O}\left( \sqrt{M}\delta_{\bbe}^2  \left\{\delta_{\bbe} + \delta_{\kappa}\right\} \right).
\end{align*}

Following similar ideas, we have
\begin{align*}
\E[(w-w_*) (\eta-\eta_z)]^2/w_* &\leq (\delta_{\bbe}^2/w_*) \E\left[(w-w_*)^2\right] \E\left[ \left\{\bbe_*'(\bX-A \bZ)\right\}^2 \right]
\end{align*}
with 
\begin{align*}
    \E\left[ \left\{\bbe_*'(\bX-A \bZ)\right\}^2 \right]
\end{align*}
bounded for small $\delta_{\kappa},\delta_{\bbe}$. Thus, 
\begin{align*}
\E[(w-w_*) (\eta-\eta_z)]^2/w_* &= \mathcal{O}\left( \sqrt{M}\delta_{\bbe}^2  \left\{\delta_{\bbe} + \delta_{\kappa}\right\}^2 \right).
\end{align*}
Hence, we have
\begin{align*}
    \left\vert \fsq - \fsqphi \right\vert = \mathcal{O}\left( \sqrt{M} \delta_{\bbe}^2 \left\{\delta_{\kappa} + \delta_{\bbe}\right\} \right)
\end{align*}

Working with the denominator, we have
\begin{align*}
    \fsqphi \geq \fsq - \left\vert \fsq - \fsqphi \right\vert,
\end{align*}
where $\fsq$ is $\Theta(\delta_{\bbe}^2)$ (Lemma~\ref{lem:zeta}) and the second term is $\mathcal{O}\left( \sqrt{M} \delta_{\bbe}^2 \left\{\delta_{\kappa} + \delta_{\bbe}\right\} \right)$. Consequently, we can choose $\delta_{\kappa}$ and $\delta_{\bbe}$ small enough to bound $\fsqphi$ below by a constant multiple of $\delta_{\bbe}^2$, yielding $\fsqphi = \Omega(\delta_{\bbe}^2)$. 

Taken together, we have shown that the numerator of the relative error is $\mathcal{O}\left( \sqrt{M} \delta_{\bbe}^2 \left\{\delta_{\kappa} + \delta_{\bbe}\right\} \right)$ and that the denominator is $\Omega(\delta_{\bbe}^2)$. We thus arrive at our desired result:
$$\left\vert \frac{ \fsq - \fsqphi }{\fsqphi} \right\vert= \mathcal{O}\left( \sqrt{M} \left\{\delta_{\kappa} + \delta_{\bbe}\right\} \right). $$
\EndPf

To analyze the relative error for our second measure of effect, $\fsqR$, we introduced another assumption to ensure that $\mu$ is sufficiently smooth so that $\mu_z$ gets close to $\mu$ relative to $v$ in expectation when $\delta_{\bbe}$ is small (Assumption~\ref{assm:smooth_mu}). Let's prove the theorem. 

\textbf{Proof of Theorem~\ref{thm:error_R}.} We start with
\begin{align*}
    (\mu - \mu_z)^2 / v,
\end{align*}
which we re-write using the remainder $\text{Rem}_{\mu}$:
\begin{align*}
\left\{\mu - \mu - \frac{\partial \mu}{\partial \eta} (\eta_z - \eta) - \text{Rem}_{\mu} \right\}^2 / v = w (\eta - \eta_z)^2 + 2 (\text{Rem}_{\mu}/\sqrt{v}) \frac{\partial \mu}{\partial \eta} (\eta_z - \eta)/\sqrt{v} + \text{Rem}_{\mu}^2 / v.
\end{align*}
Upon taking expectation, we have
\begin{align*}
    \fsqR &= \E\left[ (\mu - \mu_z)^2 / v \right] \\
    &= \fsq + 2 \E\left[(\text{Rem}_{\mu}/\sqrt{v}) \frac{\partial \mu}{\partial \eta} (\eta_z - \eta)/\sqrt{v}\right] + \E\left[ \text{Rem}_{\mu}^2 / v \right].
\end{align*}
We apply Cauchy-Schwartz twice to the middle term. We first apply Cauchy-Schwartz to get an upper bound on the absolute error in our approximation:
\begin{align*}
    \vert \fsq - \fsqR \vert \leq 2 \sqrt{\E\left[ \text{Rem}_{\mu}^2 / v \right] \fsq } + \E\left[ \text{Rem}_{\mu}^2 / v \right]
\end{align*}
We then apply Cauchy-Schwartz and use positivity of $ \E\left[ \text{Rem}_{\mu}^2 / v \right]$ to get a lower bound on $\fsqR$:
\begin{align*}
    \fsqR \geq \fsq - 2\sqrt{\E\left[ \text{Rem}_{\mu}^2 / v \right] \fsq }.
\end{align*}
Meanwhile, we have $\fsq = \Theta(\delta_{\bbe}^2)$ (Lemma~\ref{lem:zeta}) and $\E[\text{Rem}_{\mu}^2/v] = \mathcal{O}(K\delta_{\bbe}^4)$ (Assumption~\ref{assm:smooth_mu}), which means 
$$ 2 \sqrt{\E\left[ \text{Rem}_{\mu}^2 / v \right] \fsq } + \E\left[ \text{Rem}_{\mu}^2 / v \right] = \mathcal{O}( \sqrt{K} \delta_{\bbe}^3 ) $$
and
\begin{align*}
    \left\vert \fsq - 2\sqrt{\E\left[ \text{Rem}_{\mu}^2 / v \right] \fsq} \right\vert = \Omega(\delta_{\bbe}^2).
\end{align*}

Taken together, we have shown that the numerator is $\mathcal{O}\left( \sqrt{K} \delta_{\bbe}^3 \right)$ and that the denominator is $\Omega(\delta_{\bbe}^2)$ for the relative error. We thus arrive at our desired result:
$$\left\vert \frac{ \fsq - \fsqR }{\fsqR} \right\vert= \mathcal{O}\left( \sqrt{K} \delta_{\bbe} \right). $$

\newpage

\section{Plots of distributions} \label{app:distribution_plots}

\begin{figure}[H]
\centering
\begin{tikzpicture}
    \node at (0,0) {\includegraphics[width=.33\textwidth]{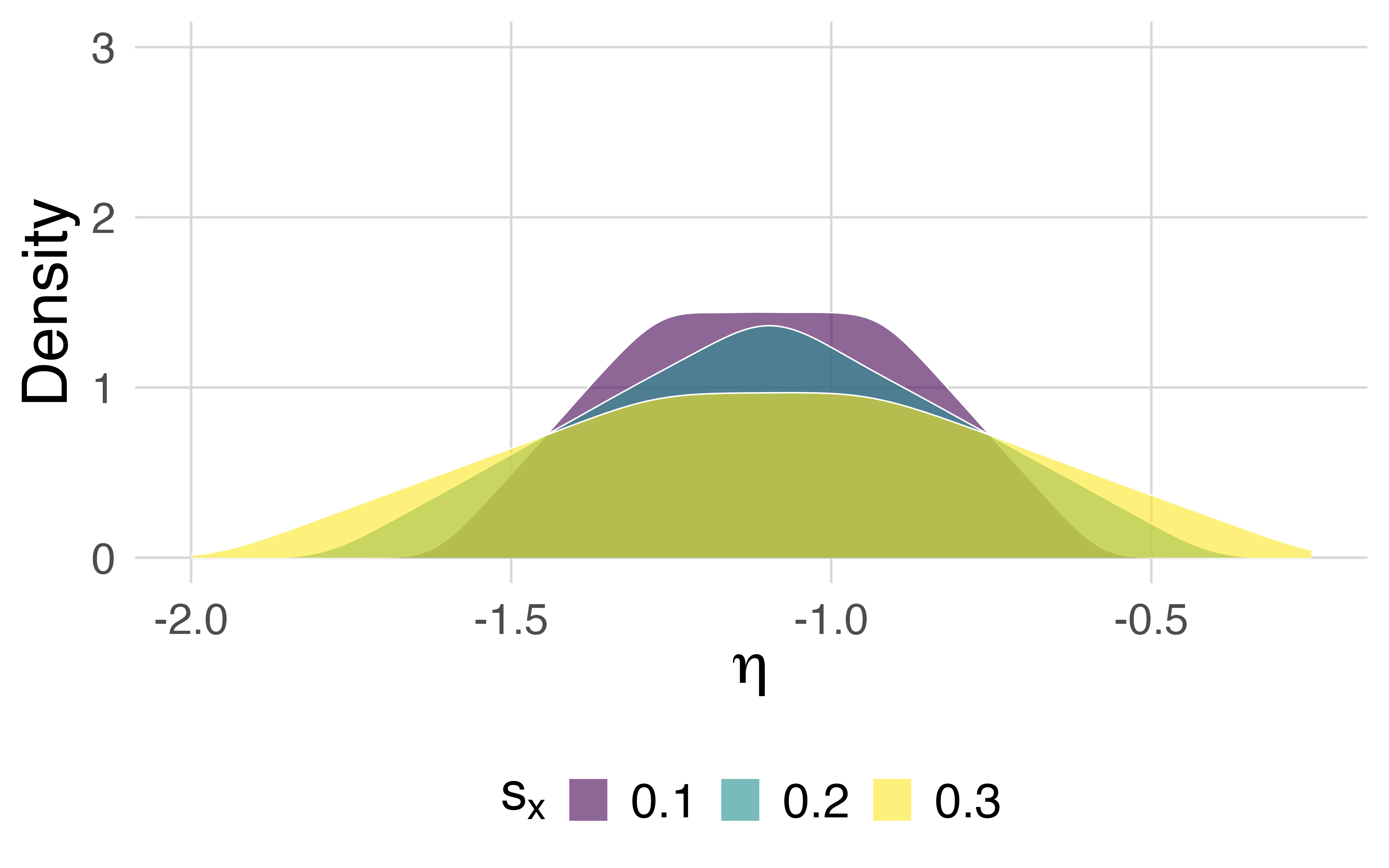}};
     \node at (11,0) {\includegraphics[width=.33\textwidth]{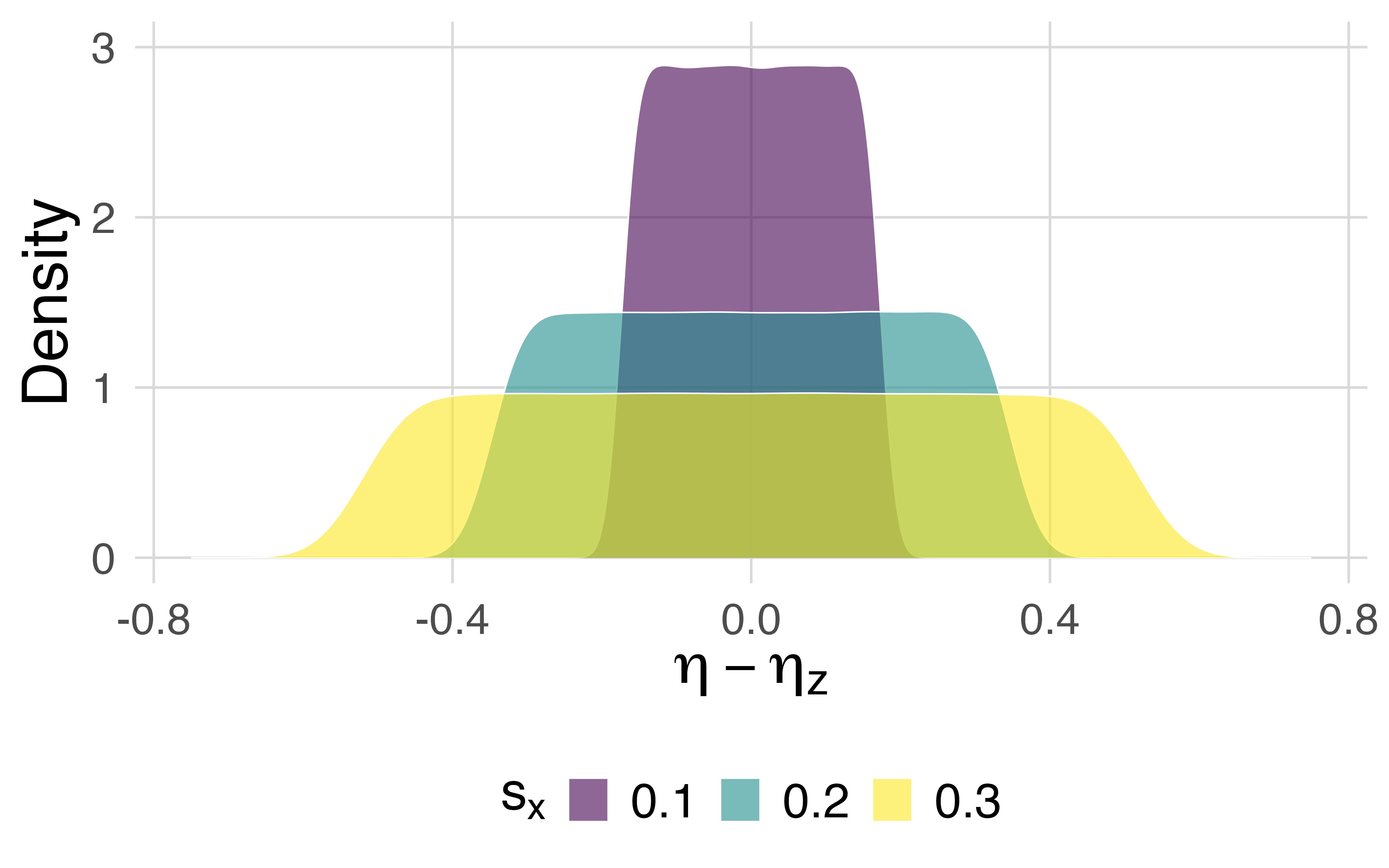}};
     \fill[white] (-2,-1.6) rectangle (13,-1.2);
    \node at (5.5,0) {\includegraphics[width=.33\textwidth]{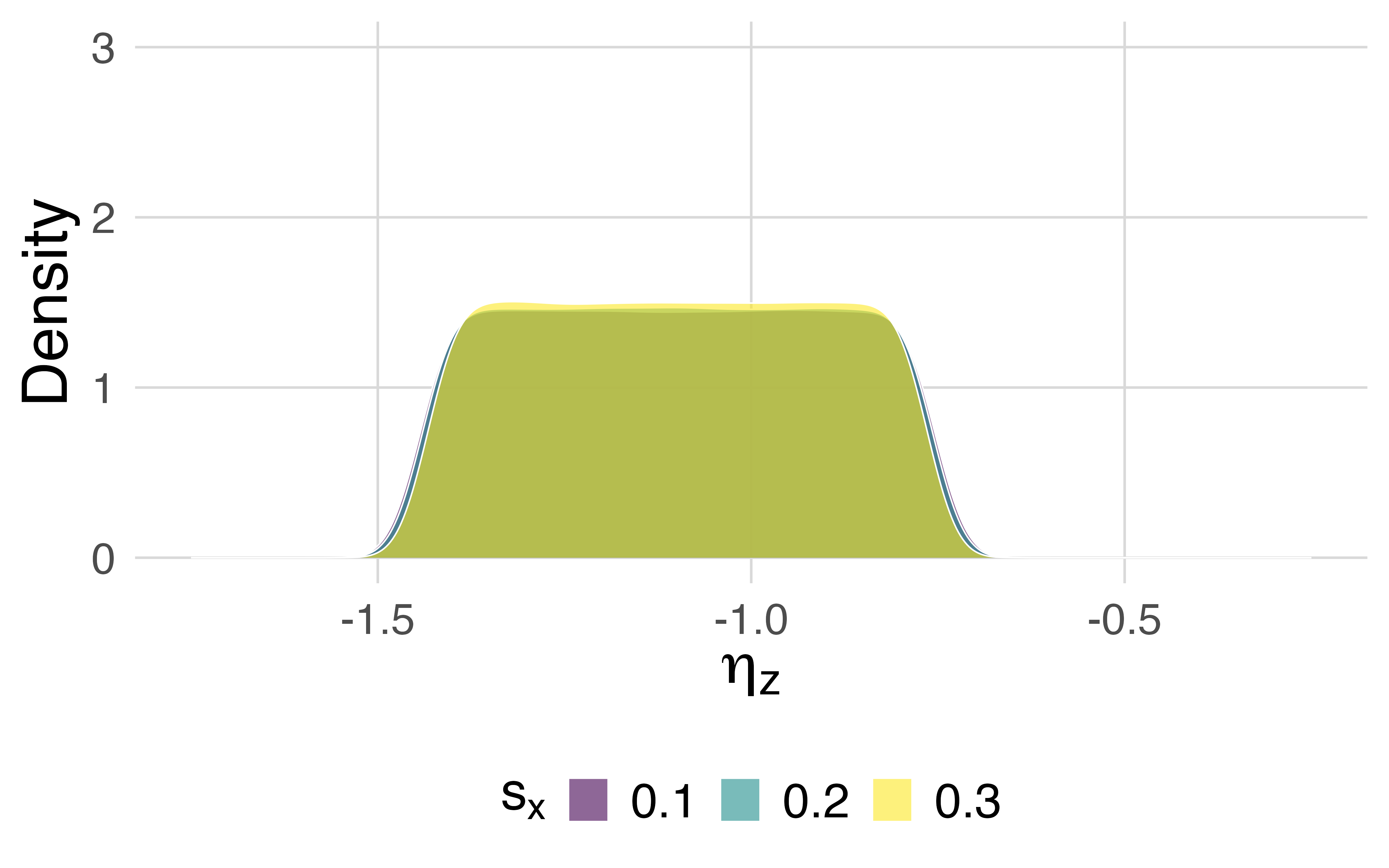}};
    \node at (0,-3.5) {\includegraphics[width=.33\textwidth]{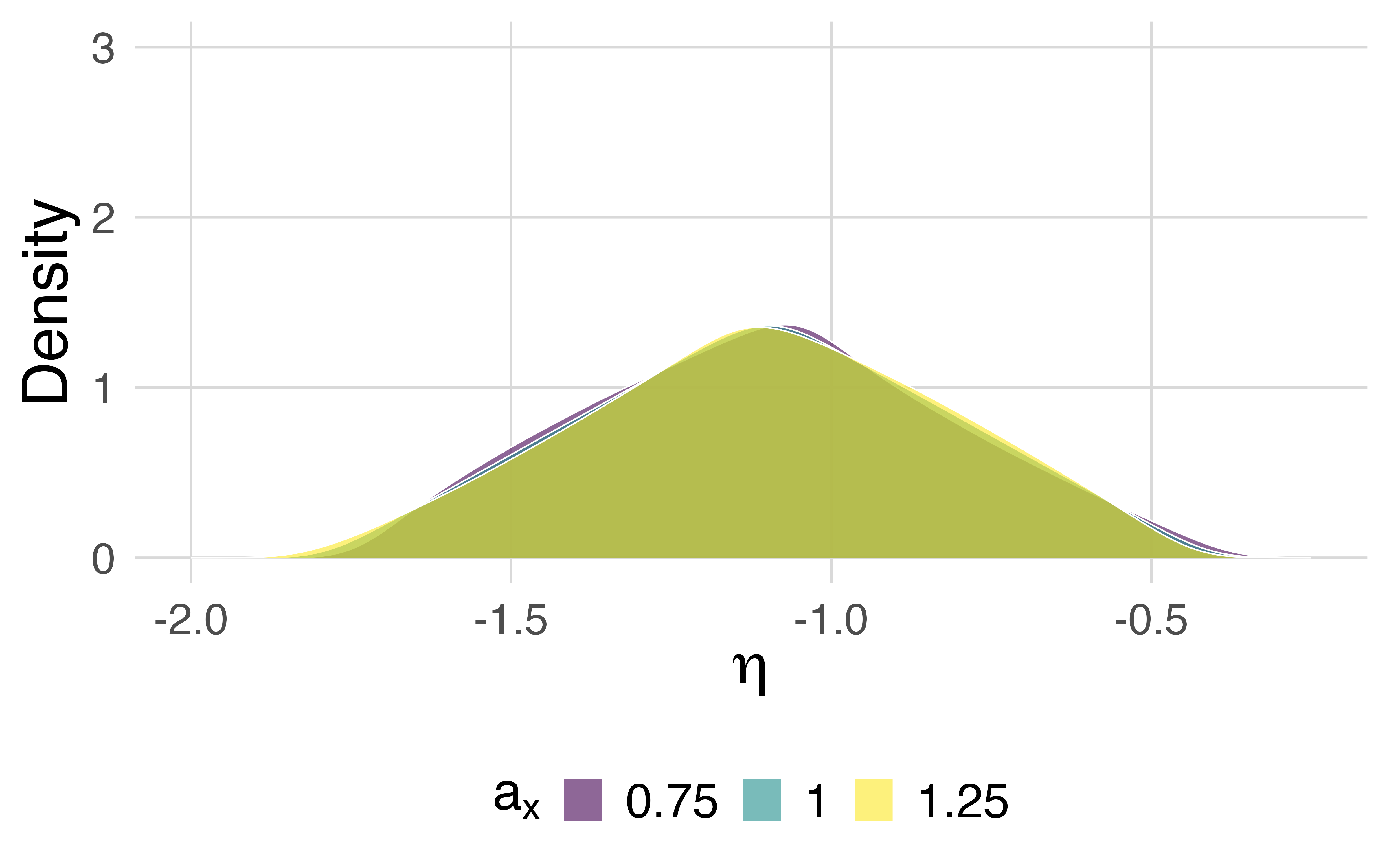}};
     \node at (11,-3.5) {\includegraphics[width=.33\textwidth]{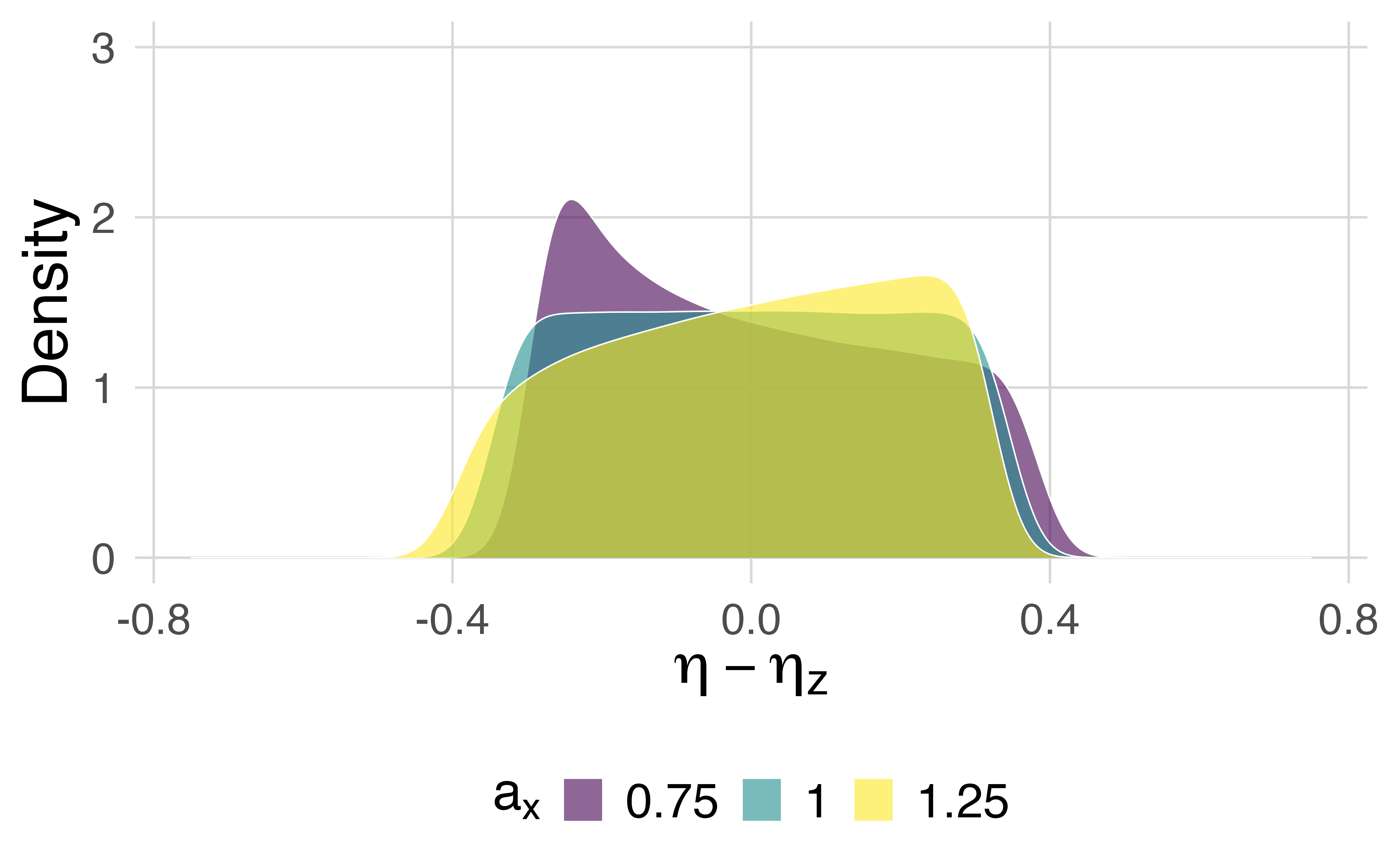}};
     \fill[white] (-2,-5.1) rectangle (13,-4.7);
    \node at (5.5,-3.5) {\includegraphics[width=.33\textwidth]{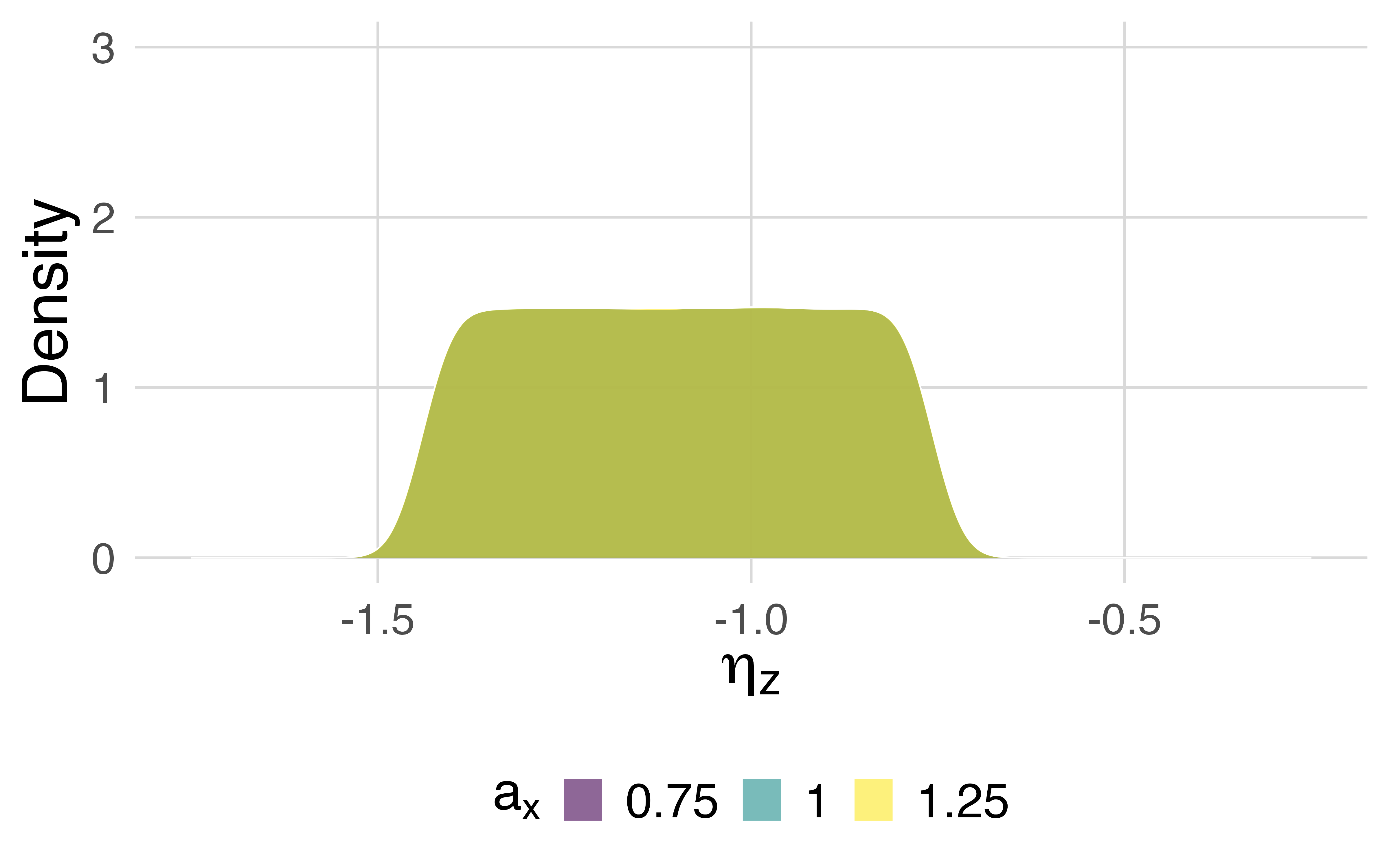}};
     \node at (0,-7) {\includegraphics[width=.33\textwidth]{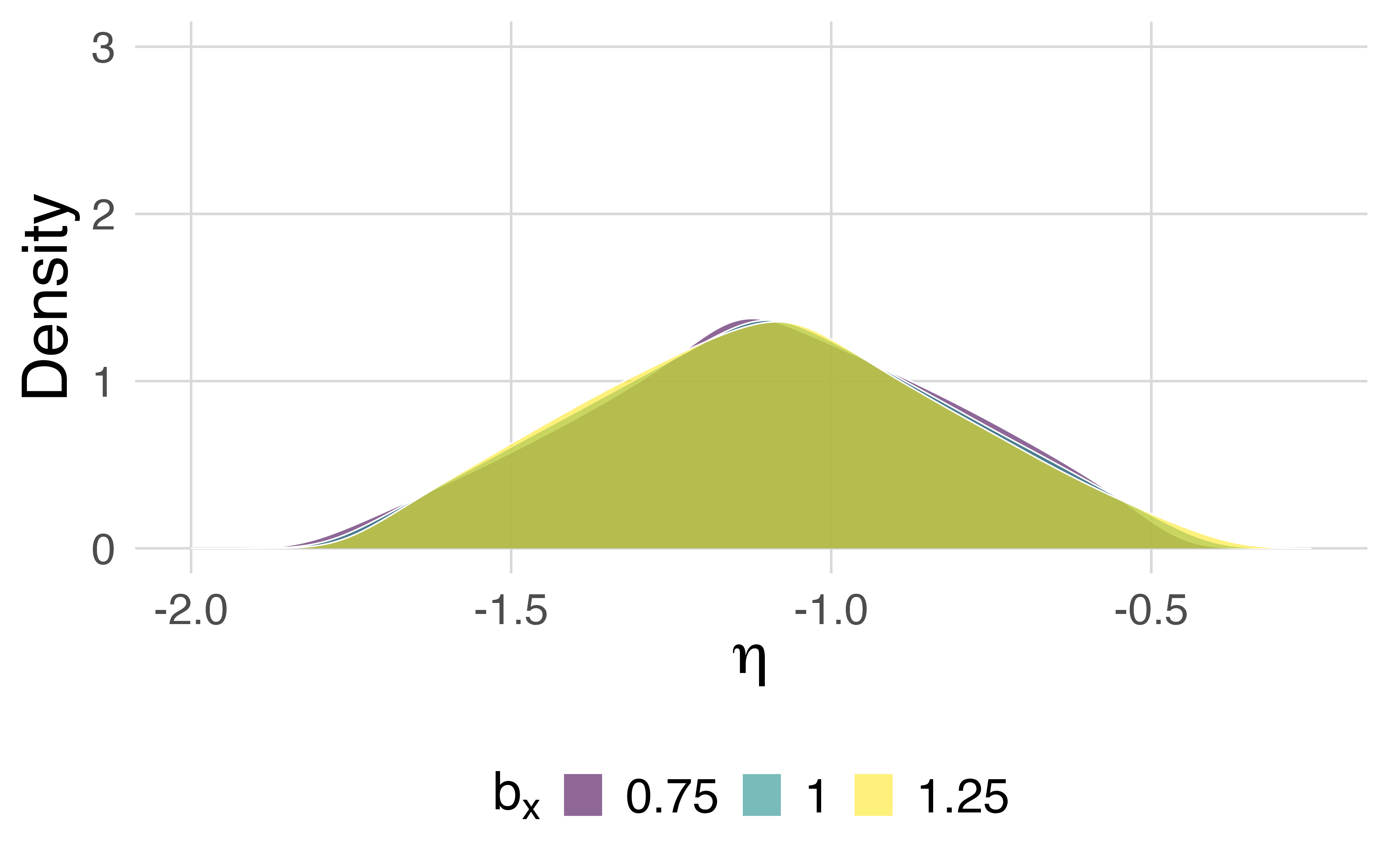}};
     \node at (11,-7) {\includegraphics[width=.33\textwidth]{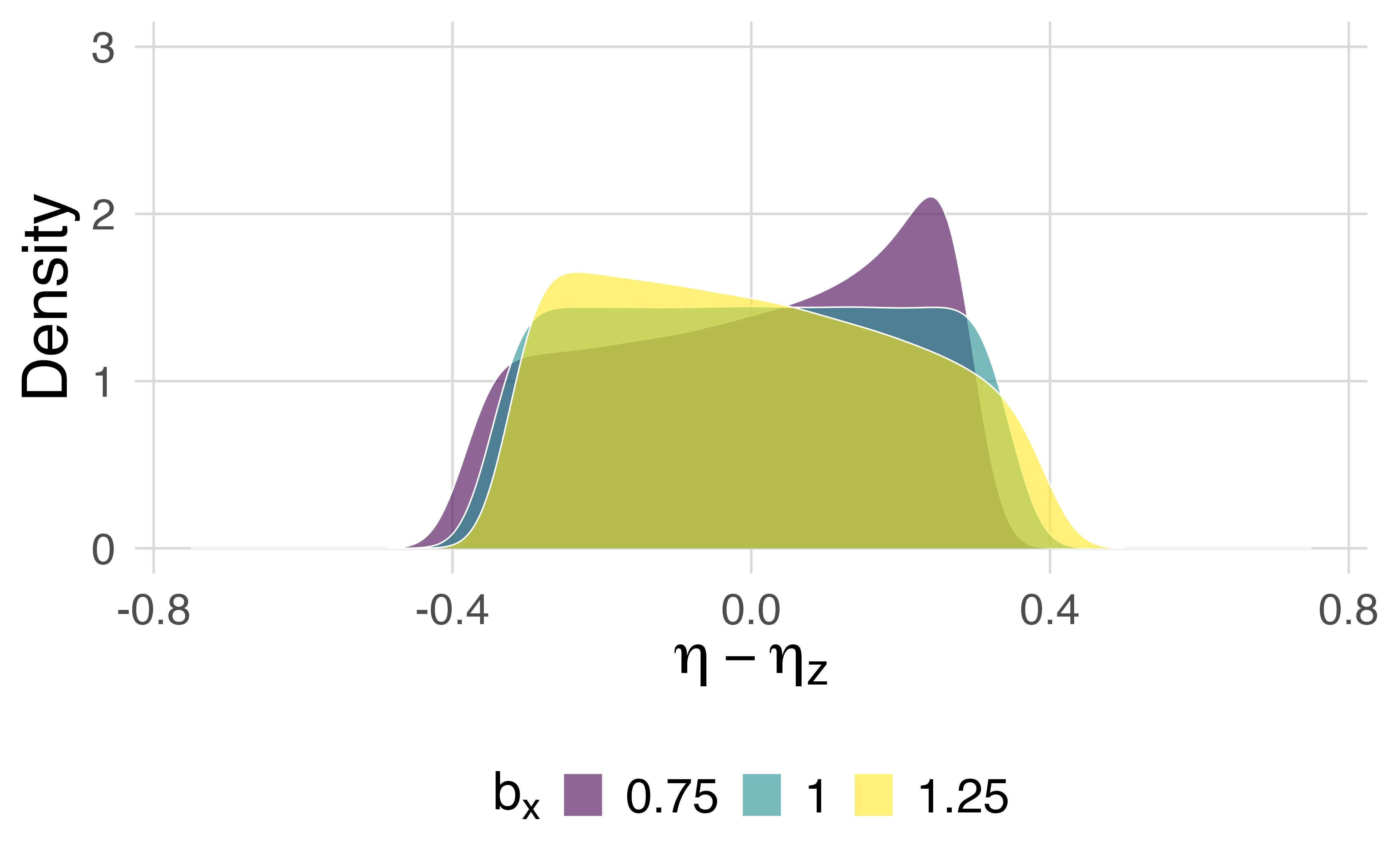}};
     \fill[white] (-2,-8.6) rectangle (13,-8.2);
    \node at (5.5,-7) {\includegraphics[width=.33\textwidth]{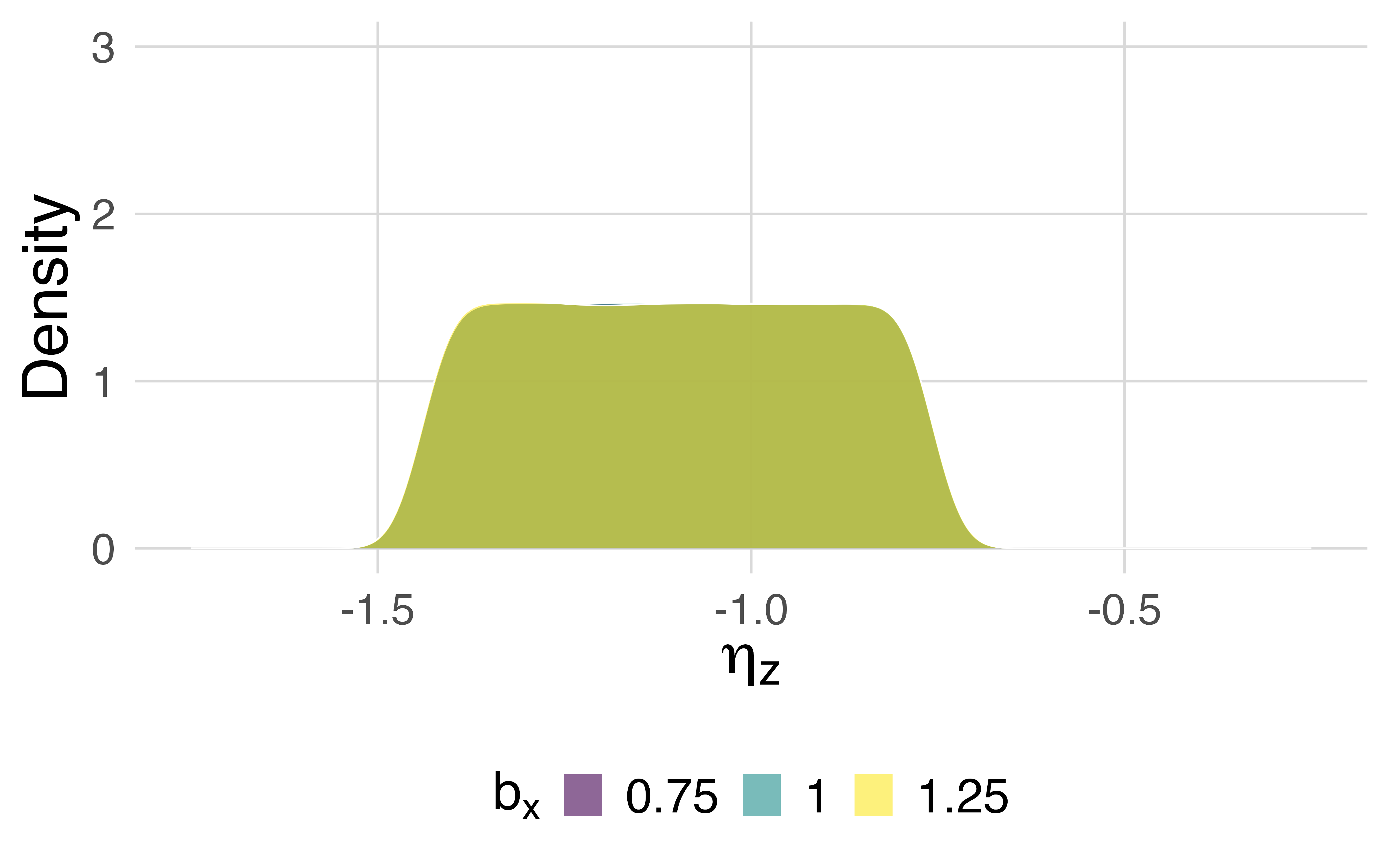}};
\end{tikzpicture}
\caption{Simulated distributions of $\eta$, $\eta_z$, and $\eta - \eta_z$ under logistic regression, as we vary three parameters that affect the distribution for $\bbe'\bX$ ($=c_2 B_x$): standard deviation $s_x$, shape parameter $a_x$, and shape parameter $b_x$. Unless otherwise noted, we fix parameters at $a_x = b_x = a_z = b_z = 1$, $s_x = s_z = 0.2$, $\rho = 0$, and $g^{-1}(\iota) = 0.25$.}
\label{fig:dist_logistic_app1}
\end{figure}

\newpage

\begin{figure}[H]
\centering
\begin{tikzpicture}
    \node at (0,0) {\includegraphics[width=.33\textwidth]{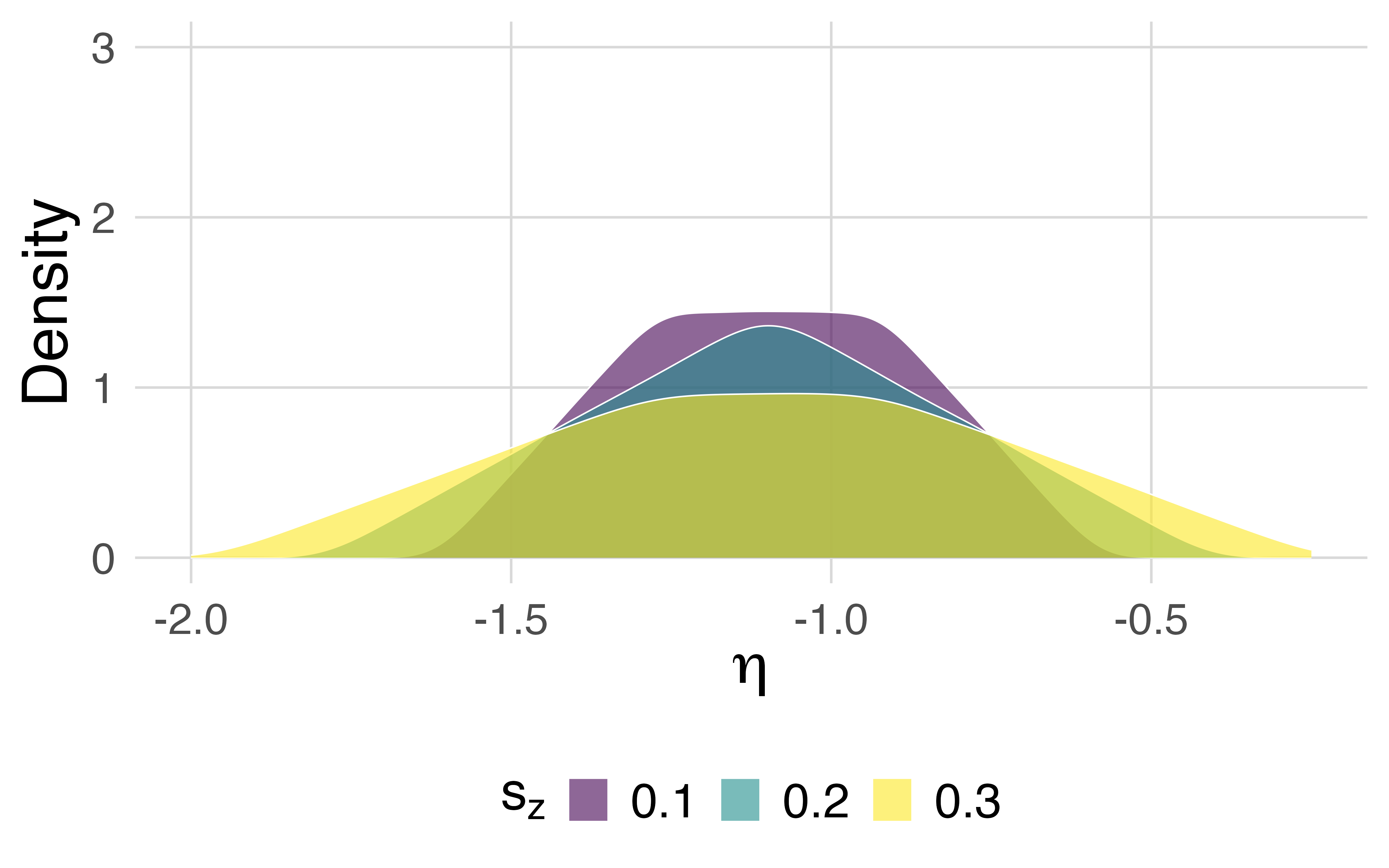}};
     \node at (11,0) {\includegraphics[width=.33\textwidth]{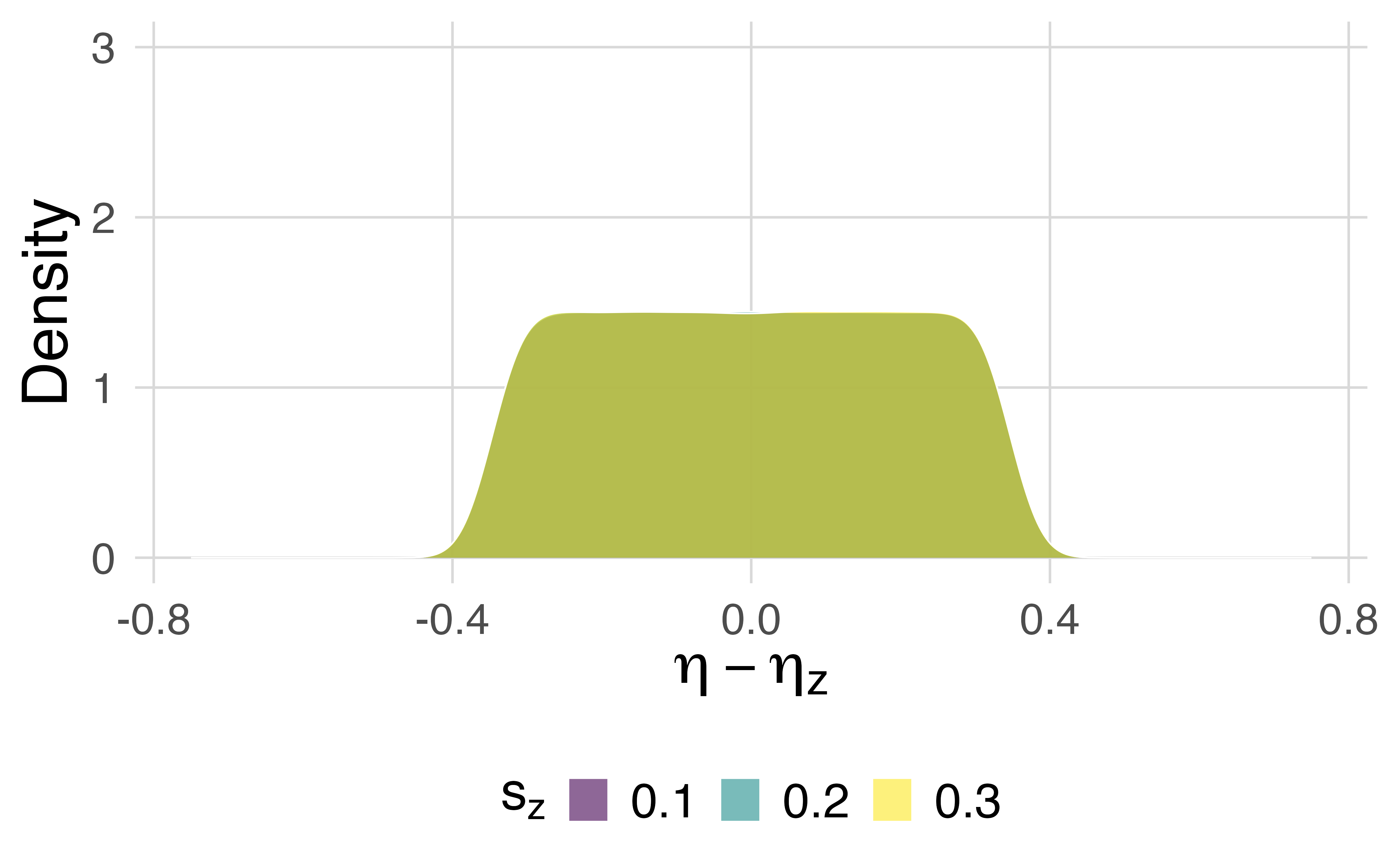}};
     \fill[white] (-2,-1.6) rectangle (13,-1.2);
    \node at (5.5,0) {\includegraphics[width=.33\textwidth]{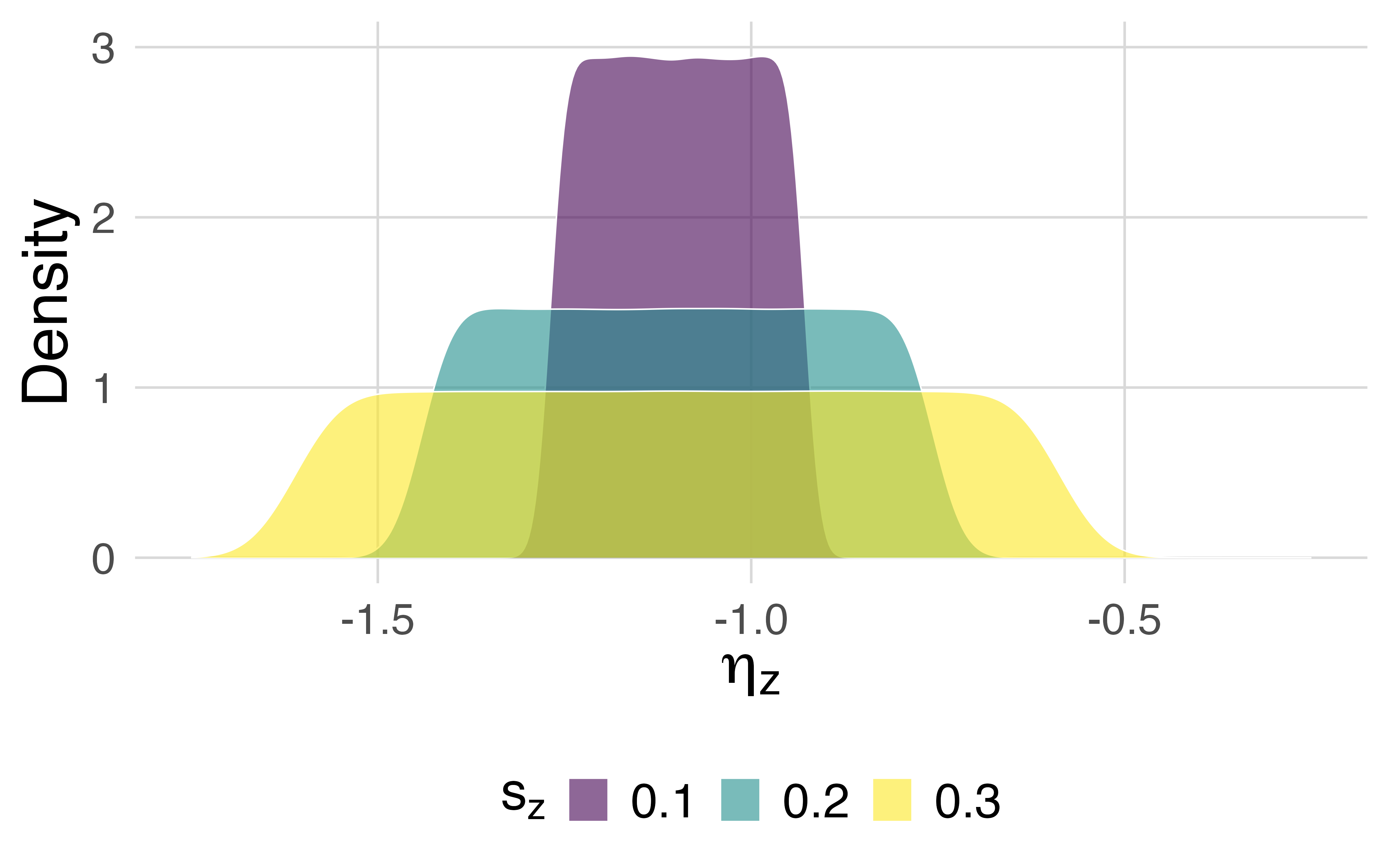}};
    \node at (0,-3.5) {\includegraphics[width=.33\textwidth]{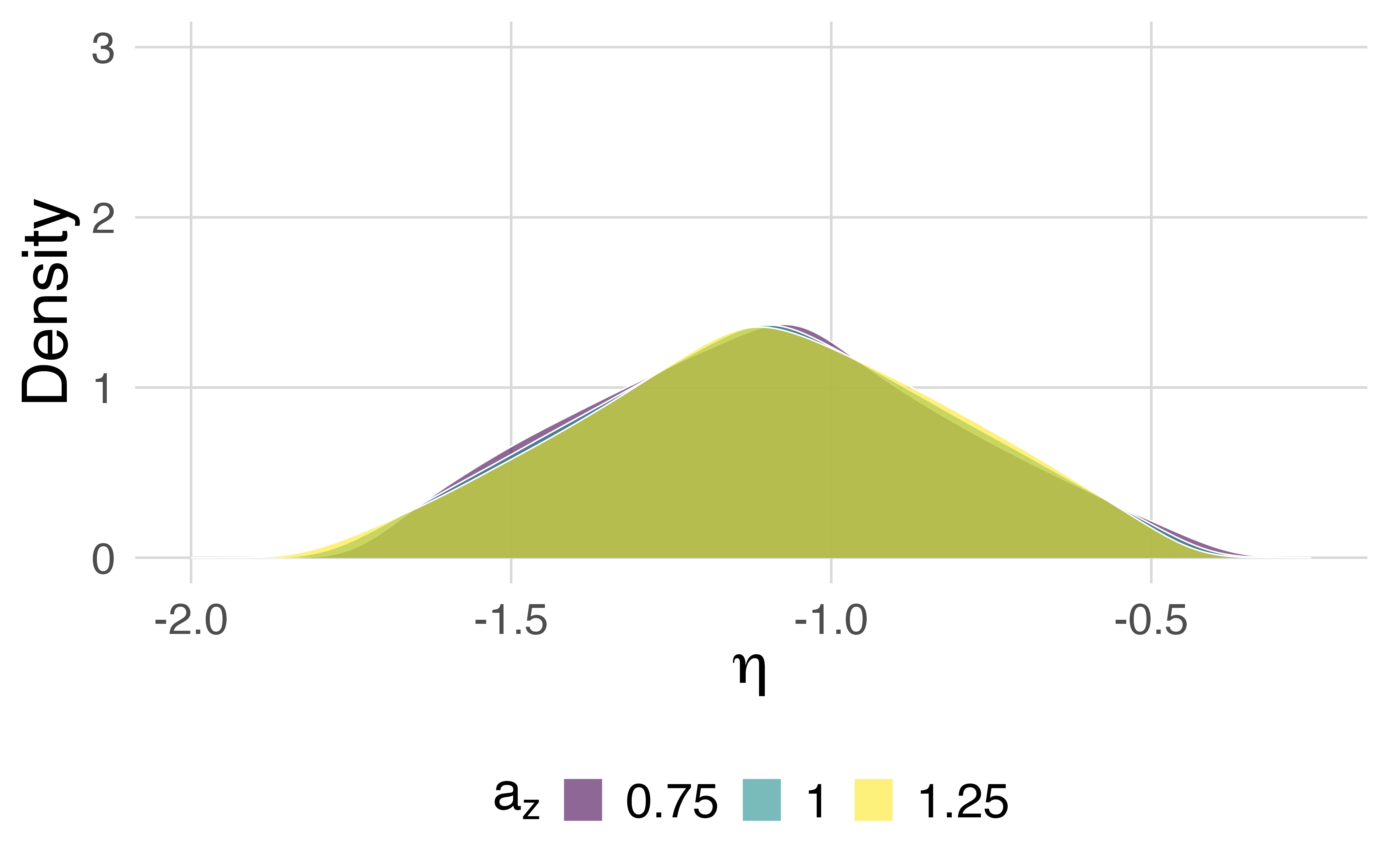}};
     \node at (11,-3.5) {\includegraphics[width=.33\textwidth]{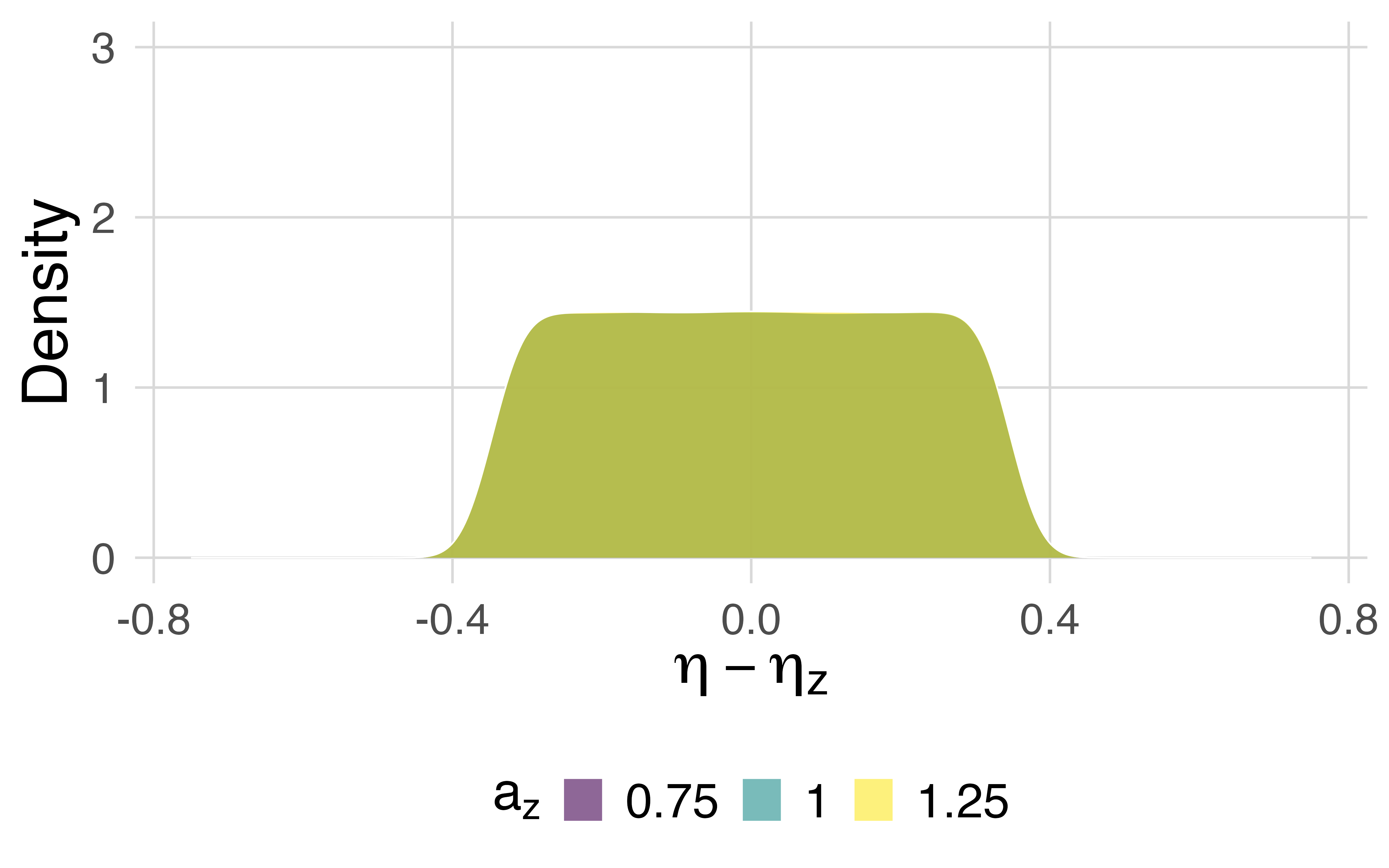}};
     \fill[white] (-2,-5.1) rectangle (13,-4.7);
    \node at (5.5,-3.5) {\includegraphics[width=.33\textwidth]{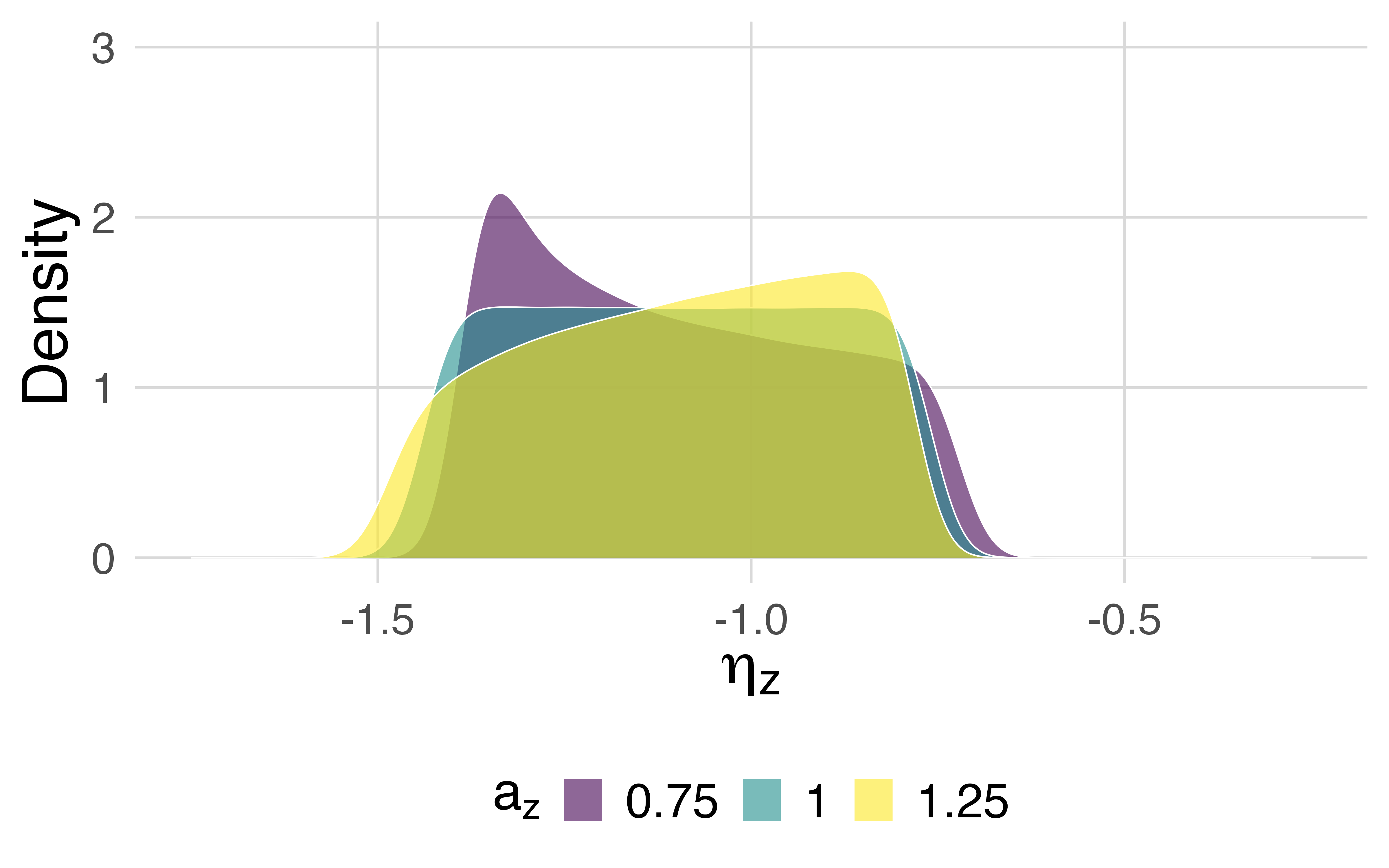}};
     \node at (0,-7) {\includegraphics[width=.33\textwidth]{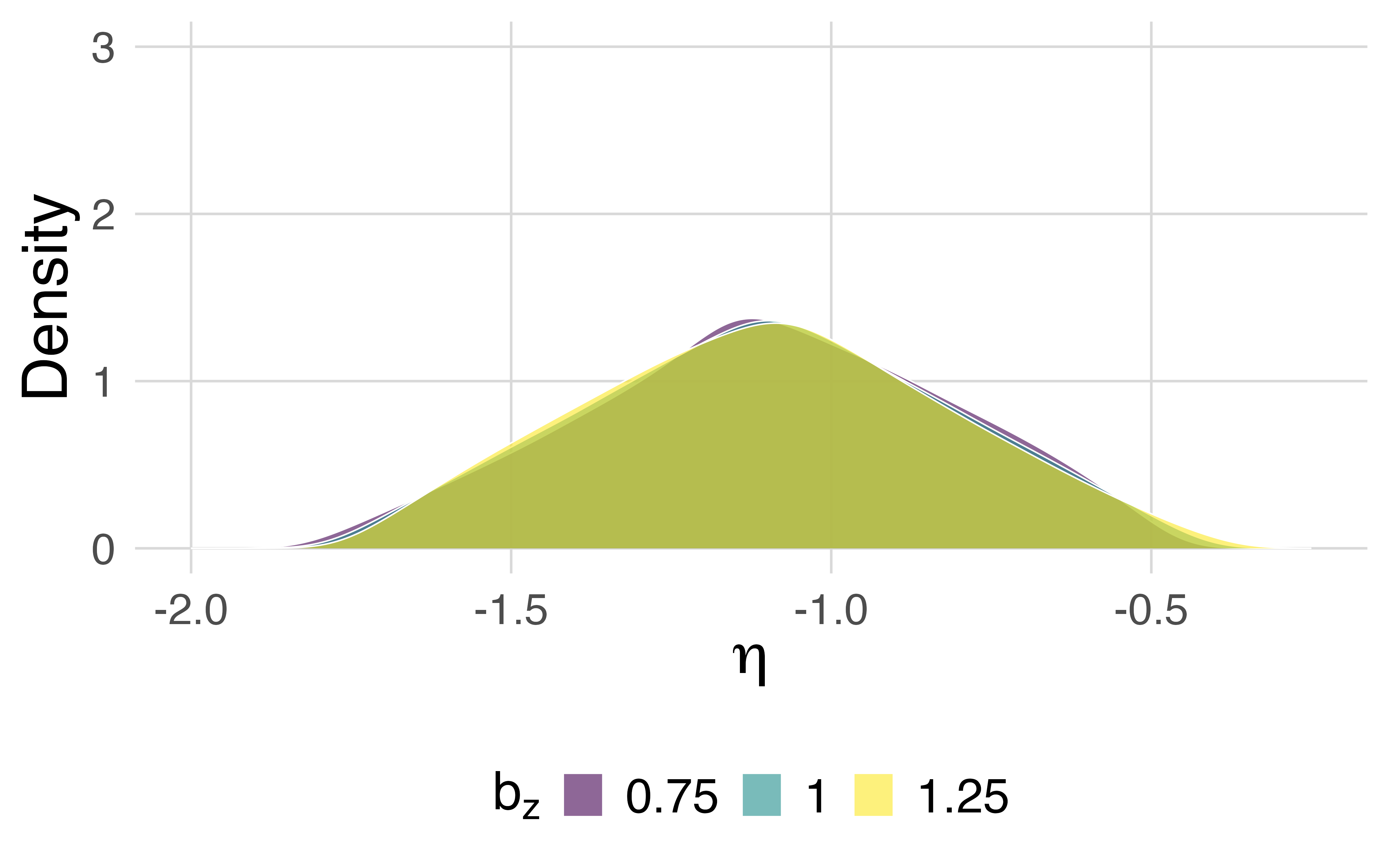}};
     \node at (11,-7) {\includegraphics[width=.33\textwidth]{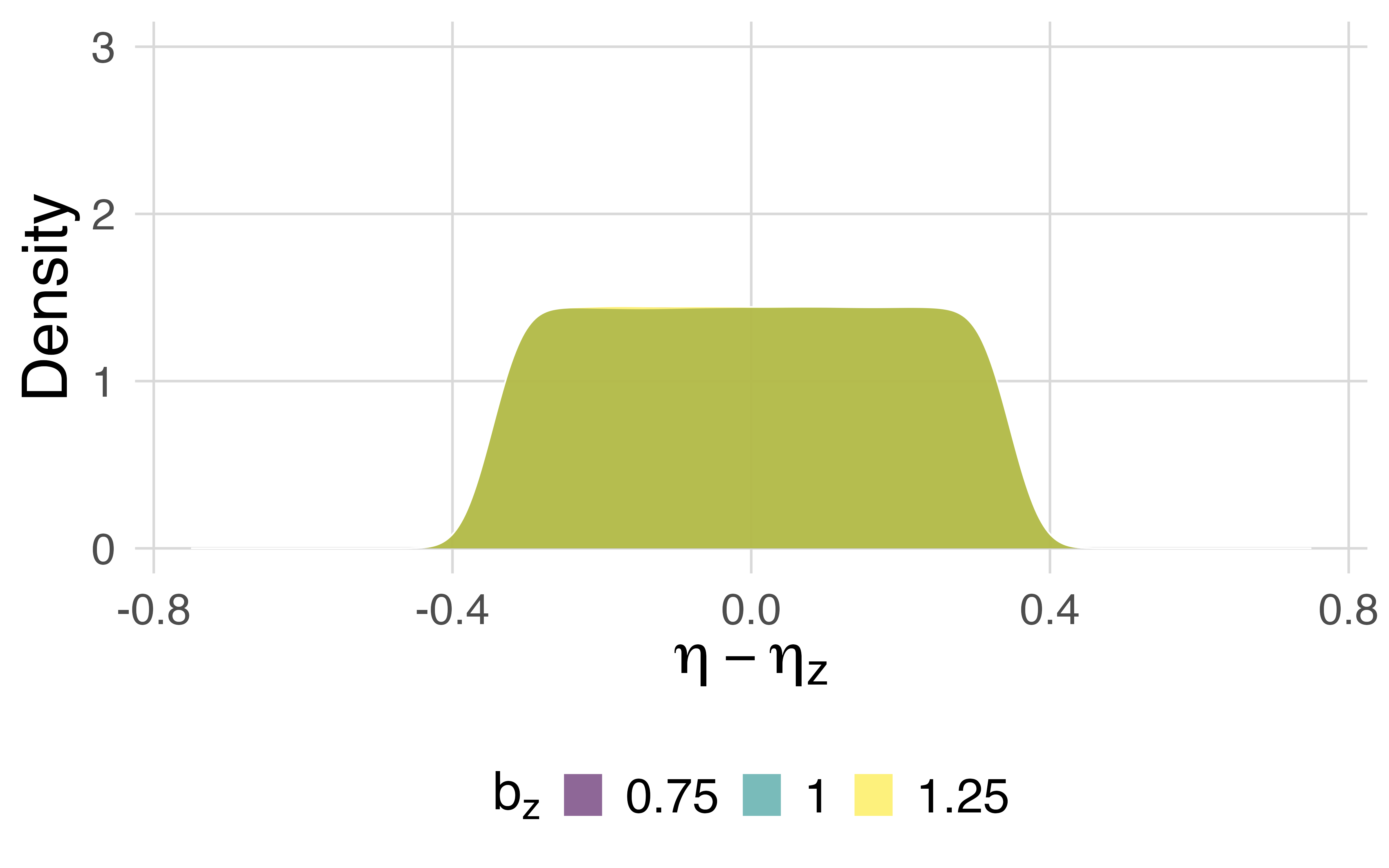}};
     \fill[white] (-2,-8.6) rectangle (13,-8.2);
    \node at (5.5,-7) {\includegraphics[width=.33\textwidth]{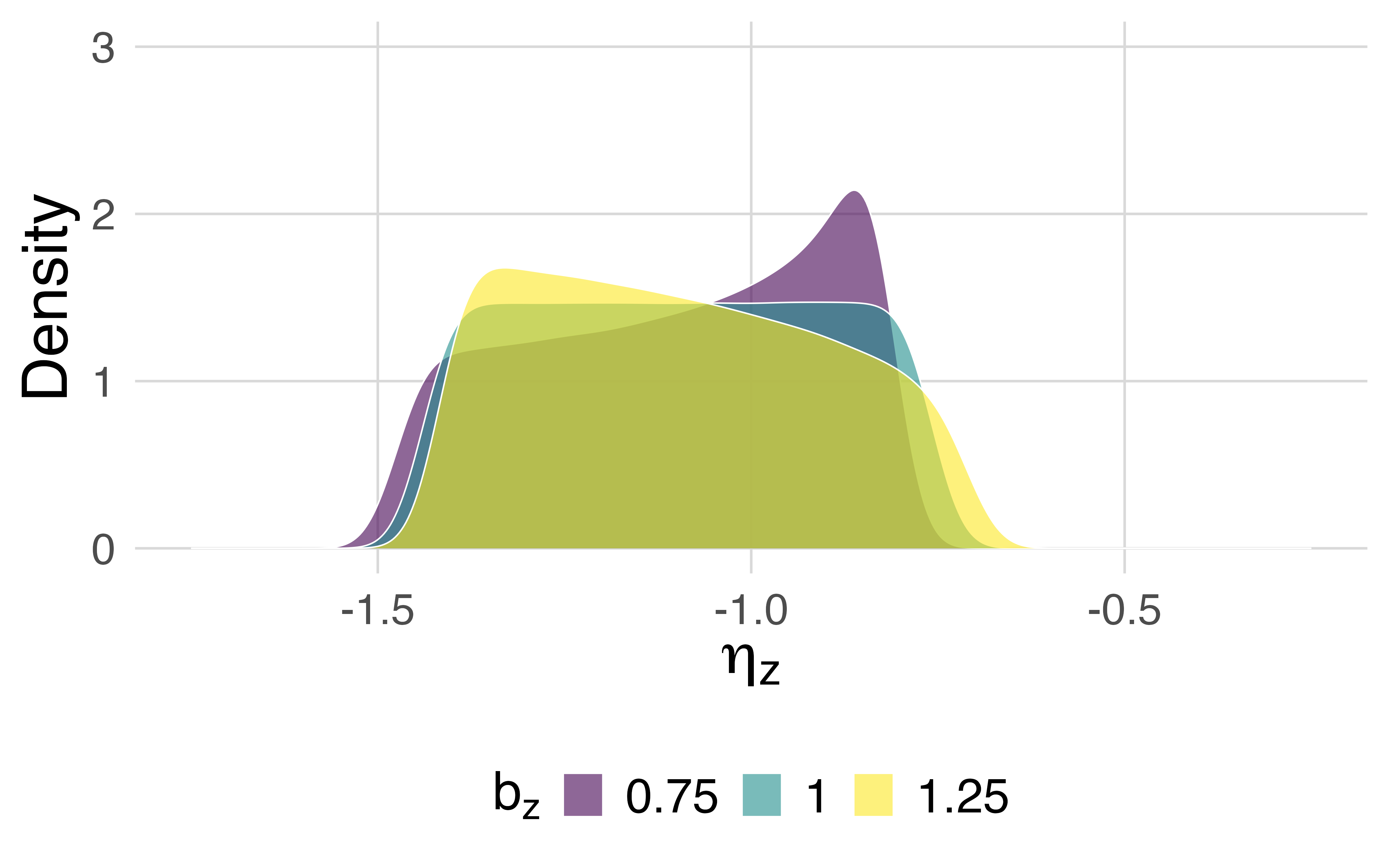}};
    \node at (0,-10.5) {\includegraphics[width=.33\textwidth]{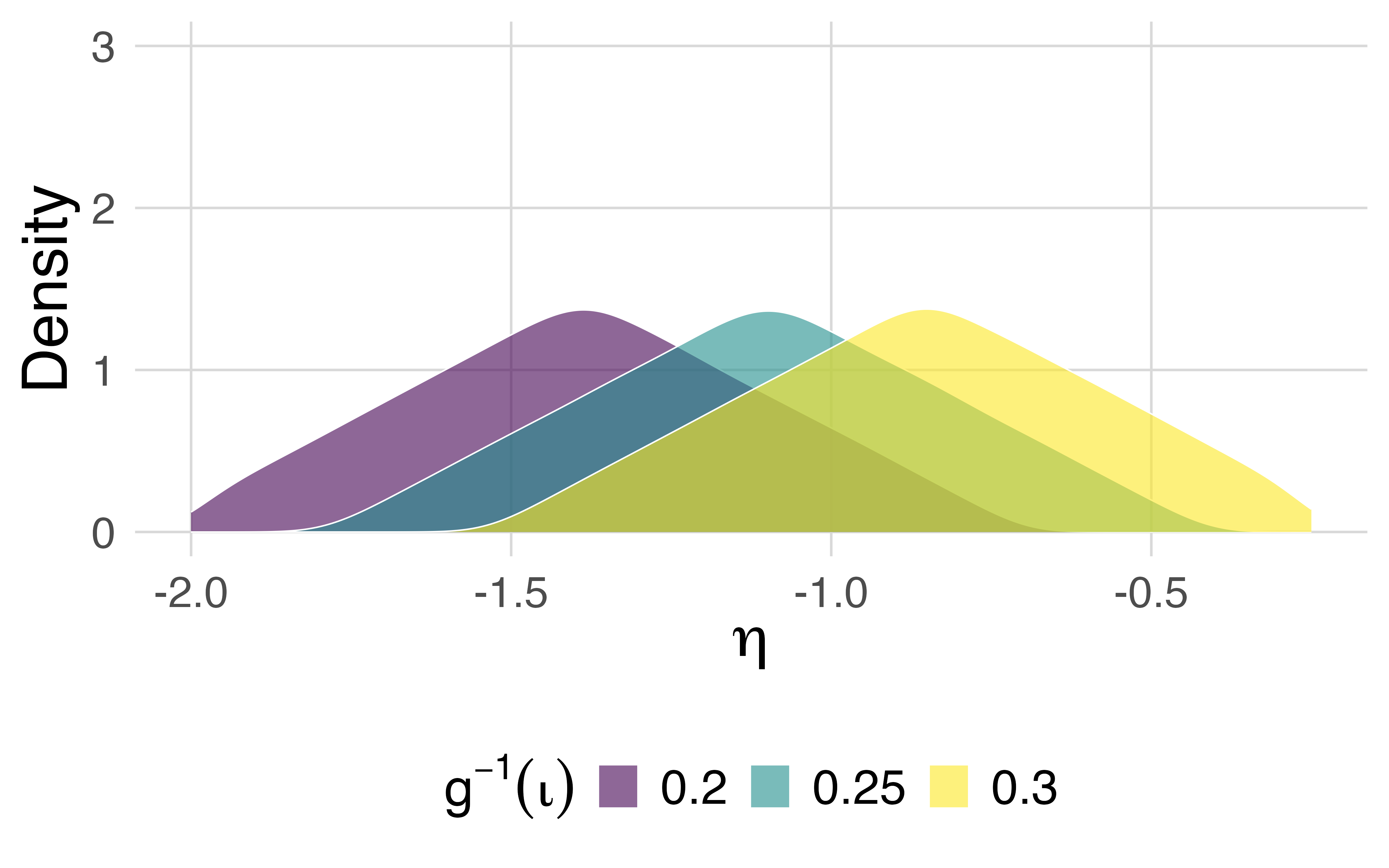}};
     \node at (11,-10.5) {\includegraphics[width=.33\textwidth]{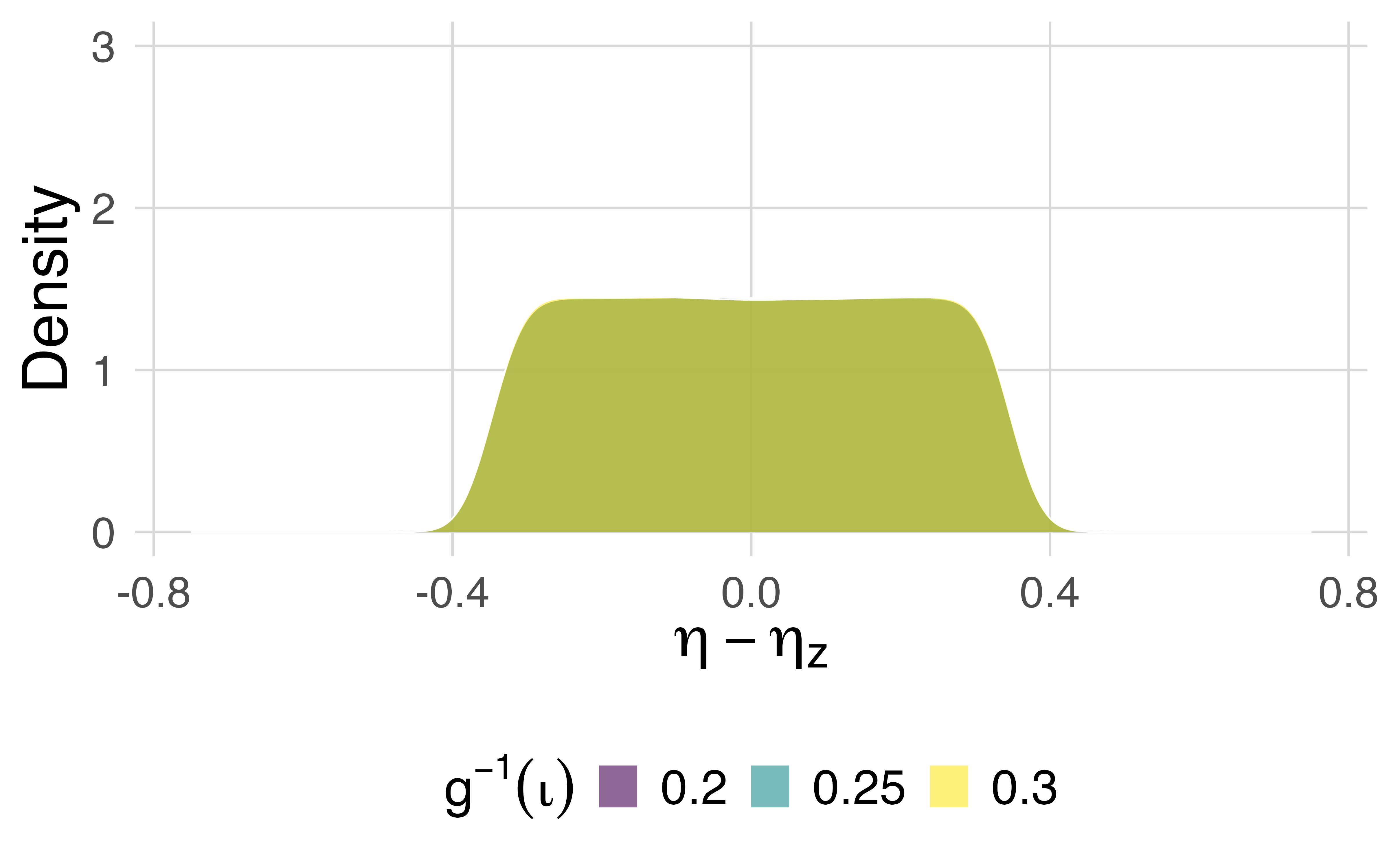}};
     \fill[white] (-2,-8.6) rectangle (13,-8.2);
    \node at (5.5,-10.5) {\includegraphics[width=.33\textwidth]{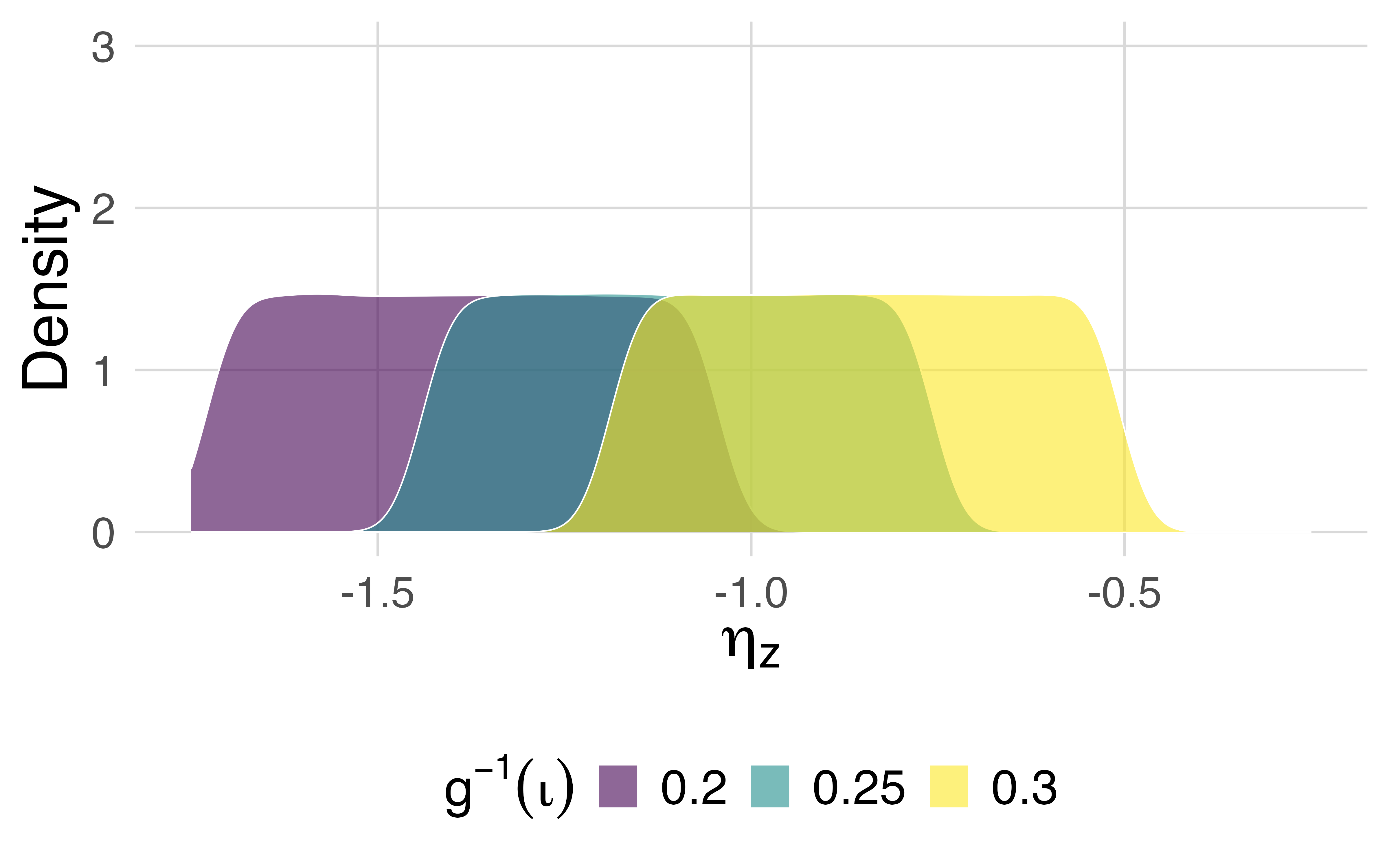}};
\node at (0,-14) {\includegraphics[width=.33\textwidth]{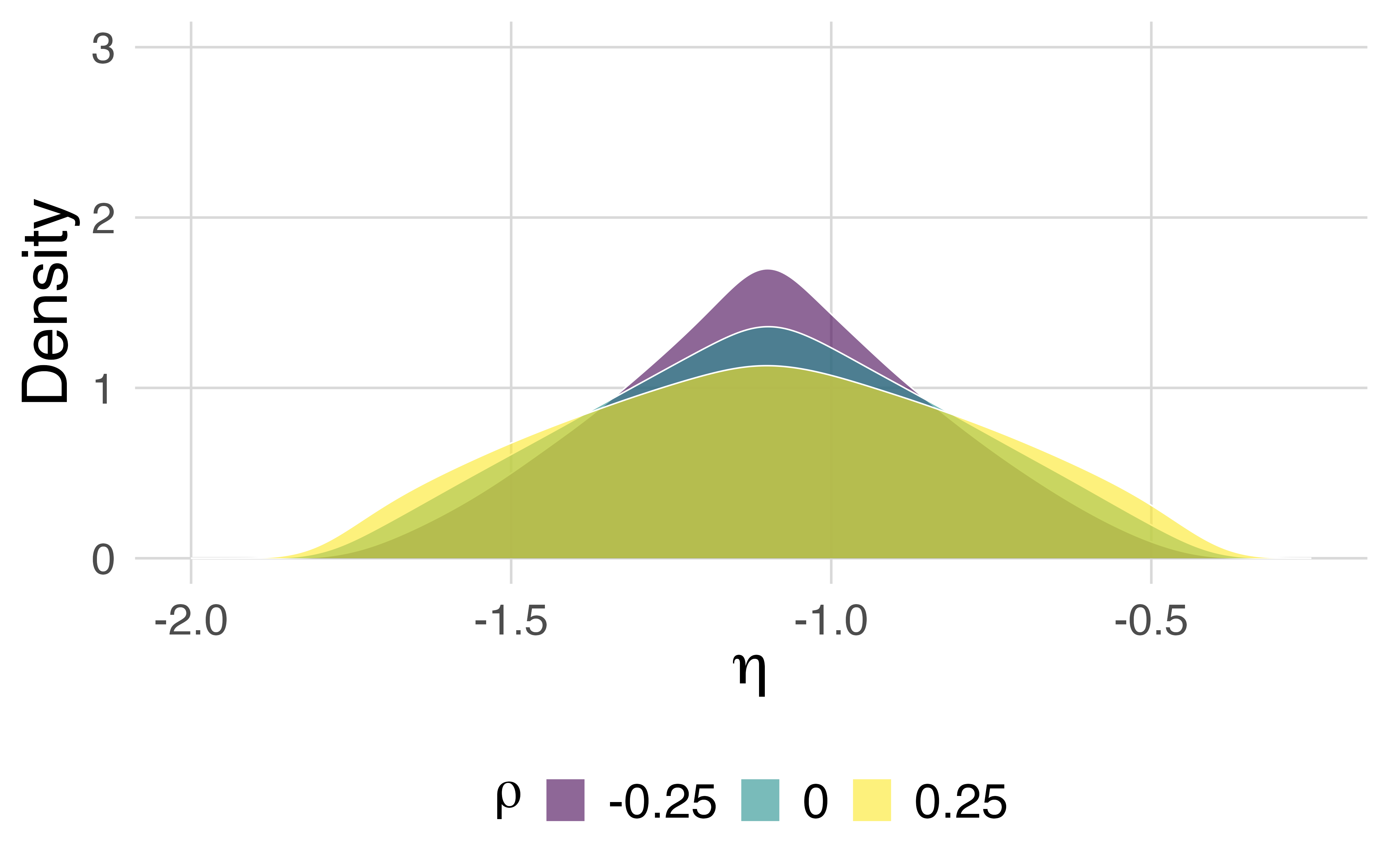}};
     \node at (11,-14) {\includegraphics[width=.33\textwidth]{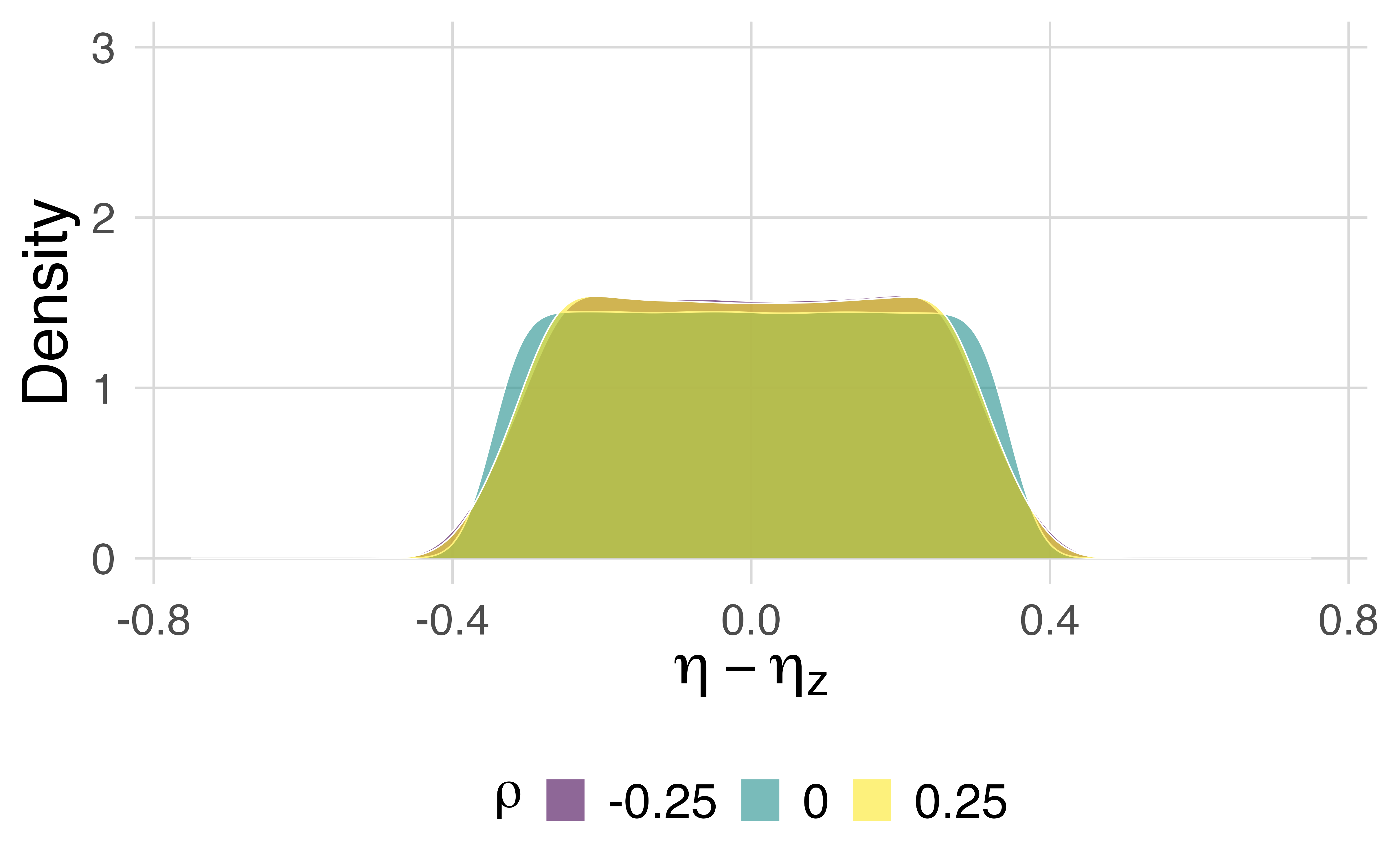}};
     \fill[white] (-2,-8.6) rectangle (13,-8.2);
    \node at (5.5,-14) {\includegraphics[width=.33\textwidth]{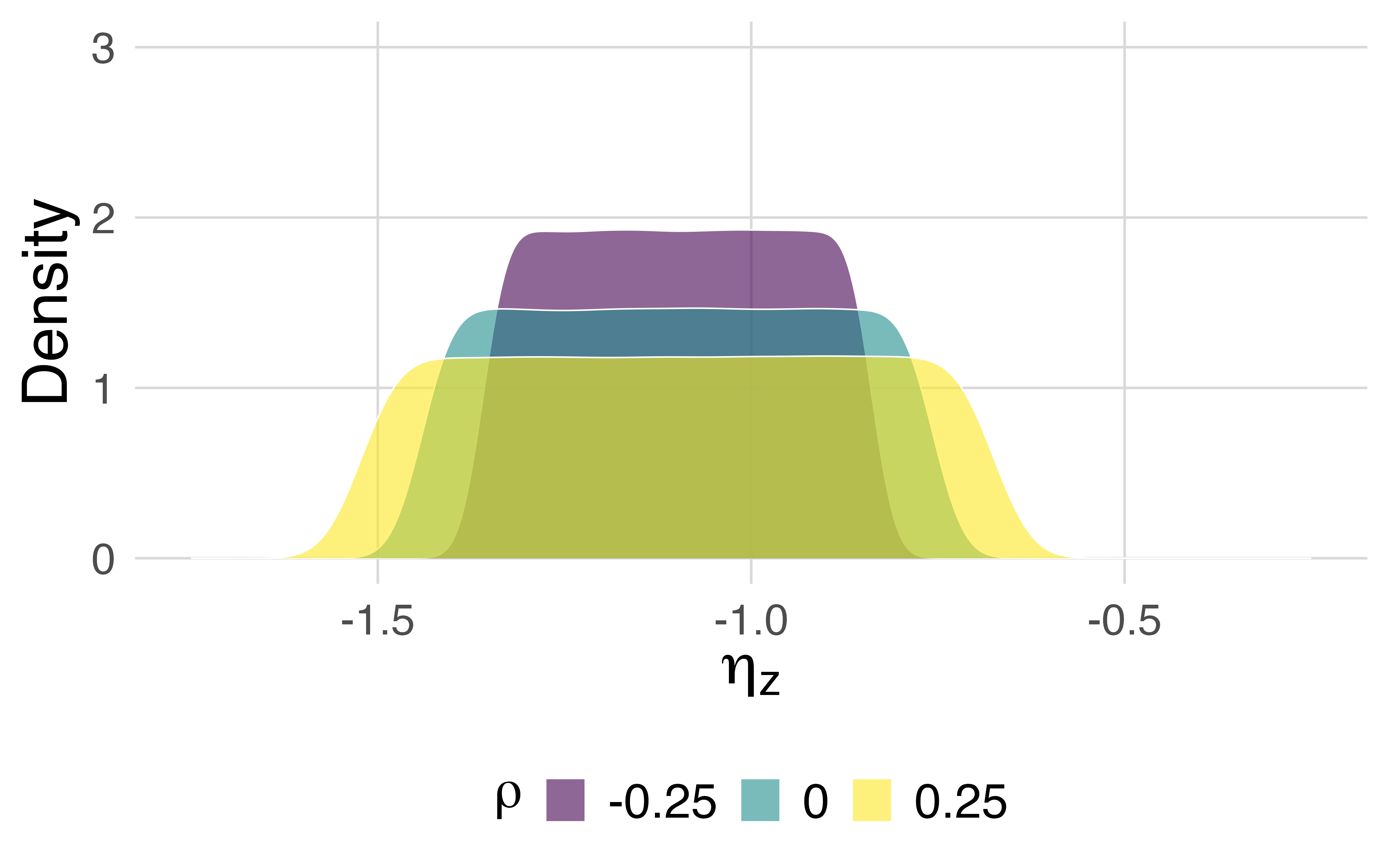}};
\end{tikzpicture}
\caption{Simulated distributions of $\eta$, $\eta_z$, and $\eta - \eta_z$ under logistic regression, as we vary five parameters: the standard deviation $s_z$, the shape parameter $a_z$, the shape parameter $b_z$, reference mean $g^{-1}(\iota)$, and the correlation $\rho$. Unless otherwise noted, we fix parameters at $a_x = b_x = a_z = b_z = 1$, $s_x = s_z = 0.2$, $\rho = 0$, and $g^{-1}(\iota) = 0.25$.}
\label{fig:dist_logistic_app2}
\end{figure}

\newpage
\section{Additional simulations}\label{app:extra_simulations}

\subsection{Logistic regression}

\begin{figure}[H]
\centering
\begin{tikzpicture}
    \node at (0,0) {\includegraphics[width=.48\textwidth]{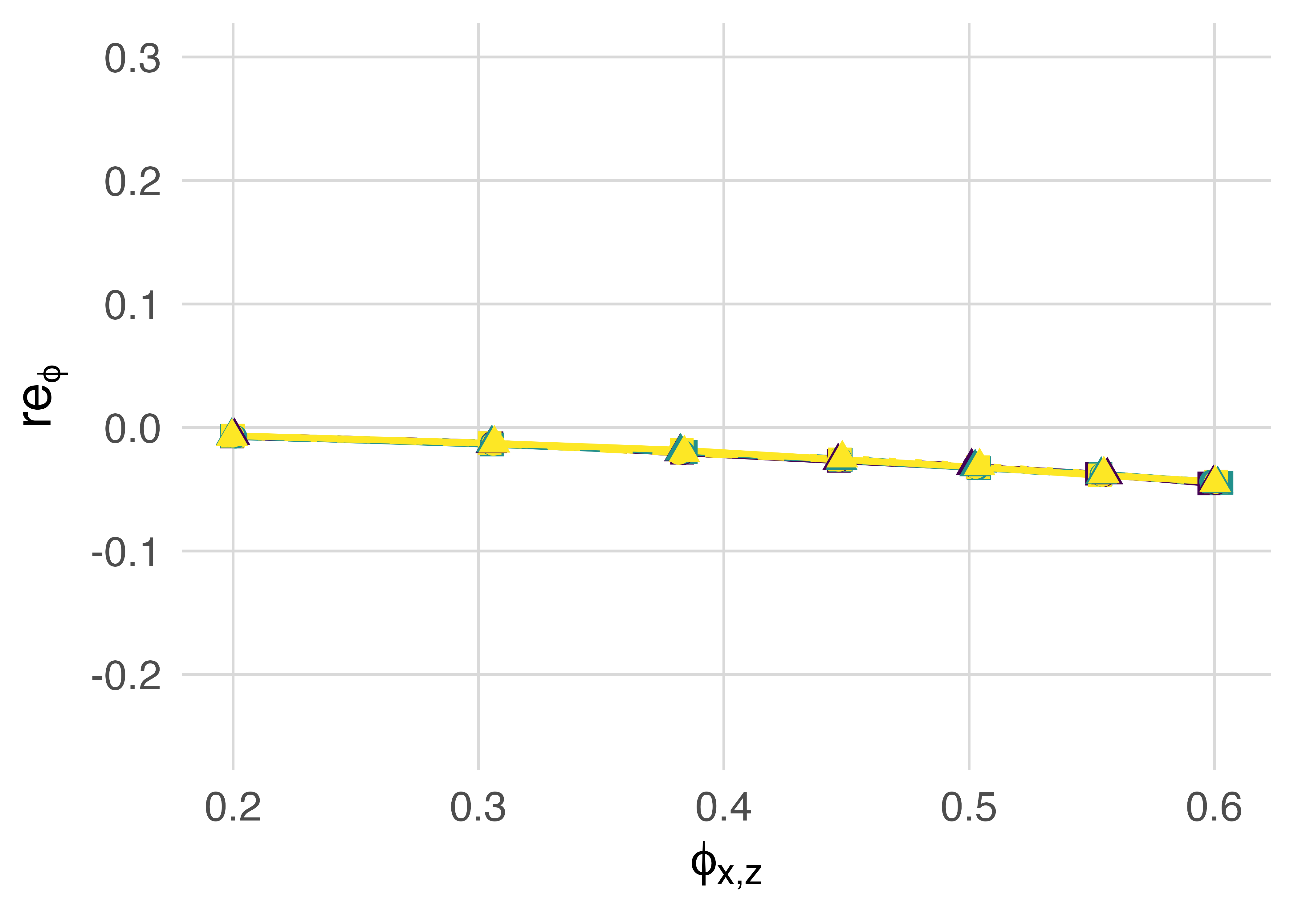}};
    \node at (8,0) {\includegraphics[width=.48\textwidth]{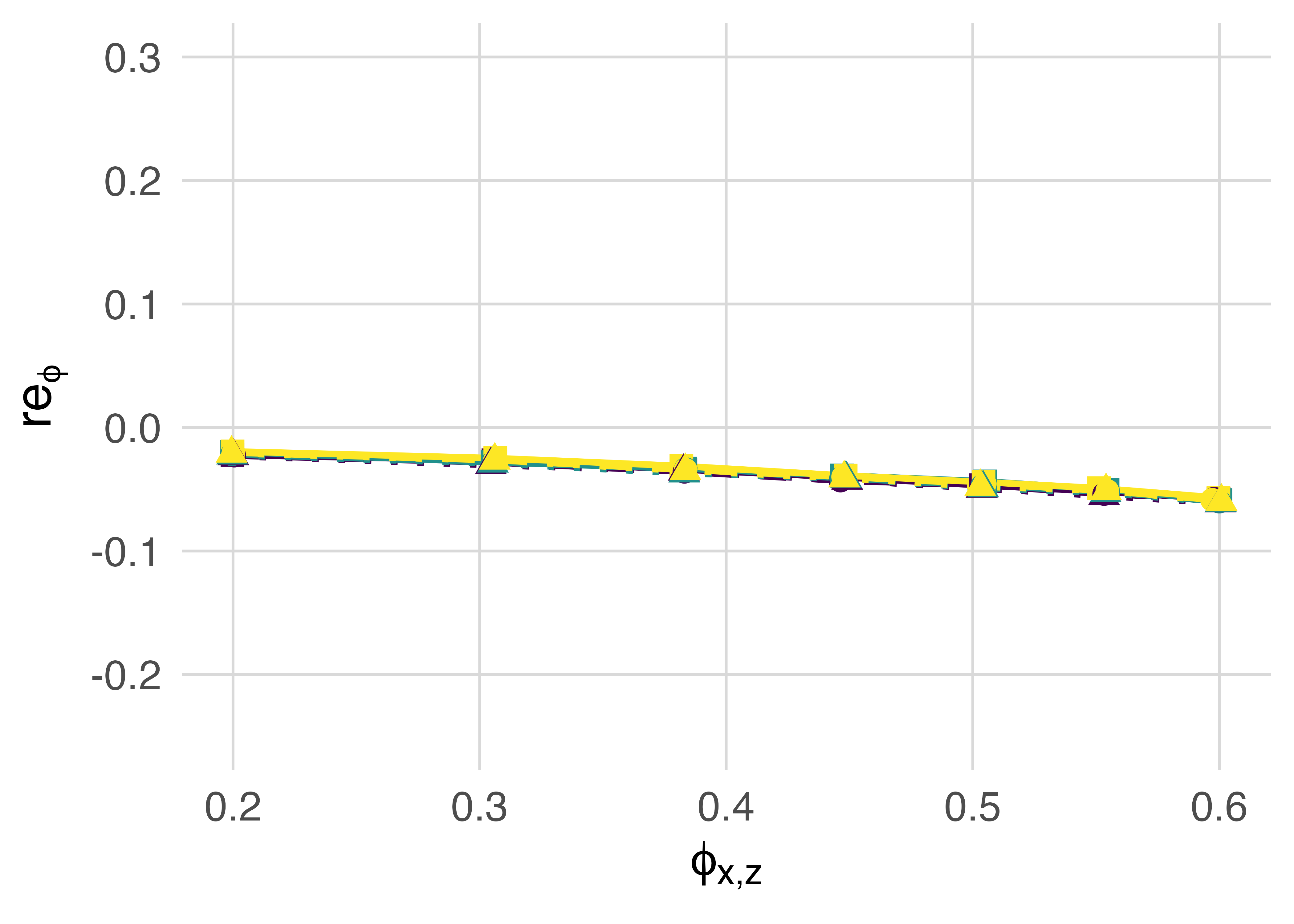}};
    \node at (0,-5.5) {\includegraphics[width=.48\textwidth]{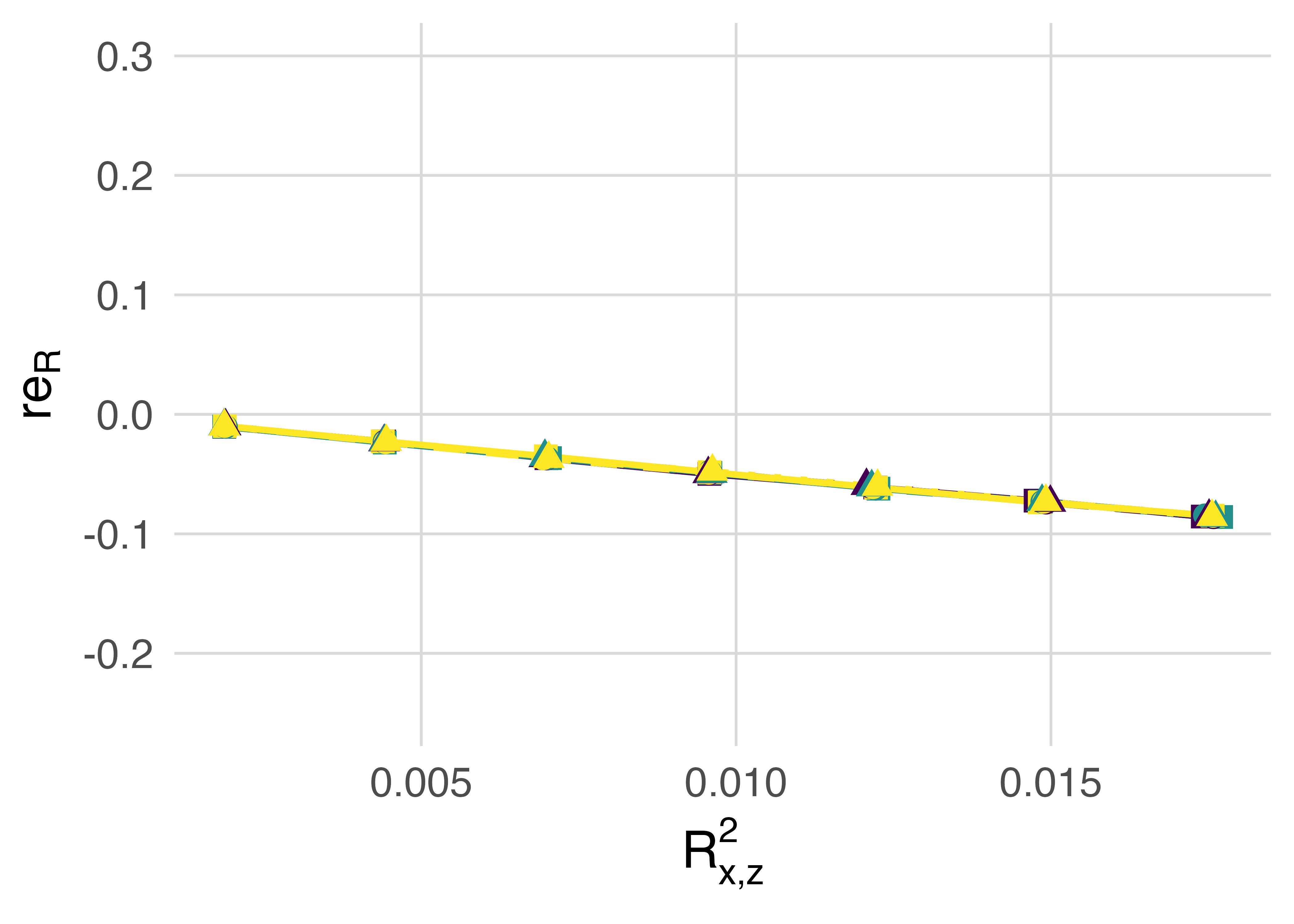}};
    \node at (8,-5.5) {\includegraphics[width=.48\textwidth]{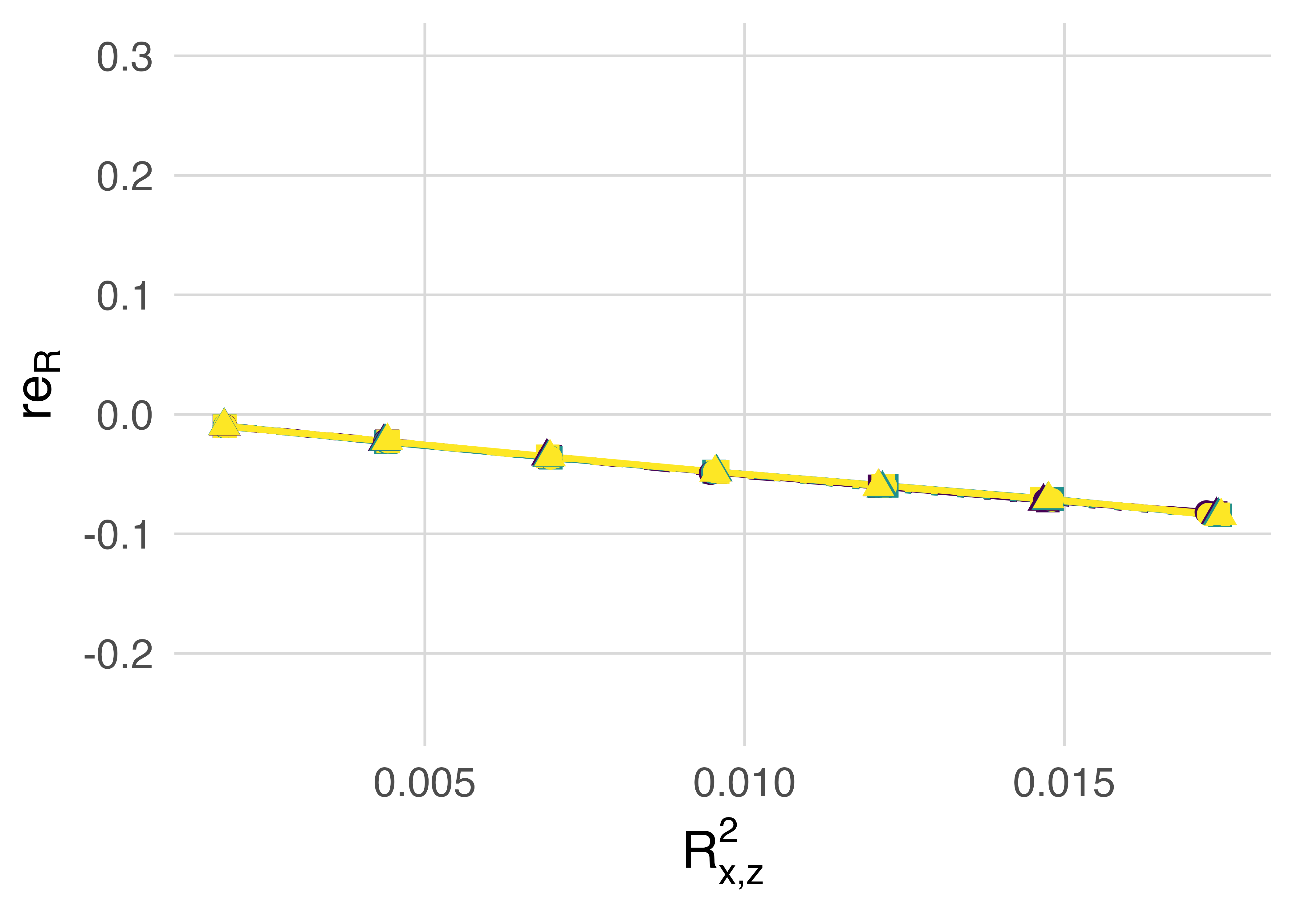}};
    \node at (4,-8.7) {\includegraphics[width=.65\textwidth,clip=true,trim=20 0 15 240]{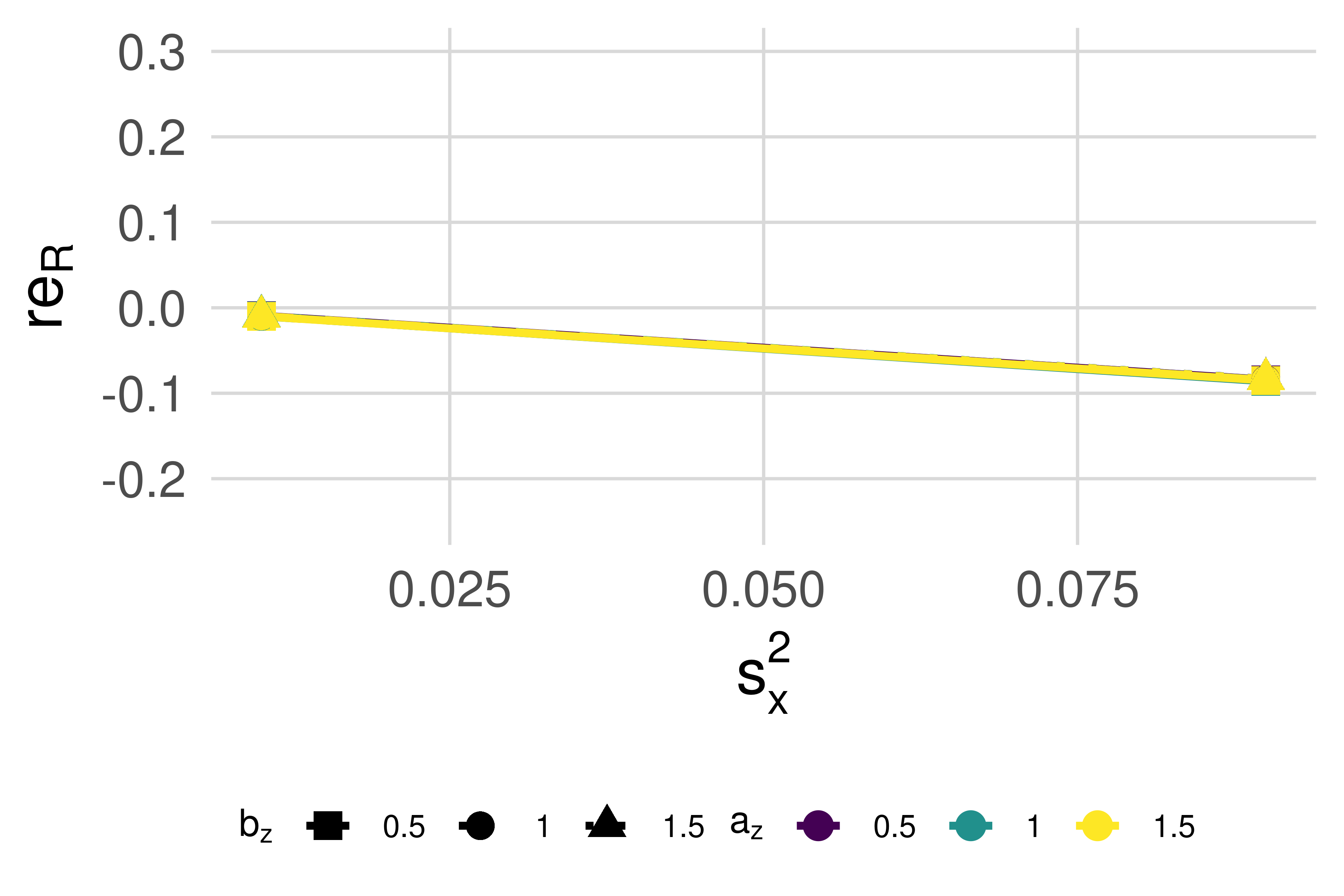}};
    \node at (0.5,3.1) {$\boldsymbol{s_z^2=0.01}$};
    \node at (8.5,3.1) {$\boldsymbol{s_z^2=0.09}$};
\end{tikzpicture}
\caption{Relative error in the approximation of $\fsq$ for logistic regression, plotted against $\phixz$ (top panels) and $\pspRsq$ (bottom panels). Left and right panels correspond to two levels of $s_z^2$ ($0.01$ and $0.09$). Within each panel, we vary $a_z$ and $b_z$ over all combinations of values in $\{0.5, 1, 1.5\}$. Although the x-axes show $\phixz$ and $\pspRsq$ directly, each point reflects an underlying value of $s_x^2$, evenly spaced from $0.01$ to $0.09$. Relative error for each measure of effect is largely insensitive to the shape parameters $a_z$ and $b_z$. Other parameters are fixed: $a_x = b_x = 1$, $\rho = 0$, and $g^{-1}(\iota) = .25$.  }
\label{fig:re_logistic_az_bz}
\end{figure}

\begin{figure}[H]
\centering
\begin{tikzpicture}
    \node at (0,0) {\includegraphics[width=.48\textwidth]{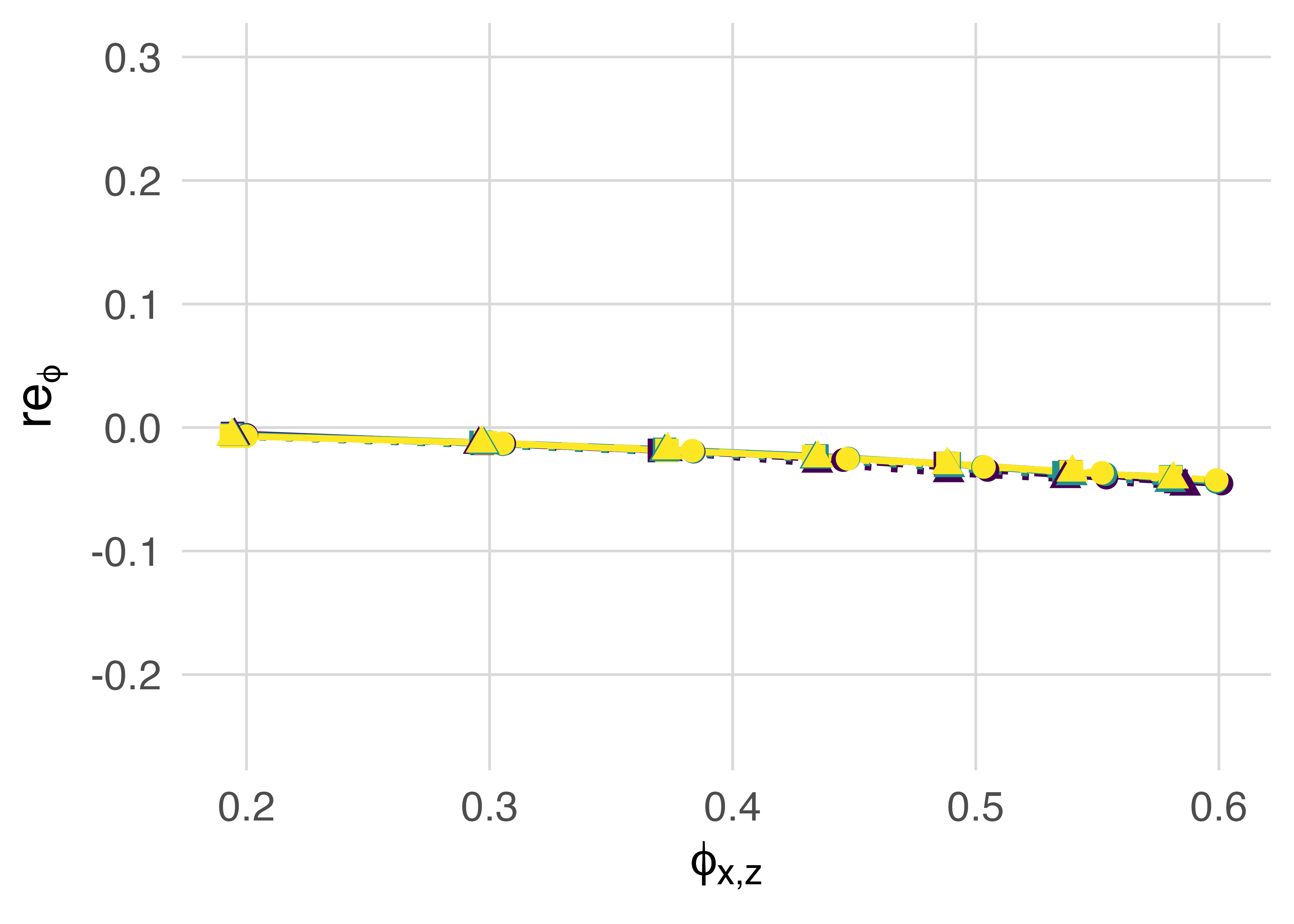}};
    \node at (8,0) {\includegraphics[width=.48\textwidth]{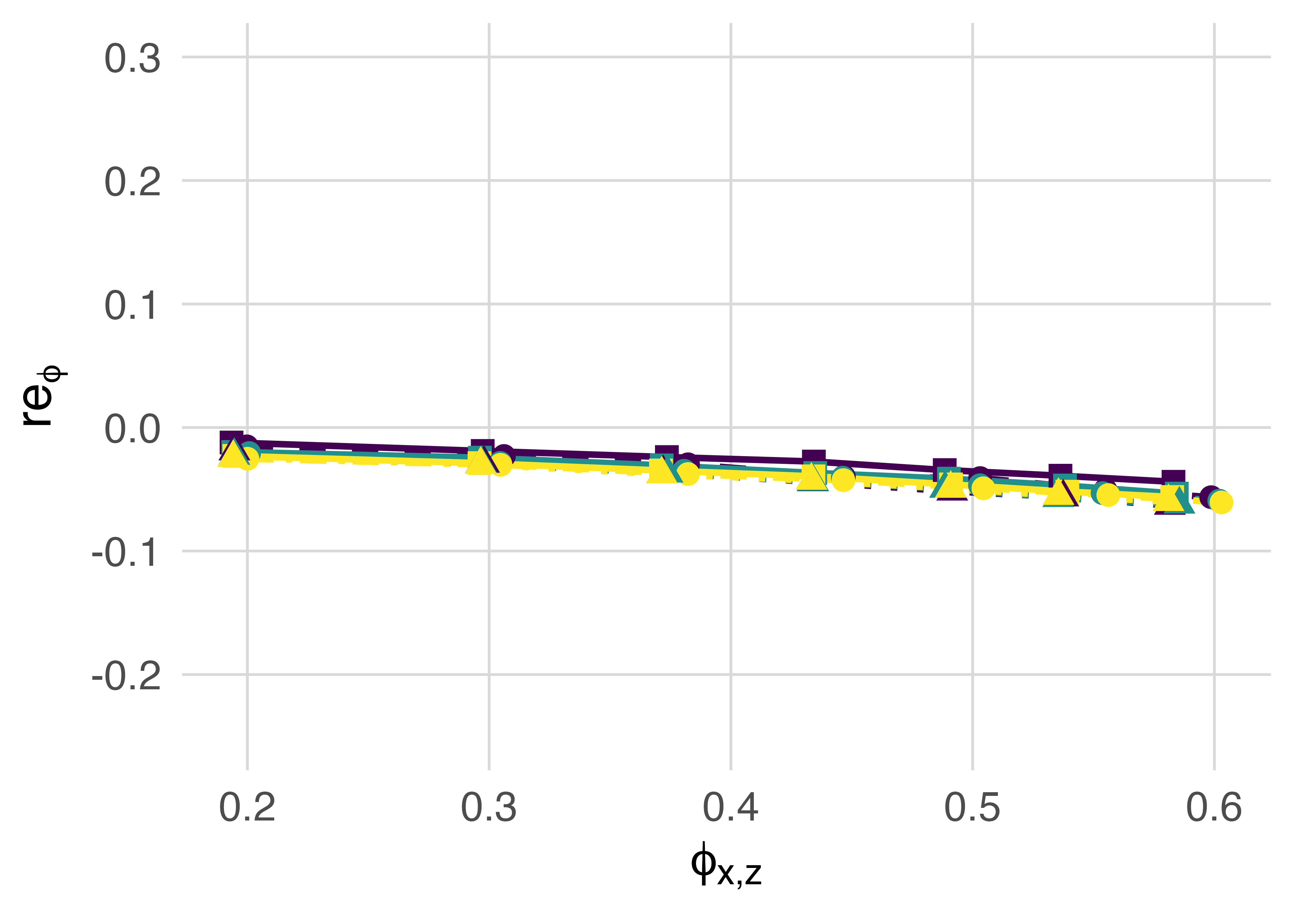}};
    \node at (0,-5.5) {\includegraphics[width=.48\textwidth]{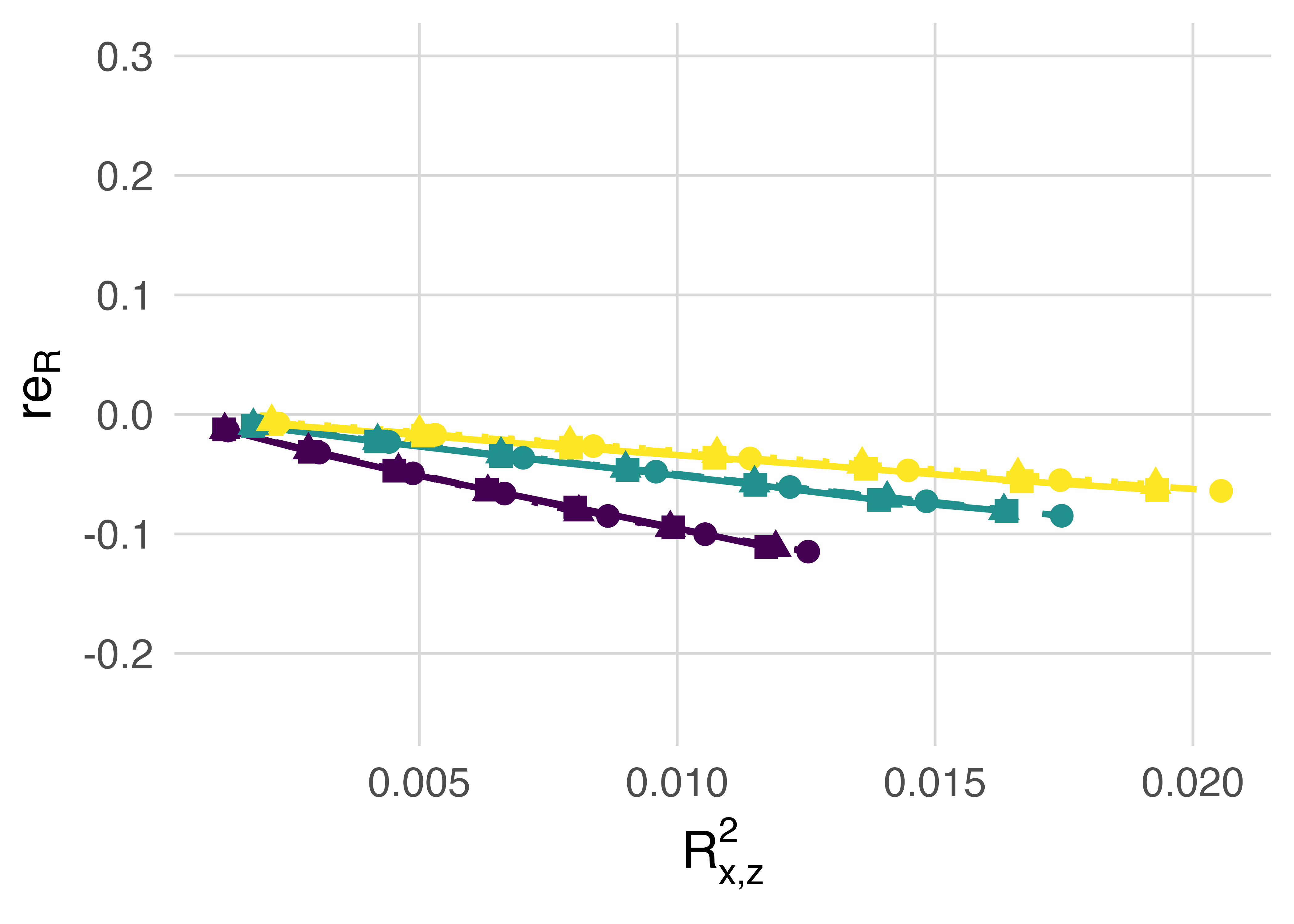}};
    \node at (8,-5.5) {\includegraphics[width=.48\textwidth]{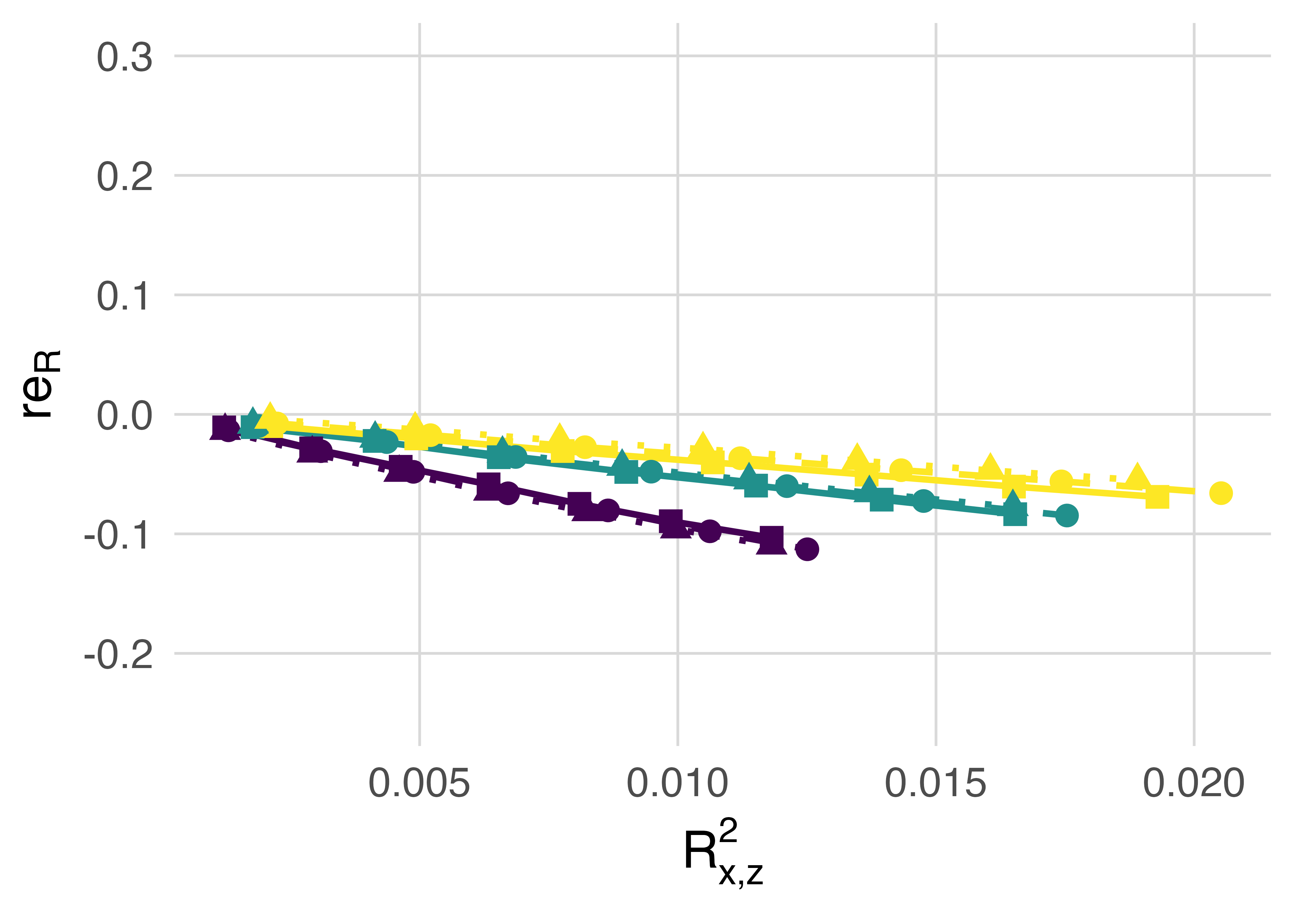}};
    \node at (4,-8.7) {\includegraphics[width=.65\textwidth,clip=true,trim=20 0 15 240]{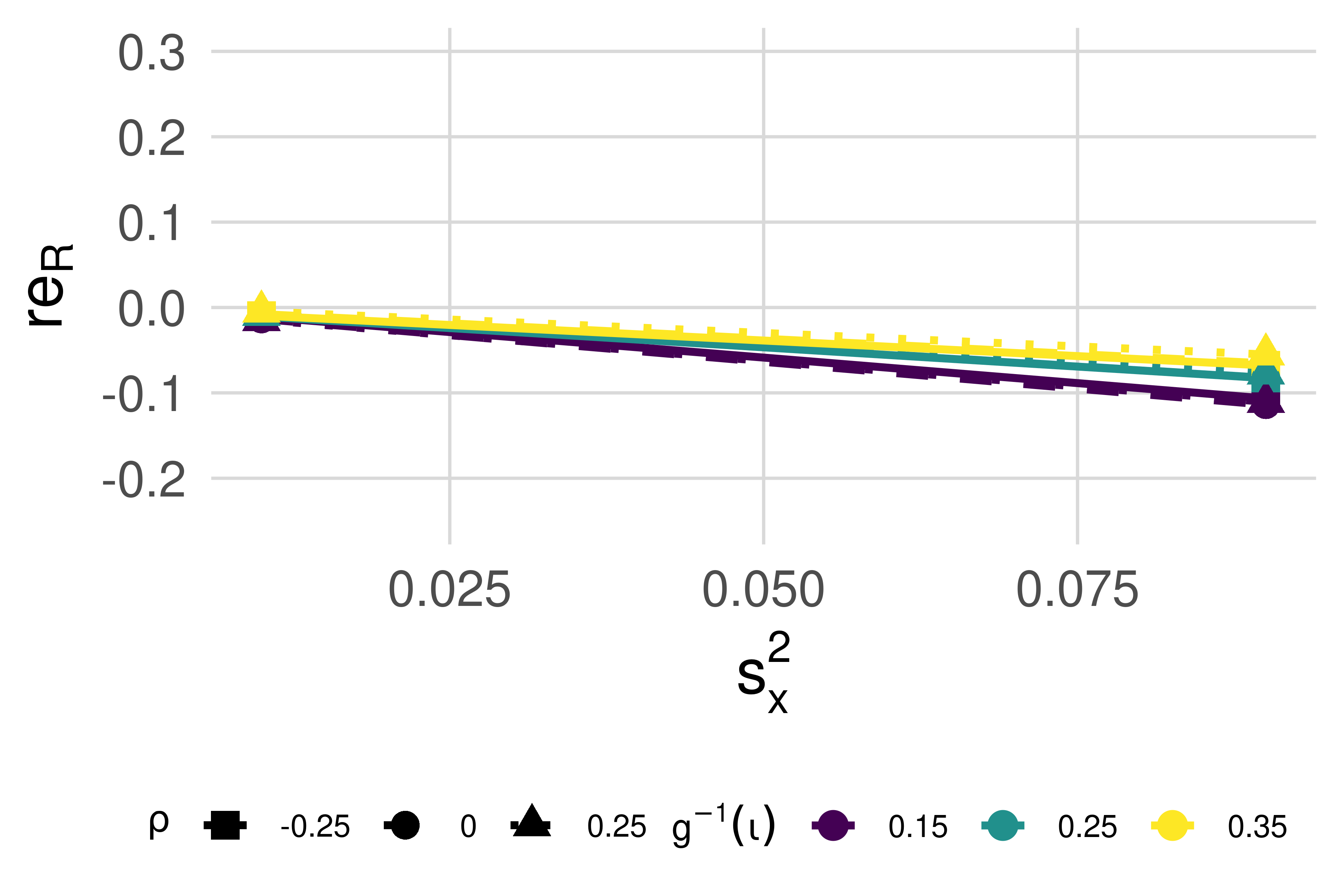}};
    \node at (0.5,3.1) {$\boldsymbol{s_z^2=0.01}$};
    \node at (8.5,3.1) {$\boldsymbol{s_z^2=0.09}$};
\end{tikzpicture}
\caption{Relative error in the approximation of $\fsq$ for logistic regression, plotted against $\phixz$ (top panels) and $\pspRsq$ (bottom panels). Left and right panels correspond to two levels of $s_z^2$ ($0.01$ and $0.09$). Within each panel, we vary $\phi$ and $g^{-1}(\iota)$. Although the x-axes show $\phixz$ and $\pspRsq$ directly, each point reflects an underlying value of $s_x^2$, evenly spaced from $0.01$ to $0.09$. Relative error for each measure of effect is largely insensitive to the correlation $\rho$ and slightly sensitive to the mean $g^{-1}(\iota)$, with $\rm{re}_R$, in particular, being biased downwards with decreasing values of $g^{-1}(\iota)$. Other parameters are fixed: $a_x = b_x = a_z = b_z = 1$.}
\label{fig:re_logistic_rho_mu0}
\end{figure}

\newpage

\subsection{Bernoulli distribution with identity link}

\begin{figure}[H]
\centering
\begin{tikzpicture}
    \node at (0,0) {\includegraphics[width=.48\textwidth]{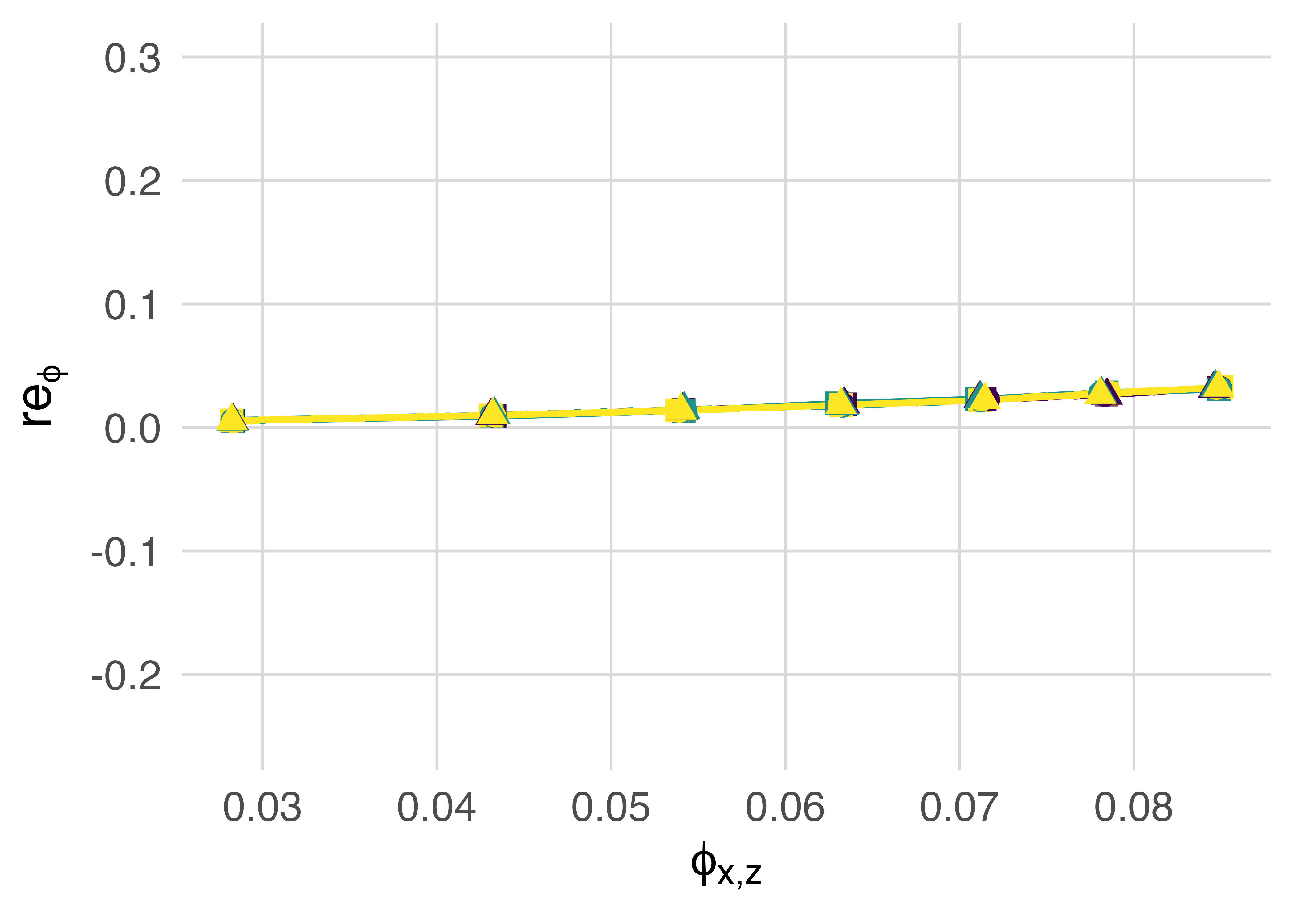}};
    \node at (8,0) {\includegraphics[width=.48\textwidth]{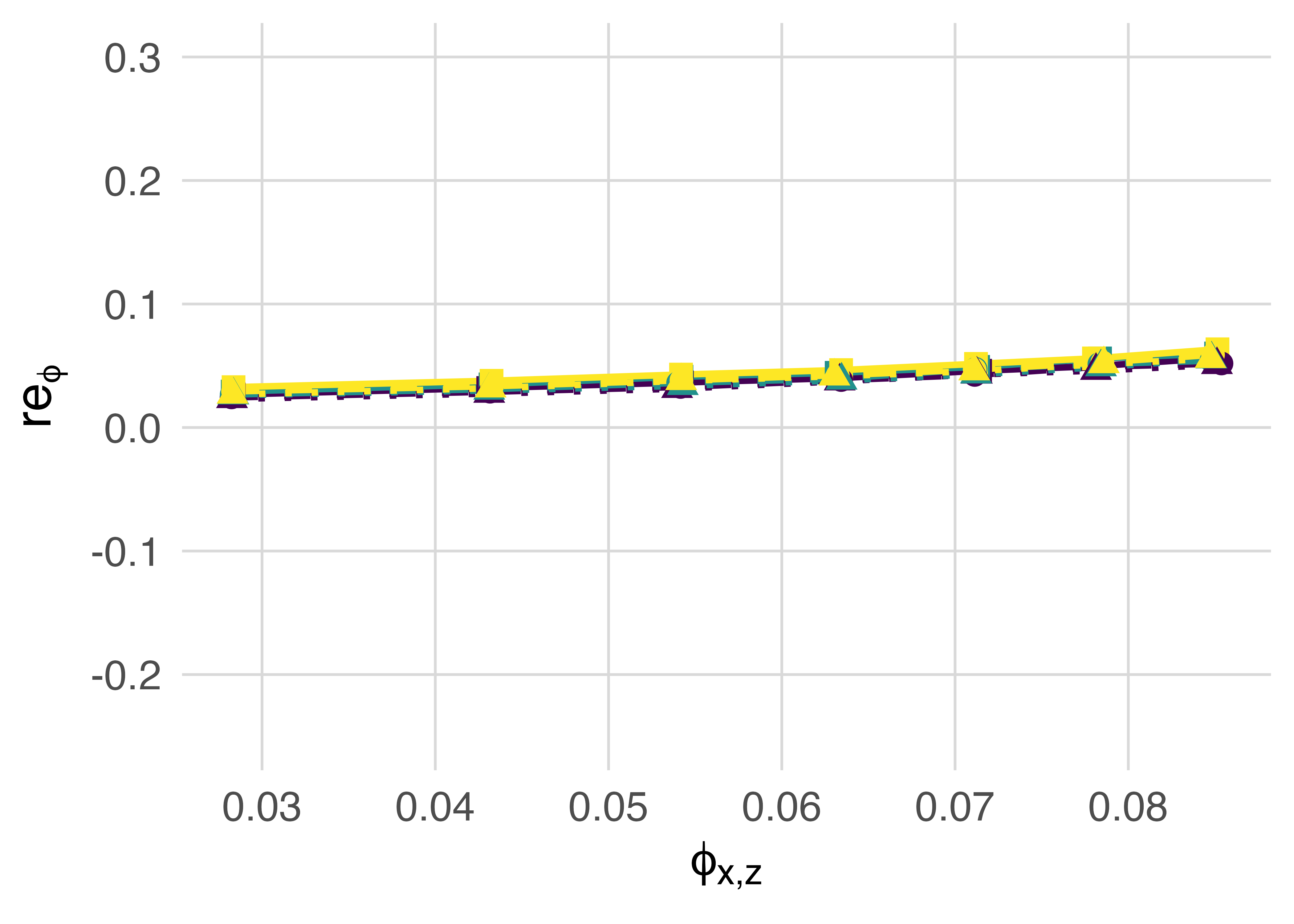}};
    \node at (4,-3.2) {\includegraphics[width=.65\textwidth,clip=true,trim=20 0 20 240]{images_new/simulation/legend_2.png}};
    \node at (0.5,3.1) {$\boldsymbol{s_z^2=0.0002}$};
    \node at (8.5,3.1) {$\boldsymbol{s_z^2=0.0018}$};
\end{tikzpicture}
\caption{Relative error $\text{re}_\phi$ in the approximation of $\fsq$ for a Bernoulli distribution and identity link (linear probability model), plotted against $\phixz$. Left and right panels correspond to two levels of $s_z^2$ ($0.0002$ and $0.0018$).  Within each panel, we vary $a_z$ and $b_z$ over all combinations of values in $\{0.5, 1, 1.5\}$. Although the x-axis shows $\phixz$ directly, each point reflects an underlying value of $s_x^2$, evenly spaced from $0.0002$ to $0.0018$. The relative error $\text{re}_R$ is omitted, since it is identically zero when an identity link is used. For $\text{re}_\phi$, relative error is largely insensitive to the shape parameters $a_z$ and $b_z$. Other parameters are fixed: $a_x = b_x = 1$, $\rho = 0$, and $g^{-1}(\iota) = .25$.}
\label{fig:re_binom_ident_az_bz}
\end{figure}

\begin{figure}[H]
\centering
\begin{tikzpicture}
    \node at (0,0) {\includegraphics[width=.48\textwidth]{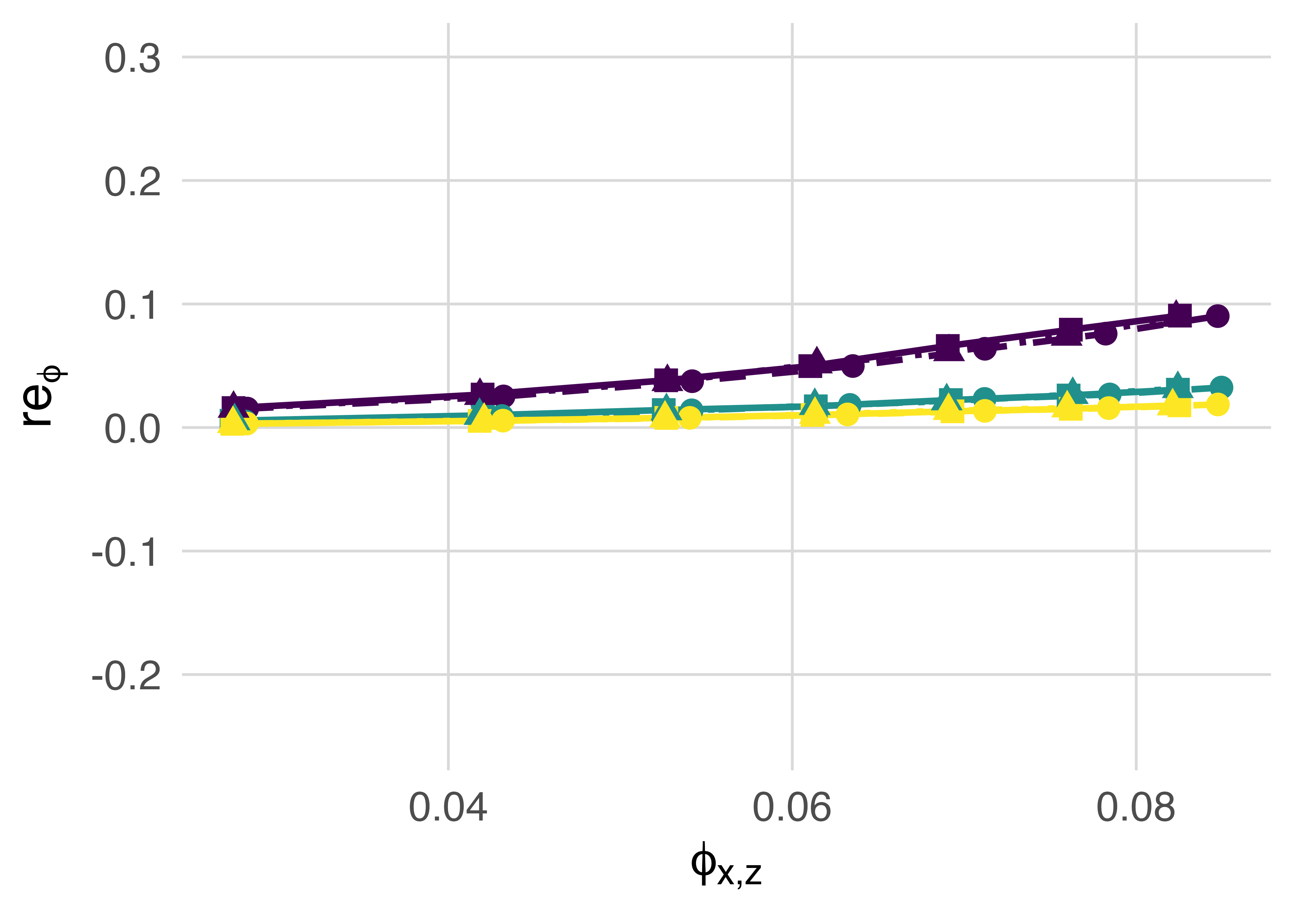}};
    \node at (8,0) {\includegraphics[width=.48\textwidth]{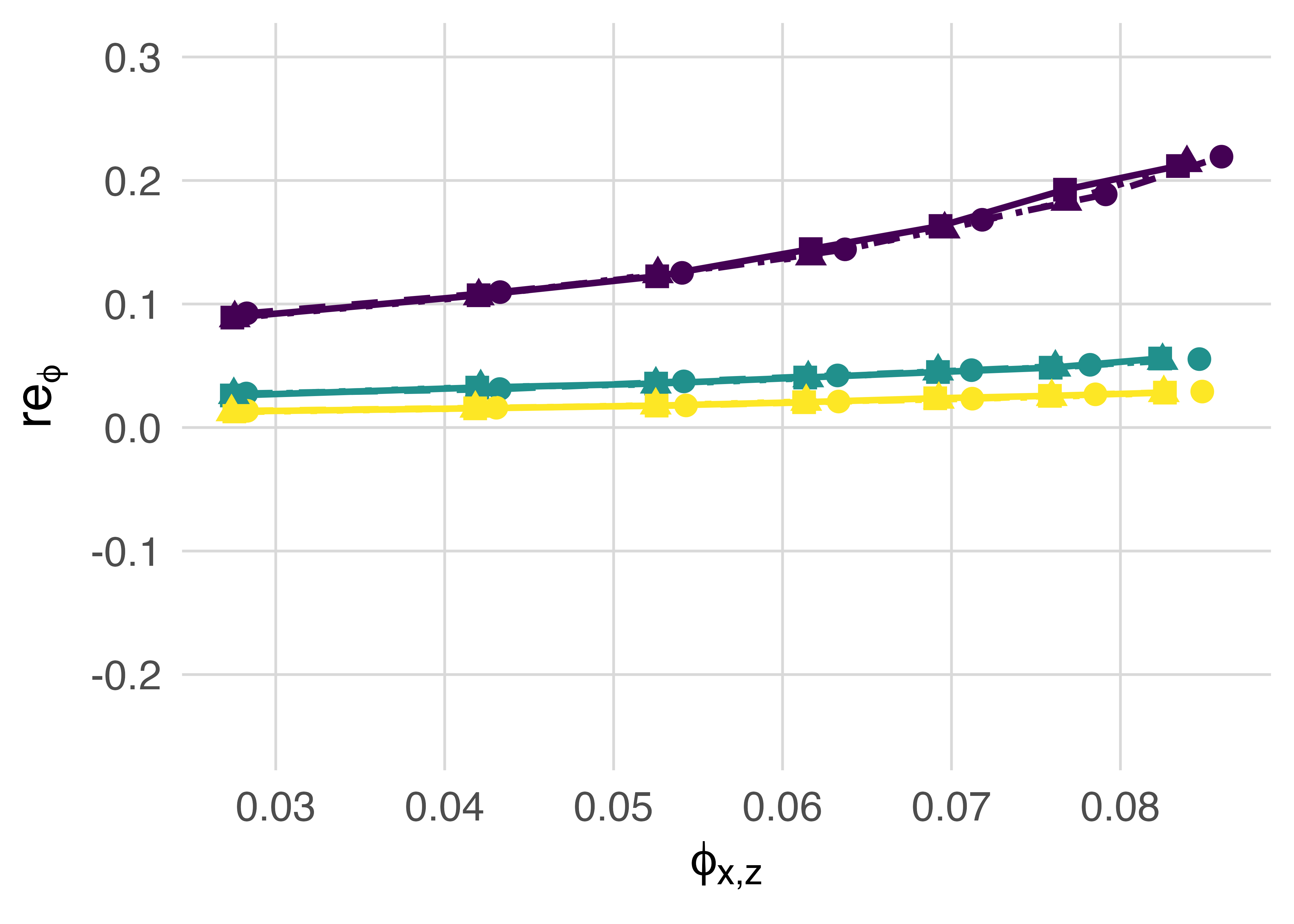}};
    \node at (4,-3.2) {\includegraphics[width=.65\textwidth,clip=true,trim=20 0 12 240]{images_new/simulation/legend_3.png}};
    \node at (0.5,3.1) {$\boldsymbol{s_z^2=0.0002}$};
    \node at (8.5,3.1) {$\boldsymbol{s_z^2=0.0018}$};
\end{tikzpicture}
\caption{Relative error $\text{re}_\phi$ in the approximation of $\fsq$ for a Bernoulli distribution and identity link (linear probability model), plotted against $\phixz$. Left and right panels correspond to two levels of $s_z^2$ ($0.0002$ and $0.0018$). Within each panel, we vary $\rho$ and $g^{-1}(\iota)$. Although the x-axis shows $\phixz$ directly, each point reflects an underlying value of $s_x^2$, evenly spaced from $0.0002$ to $0.0018$. The relative error $\text{re}_R$ is omitted, since it is identically zero when an identity link is used. For $\rm{re}_\phi$, relative error is insensitive to the correlation $\rho$, but highly sensitivity to the mean $g^{-1}(\iota)$, being biased upwards with decreasing values of $g^{-1}(\iota)$. Other parameters are fixed: $a_x = b_x = a_z = b_z = 1$.}
\label{fig:re_binom_ident_rho_mu0}
\end{figure}

\subsection{Poisson regression with log link}

\begin{figure}[H]
\centering
\begin{tikzpicture}
    \node at (0,0) {\includegraphics[width=.48\textwidth]{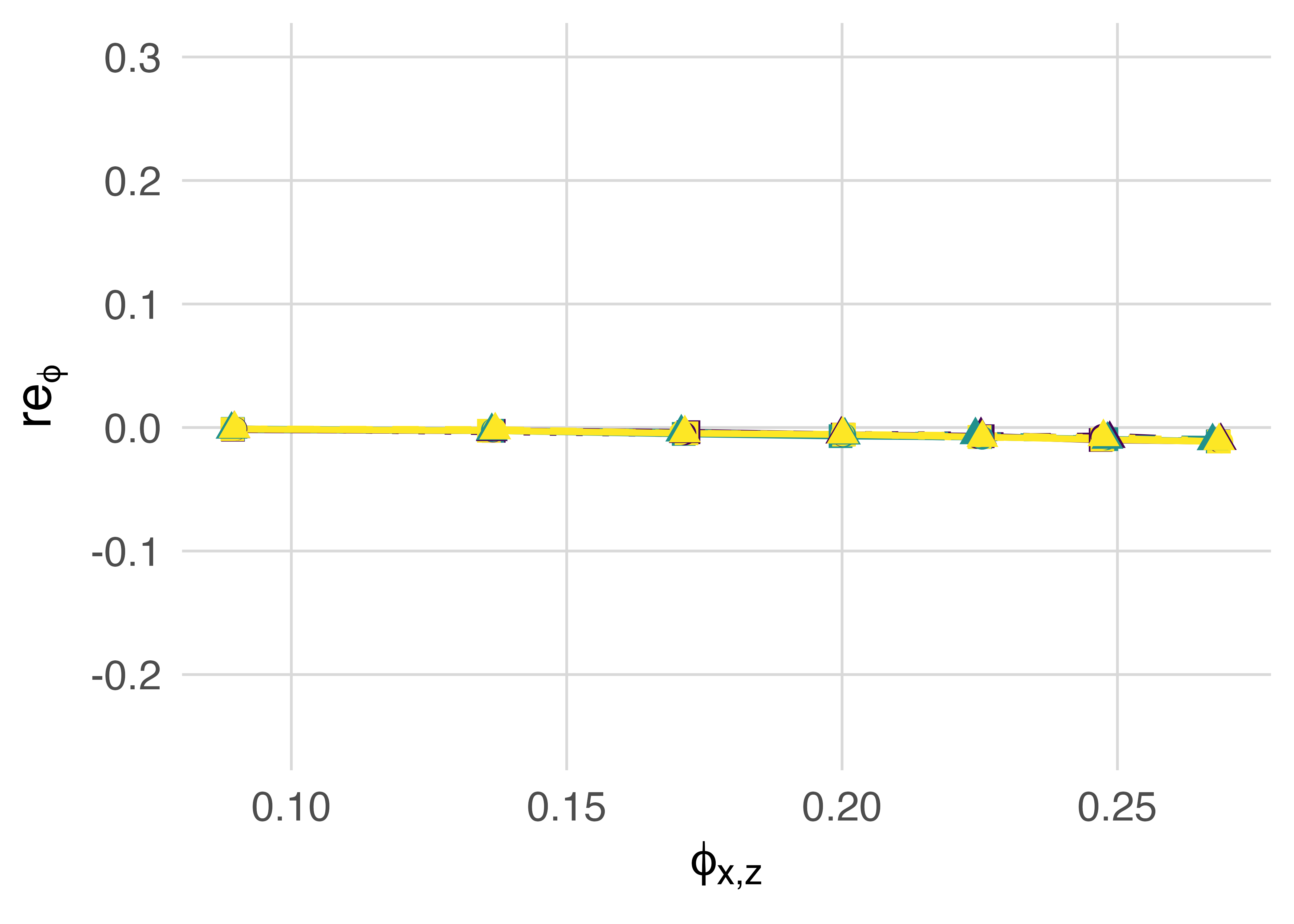}};
    \node at (8,0) {\includegraphics[width=.48\textwidth]{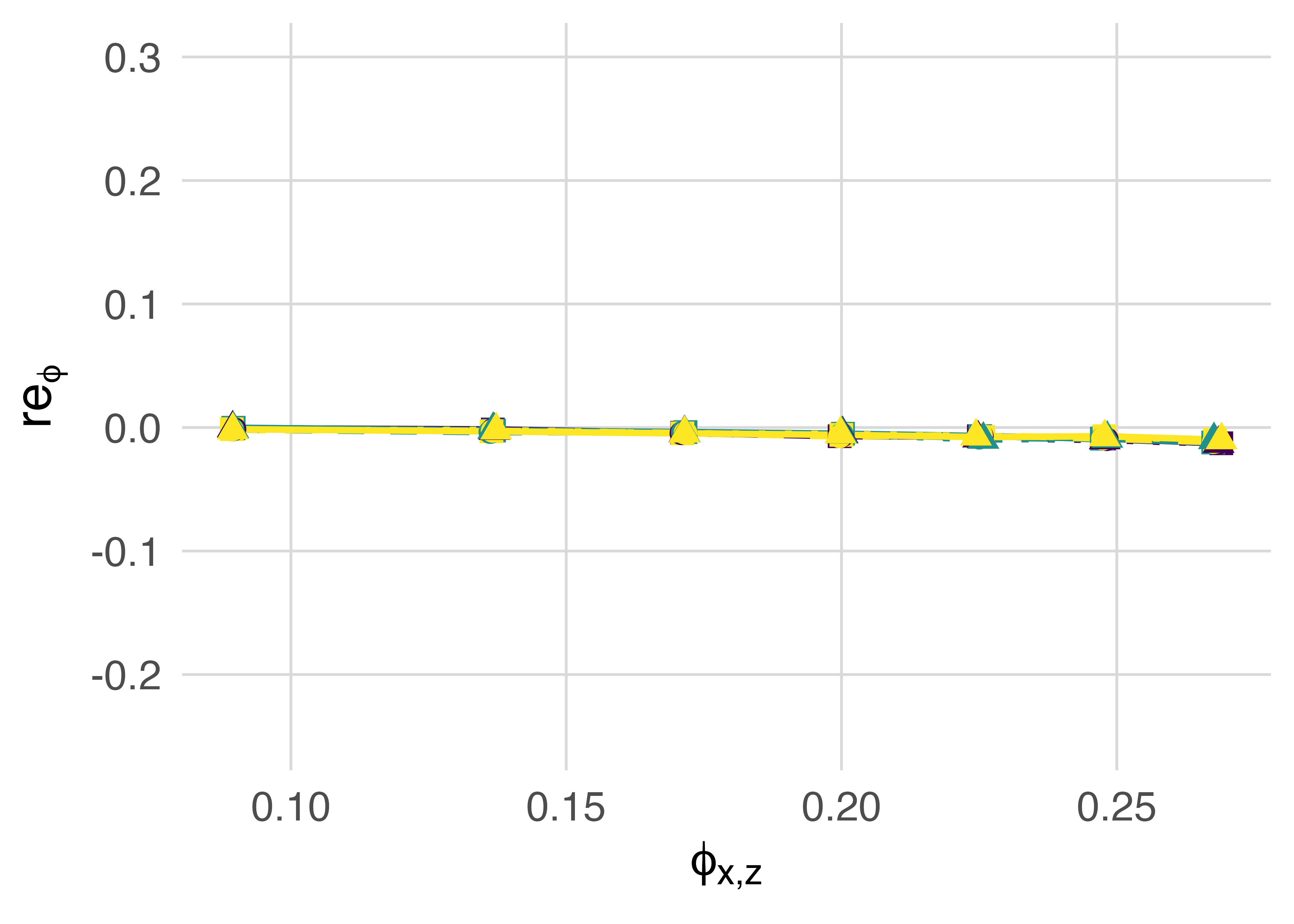}};
    \node at (0,-5.5) {\includegraphics[width=.48\textwidth]{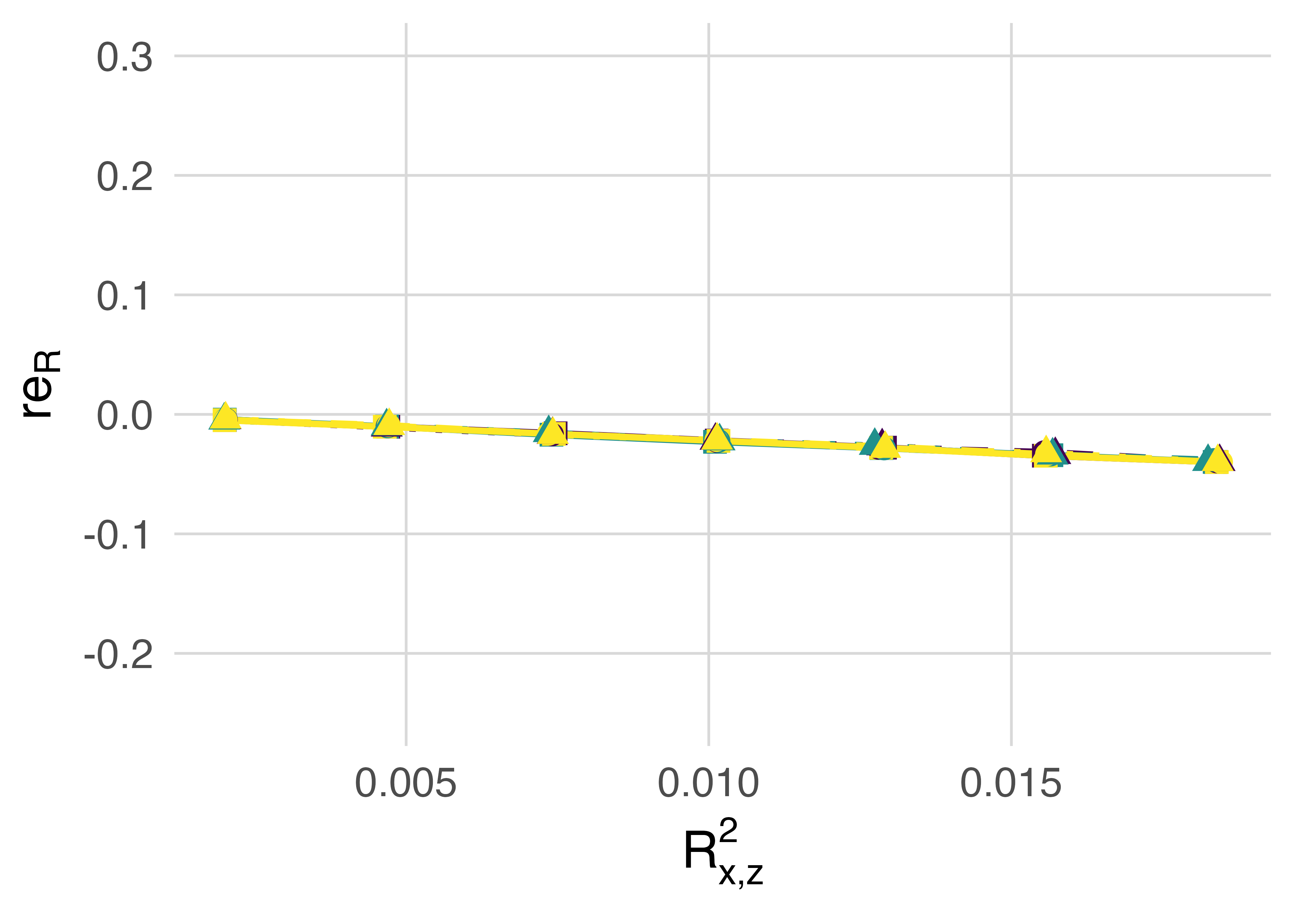}};
    \node at (8,-5.5) {\includegraphics[width=.48\textwidth]{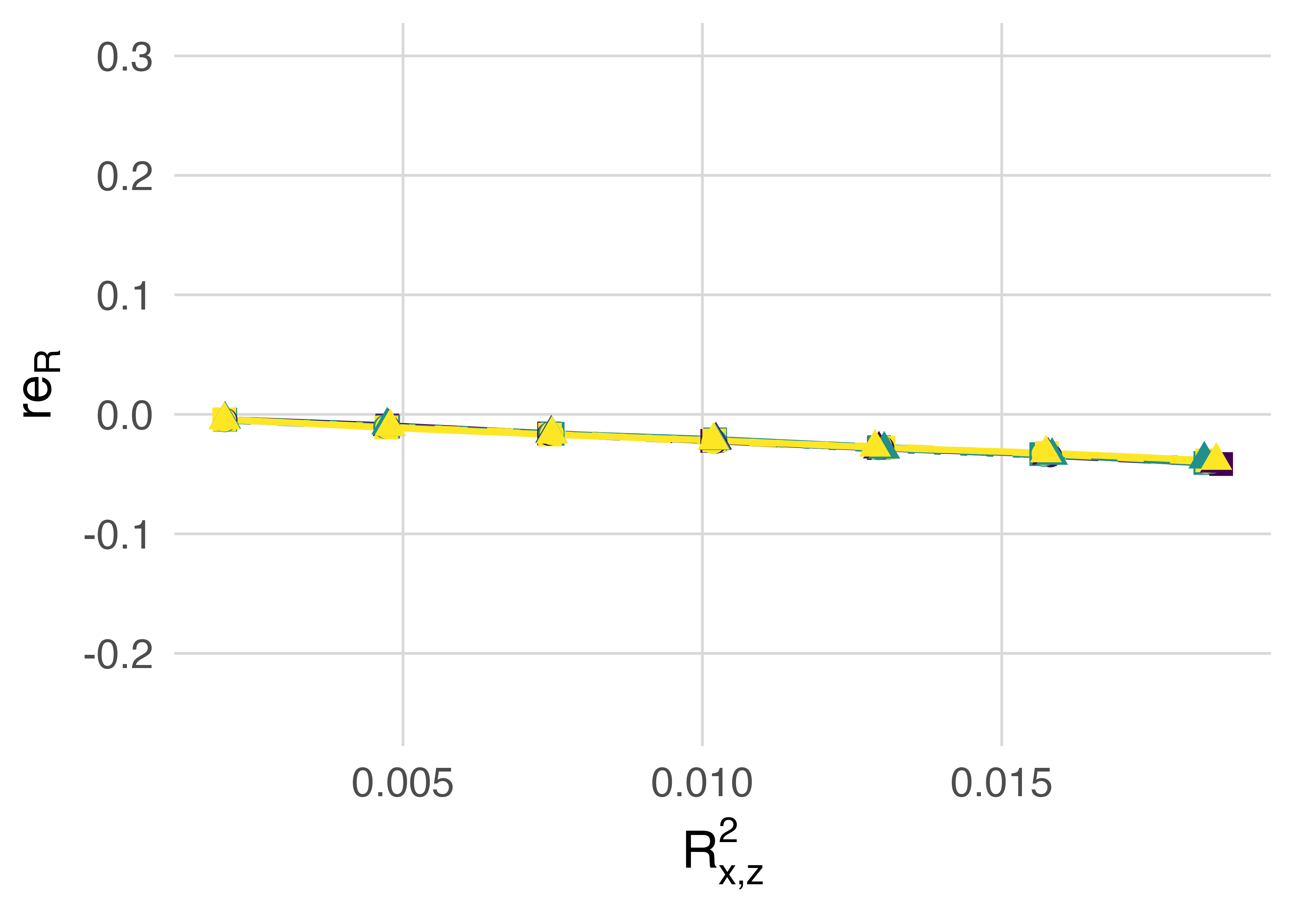}};
    \node at (4,-8.7) {\includegraphics[width=.65\textwidth,clip=true,trim=20 0 20 240]{images_new/simulation/legend_2.png}};
    \node at (0.5,3.1) {$\boldsymbol{s_z^2=0.002}$};
    \node at (8.5,3.1) {$\boldsymbol{s_z^2=0.018}$};
\end{tikzpicture}
\caption{Relative error in the approximation of $\fsq$ for a Poisson distribution with a log link, plotted against $\phixz$ (top panels) and $\pspRsq$ (bottom panels). Left and right panels correspond to two levels of $s_z^2$ ($0.002$ and $0.018$). Within each panel, we vary $a_z$ and $b_z$ over all combinations of values in $\{0.5, 1, 1.5\}$. Although the x-axes show $\phixz$ and $\pspRsq$ directly, each point reflects an underlying value of $s_x^2$, evenly spaced from $0.002$ to $0.018$. Relative error for each measure of effect is largely insensitive to the shape parameters $a_z$ and $b_z$. Other parameters are fixed: $a_x = b_x = 1$, $\rho = 0$, and $g^{-1}(\iota) = 1$.}
\label{fig:re_poisson_az_bz}
\end{figure}

\begin{figure}[H]
\centering
\begin{tikzpicture}
    \node at (0,0) {\includegraphics[width=.48\textwidth]{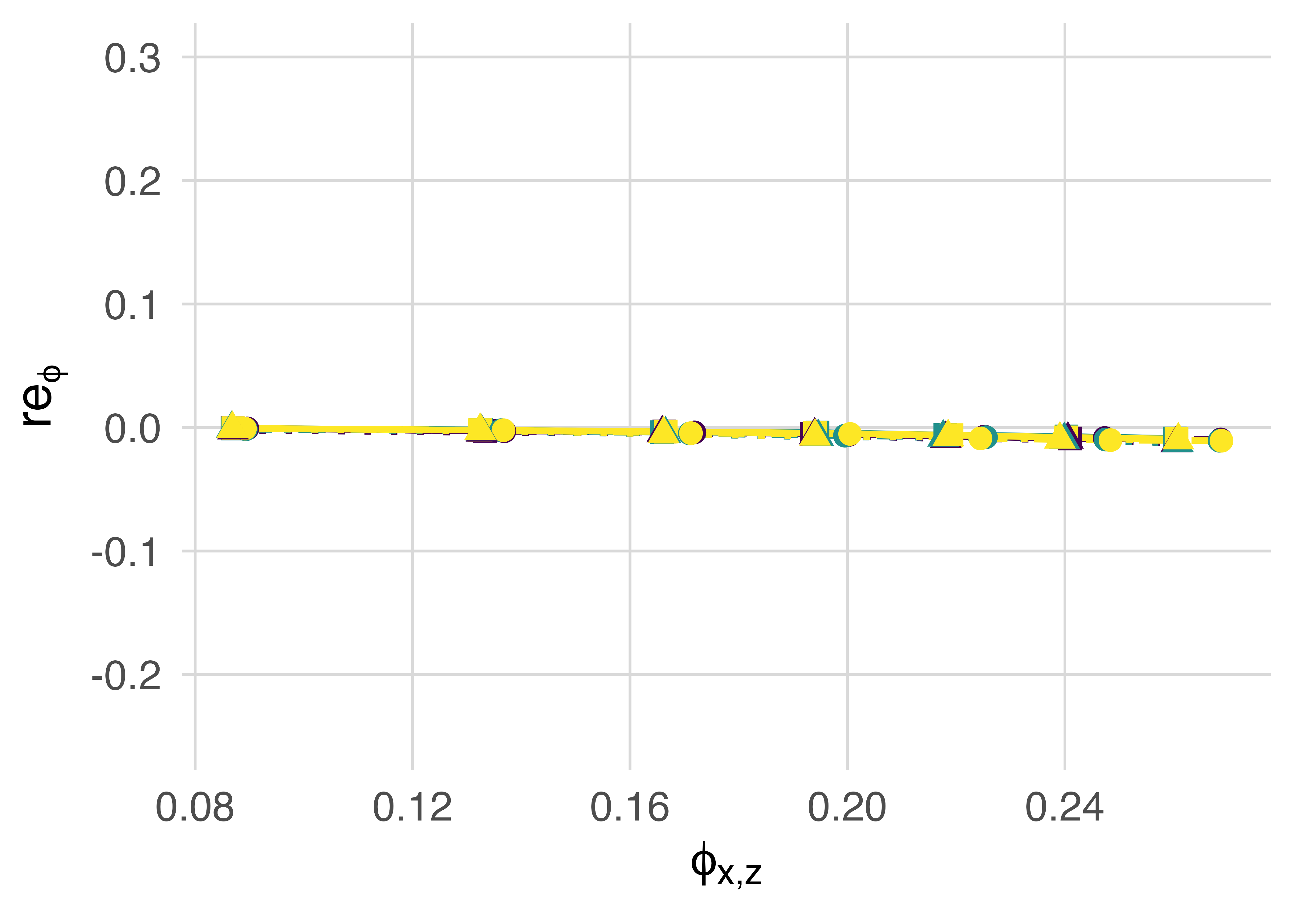}};
    \node at (8,0) {\includegraphics[width=.48\textwidth]{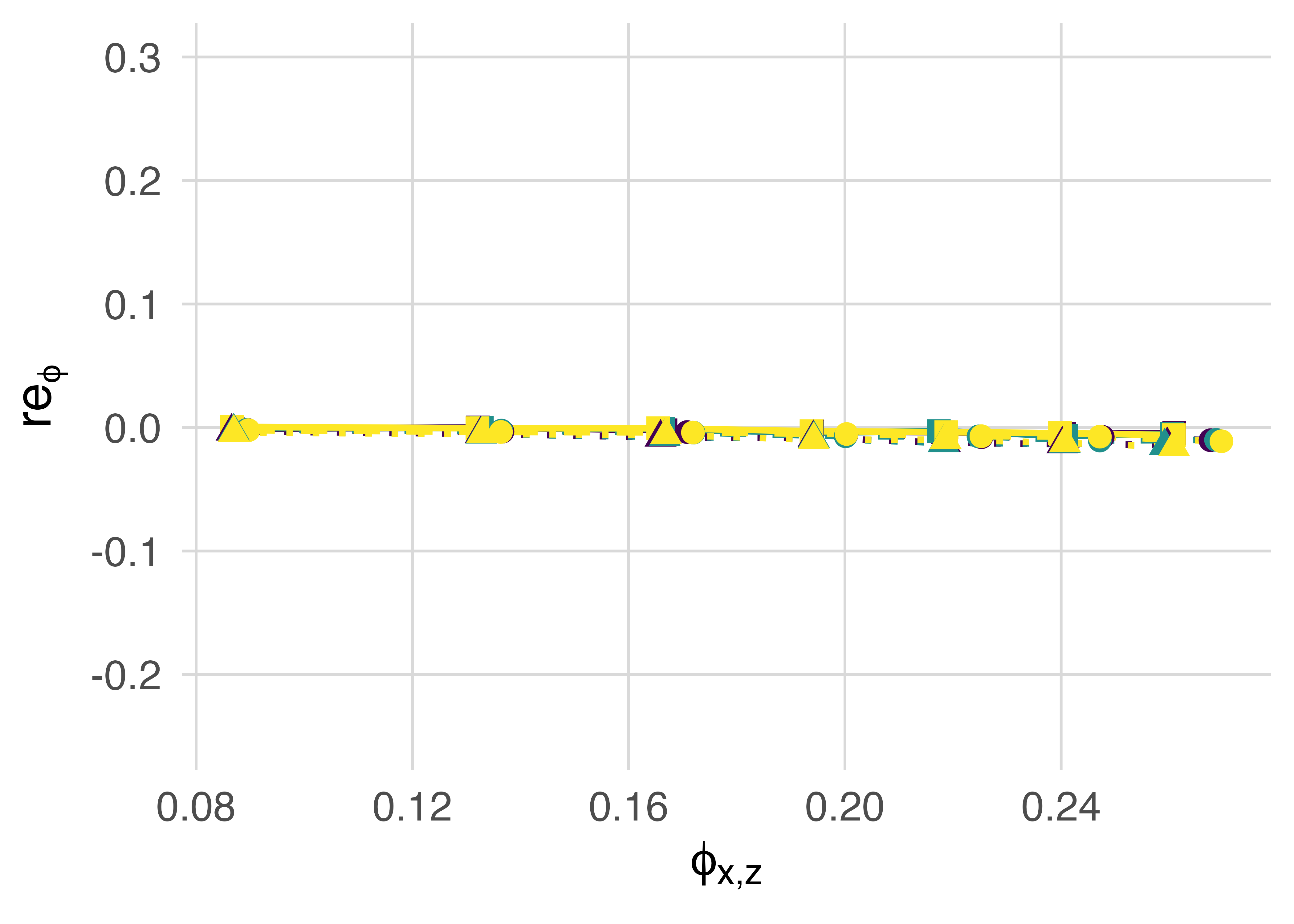}};
    \node at (0,-5.5) {\includegraphics[width=.48\textwidth]{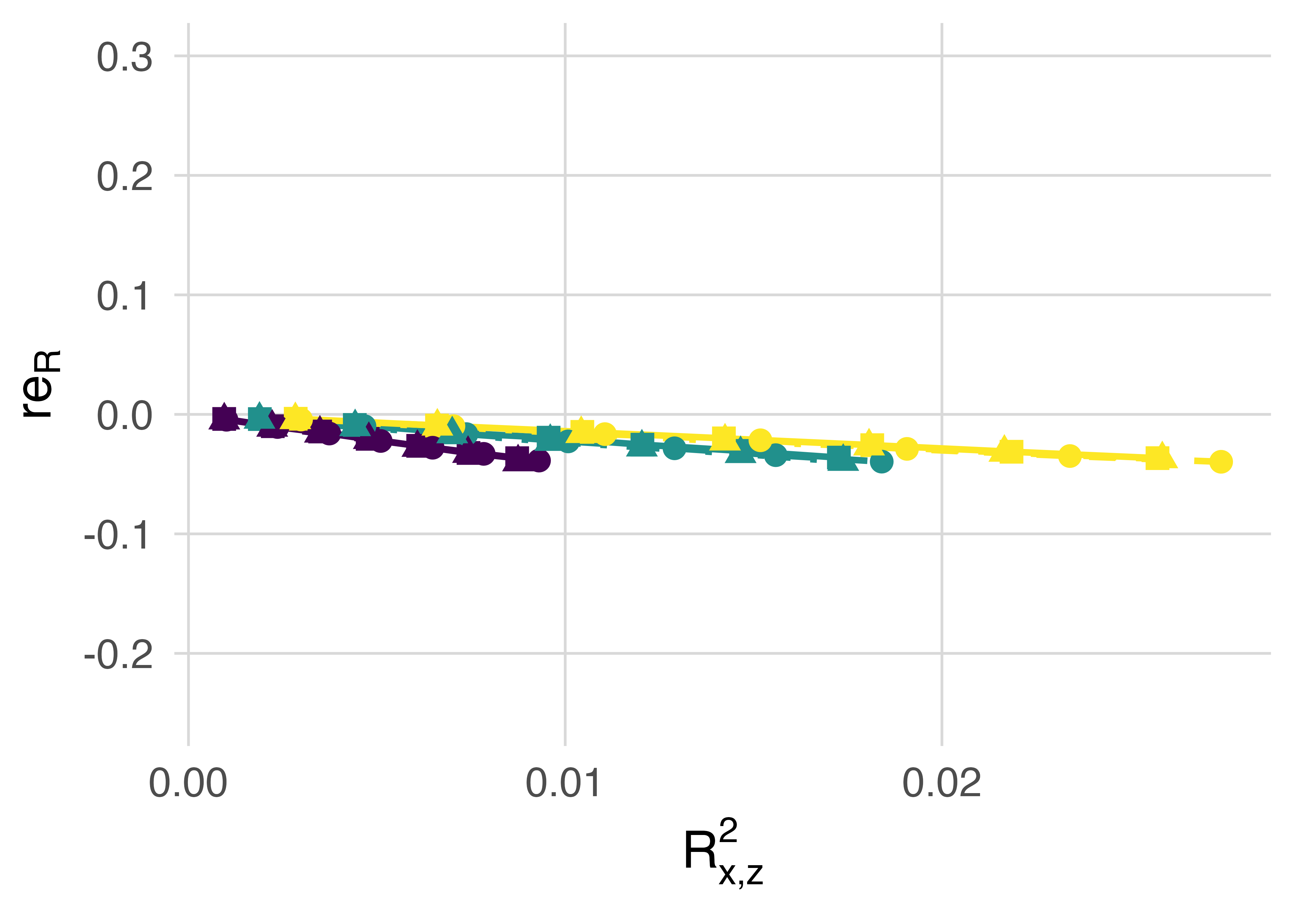}};
    \node at (8,-5.5) {\includegraphics[width=.48\textwidth]{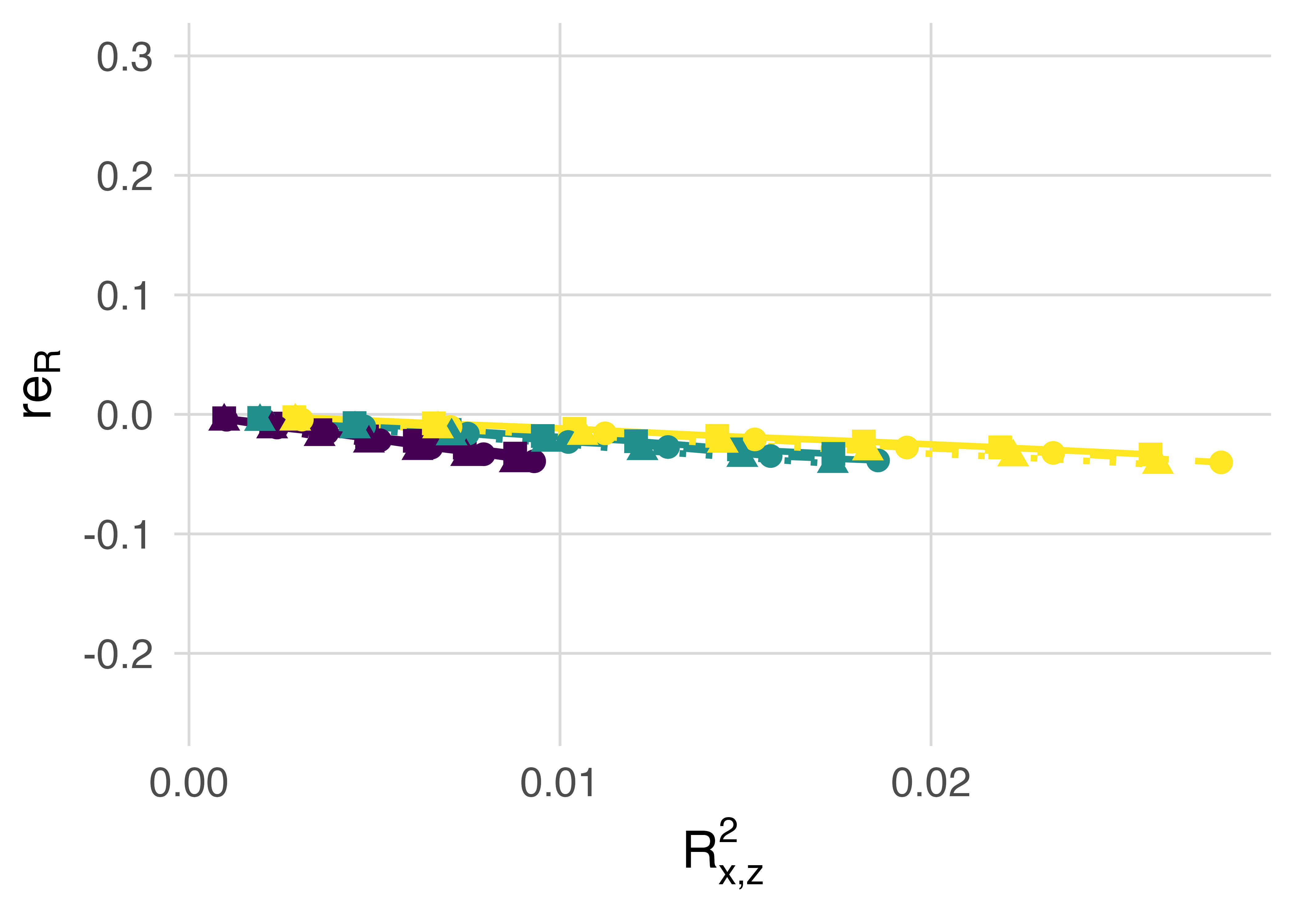}};
    \node at (4,-8.75) {\includegraphics[width=.65\textwidth,clip=true,trim=250 80 250 2020]{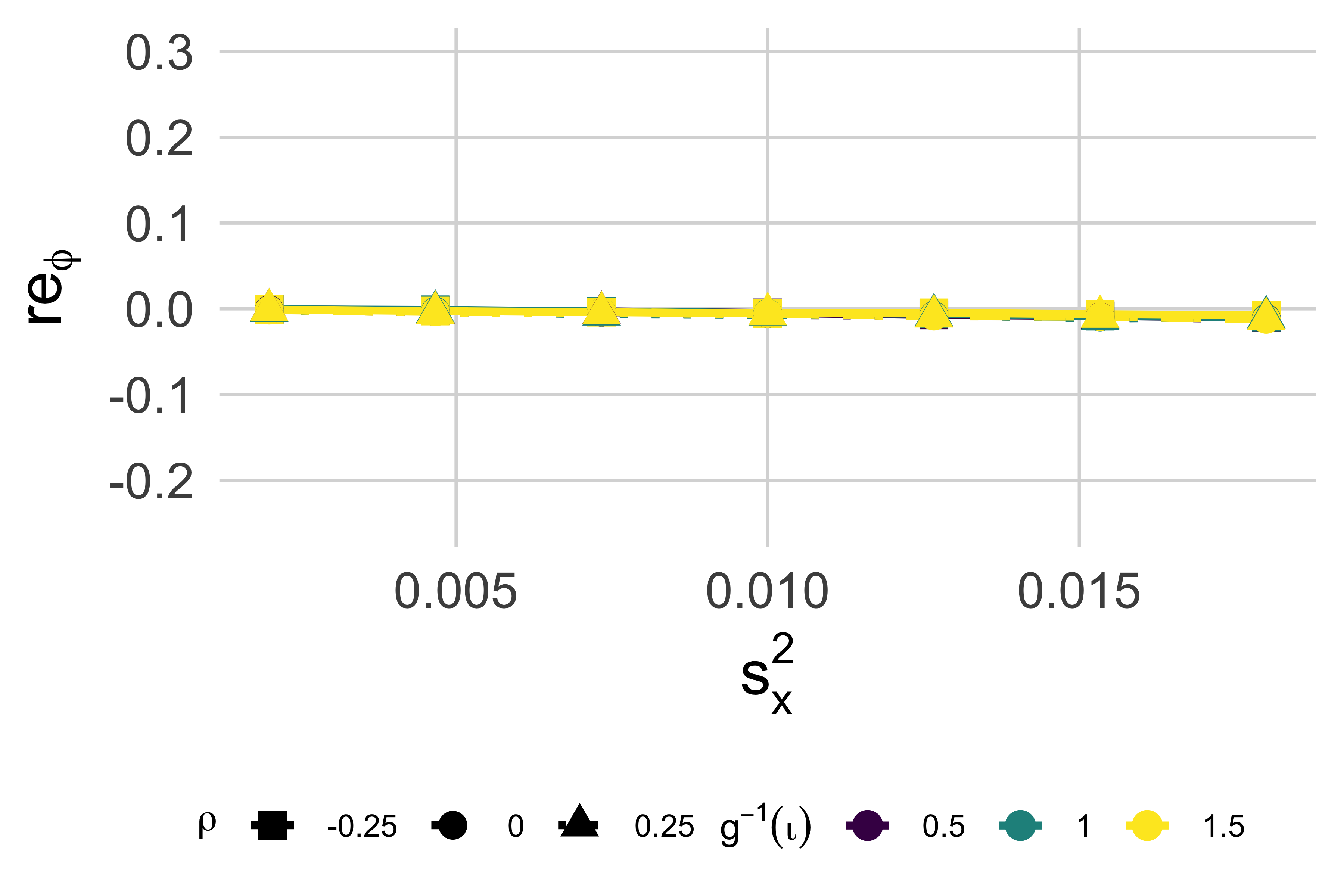}};
    \node at (0.5,3.1) {$\boldsymbol{s_z^2=0.002}$};
    \node at (8.5,3.1) {$\boldsymbol{s_z^2=0.018}$};
\end{tikzpicture}
\caption{Relative error in the approximation of $\fsq$ for a Poisson distribution with a log link, plotted against $\phixz$ (top panels) and $\pspRsq$ (bottom panels). Left and right panels correspond to two levels of $s_z^2$ ($0.002$ and $0.018$). Within each panel, we vary $\rho$, and $g^{-1}(\iota)$. Although the x-axes show $\phixz$ and $\pspRsq$ directly, each point reflects an underlying value of $s_x^2$, evenly spaced from $0.002$ to $0.018$. Relative error for each measure of effect is largely insensitive to mean $g^{-1}(\iota)$ but slightly sensitive to the correlation $\rho$. Other parameters are fixed: $a_x = b_x = a_z = b_z = 1$. }
\label{fig:re_poisson_rho_mu0}
\end{figure}

\subsection{Gamma regression with log link}

\begin{figure}[H]
\centering
\begin{tikzpicture}
    \node at (0,0) {\includegraphics[width=.48\textwidth]{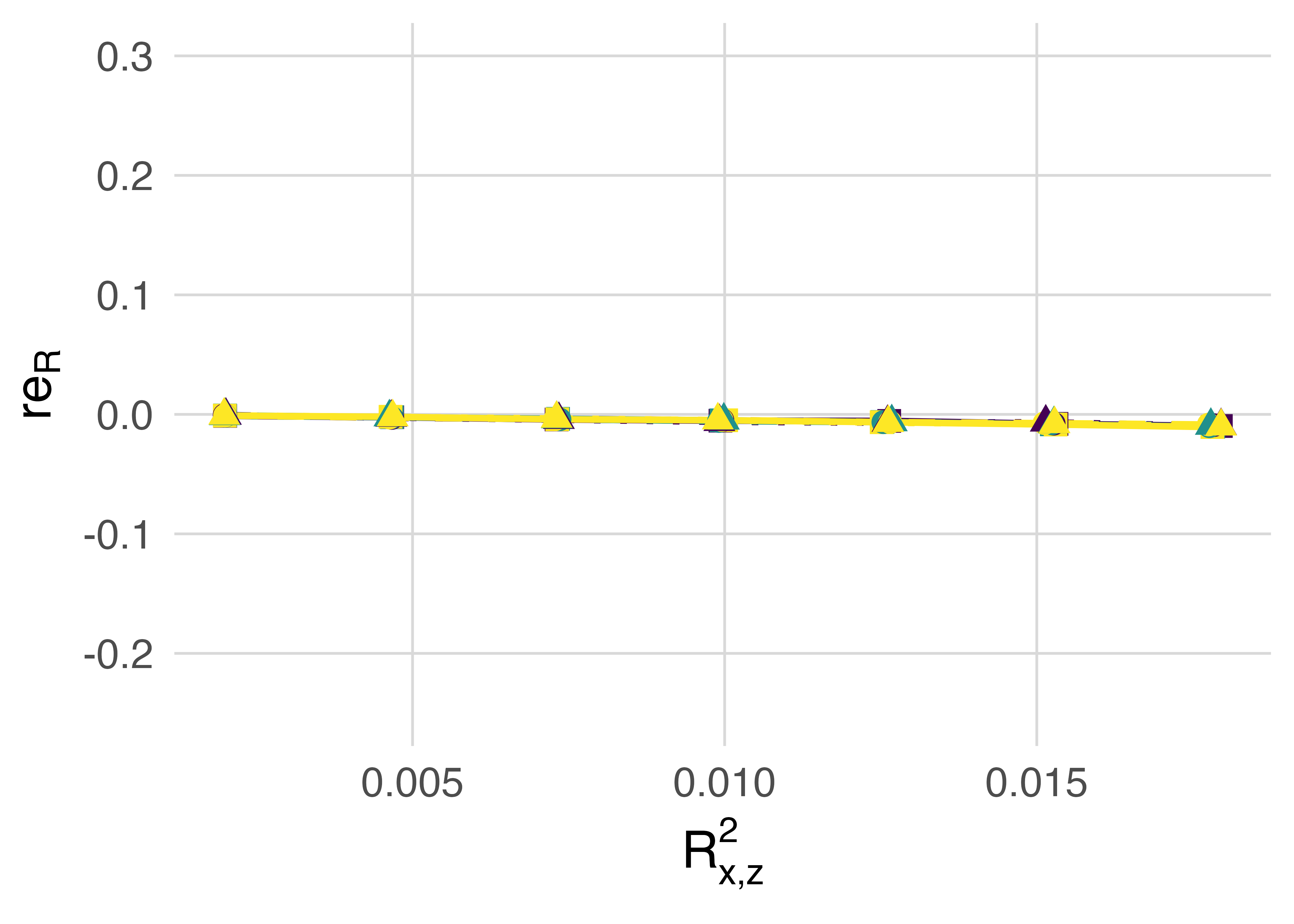}};
    \node at (8,0) {\includegraphics[width=.48\textwidth]{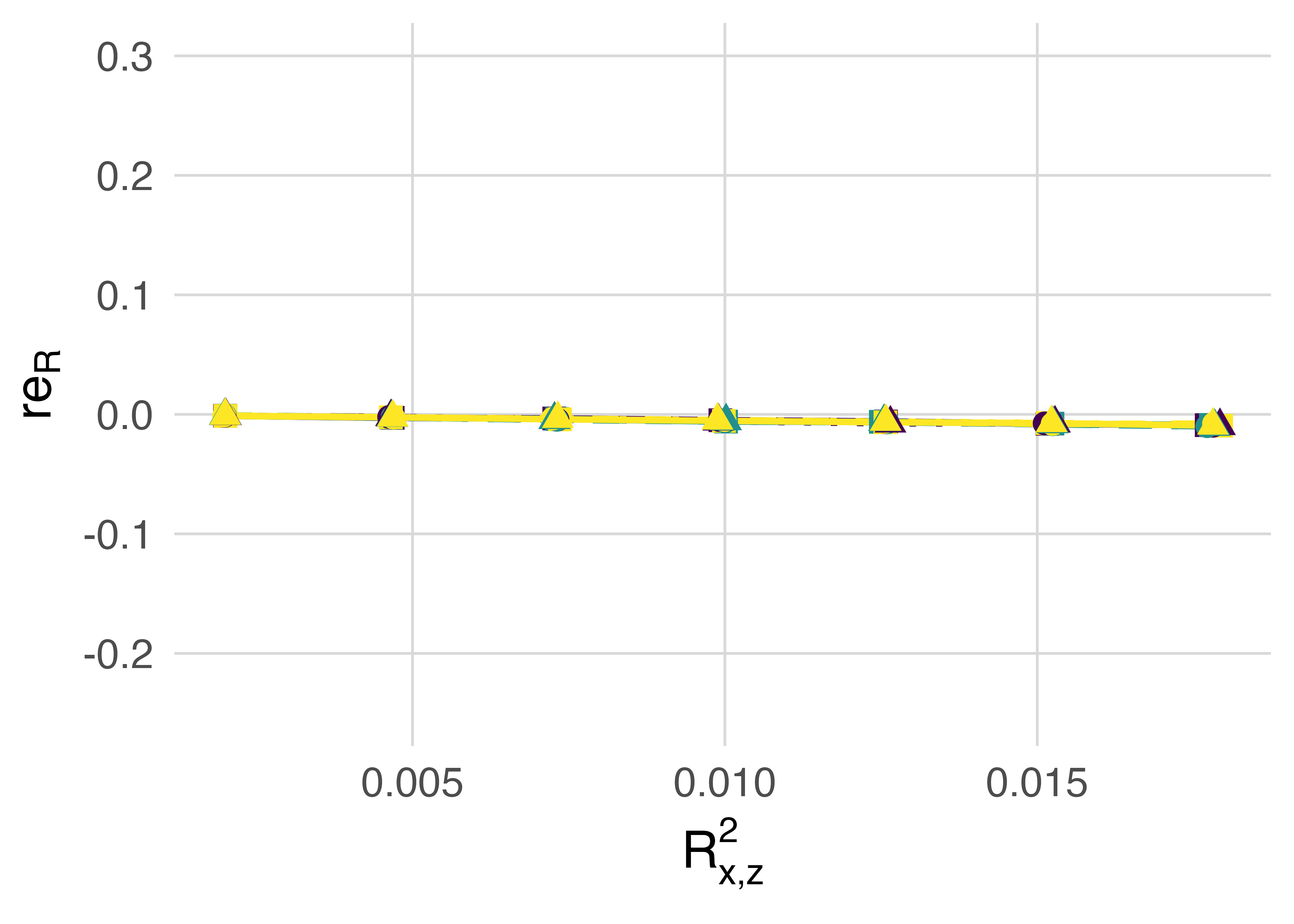}};
    \node at (4,-3.2) {\includegraphics[width=.65\textwidth,clip=true,trim=20 0 20 240]{images_new/simulation/legend_2.png}};
    \node at (0.5,3.1) {$\boldsymbol{s_z^2=0.001}$};
    \node at (8.5,3.1) {$\boldsymbol{s_z^2=0.009}$};
\end{tikzpicture}
\caption{Relative error $\text{re}_R$ in the approximation of $\fsq$ for a gamma distribution with a log link, plotted against $\phixz$ (top panels) and $\pspRsq$ (bottom panels). Relative error $\text{re}_\phi$ is not plotted, as it is identically zero. Left and right panels correspond to two levels of $s_z^2$ ($0.001$ and $0.009$). Within each panel, we vary $a_z$ and $b_z$ over all combinations of values in $\{0.5, 1, 1.5\}$. Although the x-axes show $\phixz$ and $\pspRsq$ directly, each point reflects an underlying value of $s_x^2$, evenly spaced from $0.001$ to $0.009$. Relative error $\rm{re}_R$ is largely insensitive to the shape parameters $a_z$ and $b_z$. Other parameters are fixed: $a_x = b_x = 1$, $\rho = 0$, $g^{-1}(\iota) = 4$, and Gamma shape parameter is $2$.}
\label{fig:re_gamma_az_bz}
\end{figure}

\begin{figure}[H]
\centering
\begin{tikzpicture}
    \node at (0,0) {\includegraphics[width=.48\textwidth]{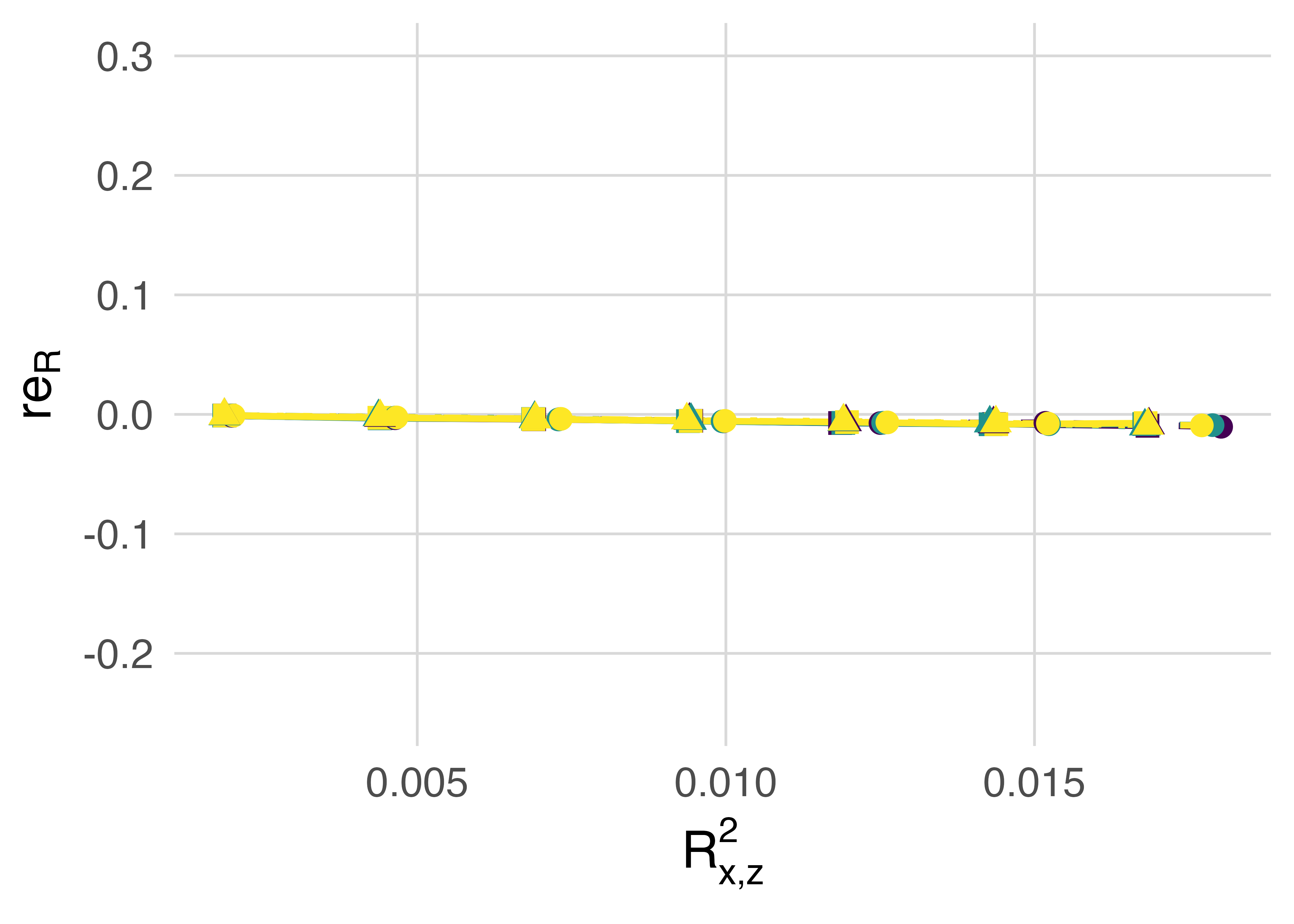}};
    \node at (8,0) {\includegraphics[width=.48\textwidth]{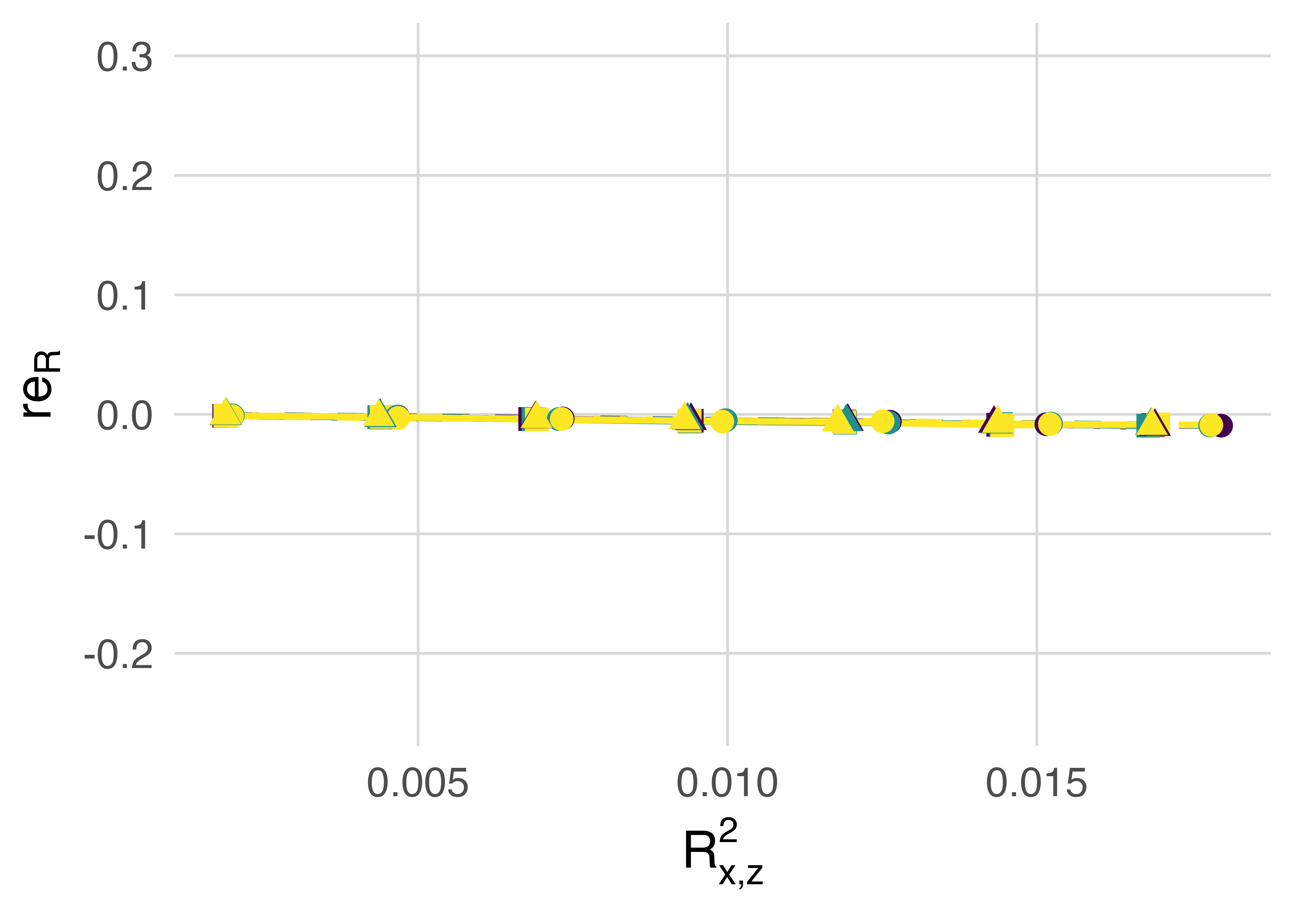}};
    \node at (4,-3.25) {\includegraphics[width=.65\textwidth,clip=true,trim=250 80 250 2020]{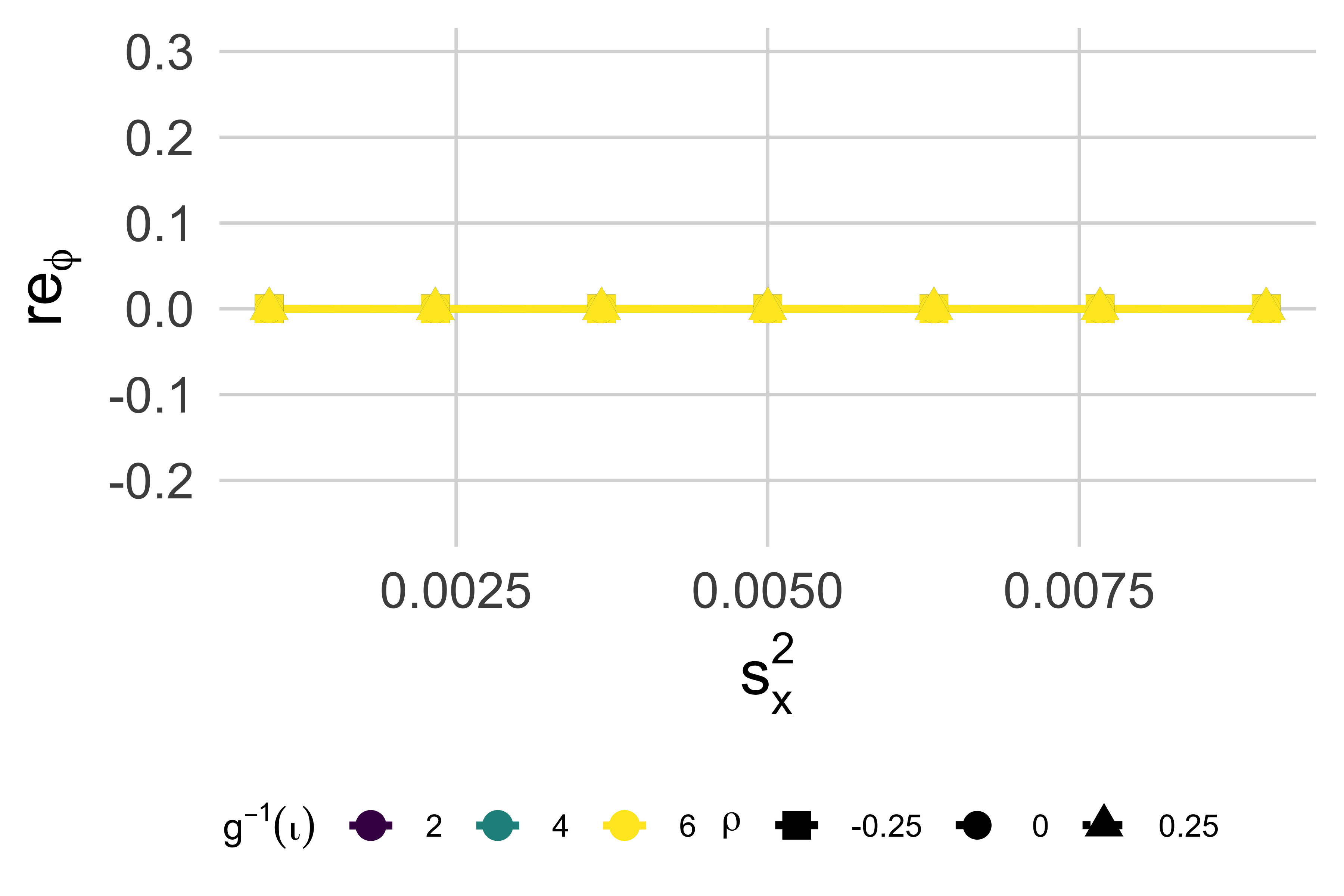}};
    \node at (0.5,3.1) {$\boldsymbol{s_z^2=0.001}$};
    \node at (8.5,3.1) {$\boldsymbol{s_z^2=0.009}$};
\end{tikzpicture}
\caption{Relative error $\text{re}_R$ in the approximation of $\fsq$ for a gamma distribution with a log link, plotted against $\phixz$ (top panels) and $\pspRsq$ (bottom panels). Relative error $\text{re}_\phi$ is not plotted, as it is identically zero. Left and right panels correspond to two levels of $s_z^2$ ($0.001$ and $0.009$). Within each panel, we vary $\rho$, and $g^{-1}(\iota)$. Although the x-axes show $\phixz$ and $\pspRsq$ directly, each point reflects an underlying value of $s_x^2$, evenly spaced from $0.001$ to $0.009$. Relative error $\rm{re}_R$ is largely insensitive to the correlation $\rho$ and mean $g^{-1}(\iota)$. Other parameters are fixed: $a_x = b_x = a_z = b_z = 1$ and Gamma shape parameter is $2$.}
\label{fig:re_gamma_rho_mu0}
\end{figure}

\newpage

\subsection{Sensitivity analysis} \label{app:sensitivity}

We performed a sensitivity analysis to see how the relative error in $\fsq$ changes with the distributions of $\eta$ and $\eta_z$ across different GLMs. For each GLM, we drew 1000 Latin hypercube samples of the 8 parameters. For each set, we measured the relative error in $\fsq$ for our two approximations ($\fsqphi$ and $\fsqR$).  This approach allows us to explore interactions between parameters that may not have been evident in previous simulations, where only subsets of parameters were varied at a time. Table~\ref{tab:parameters} provides ranges of the various parameters. The shape parameter for the gamma distribution is fixed at 2.

\begin{table}[H]
    \centering  {
    \caption{Parameters controlling the distribution of $\eta$ and $\eta_z$ for different GLMs}

    \medskip
    
    \label{tab:parameters}
    \begin{tabular}{c c l}  
    \toprule
        \textbf{Parameter} &  \textbf{Description} & \textbf{Range} \\
    \midrule 
        $a_x$ & Shape parameter 1 for $x_0$ & [0.5, 1.5] \\
        $b_x$ & Shape parameter 2 for $x_0$ & [0.5, 1.5] \\
        $a_z$ & Shape parameter 1 for $z_0$ & [0.5, 1.5] \\
        $b_z$ & Shape parameter 2 for $z_0$ & [0.5, 1.5] \\
        $s_x$ & Standard deviation for $B_x$ & 
        Binomial (logit): $[\sqrt{0.01}, \sqrt{0.09}]$ \\ 
        && Binomial (identity): $[\sqrt{0.0002}, \sqrt{0.0018}]$ \\ 
        && Poisson (log): $[\sqrt{0.002}, \sqrt{0.018}]$ \\ 
        && Gamma (log): $[\sqrt{0.001}, \sqrt{0.009}]$ \\
        $s_z$ & Standard deviation for $B_z$ & 
        Binomial (logit): $[\sqrt{0.01}, \sqrt{0.09}]$ \\ 
        && Binomial (identity): $[\sqrt{0.0002}, \sqrt{0.0018}]$ \\ 
        && Poisson (log): $[\sqrt{0.002}, \sqrt{0.018}]$ \\ 
        && Gamma (log): $[\sqrt{0.001}, \sqrt{0.009}]$ \\
        $g^{-1}(\iota)$ & Reference mean & 
        Binomial (logit): $[0.15, 0.35]$ \\ 
        && Binomial (identity): $[0.15, 0.35]$ \\ 
        && Poisson (log): $[0.5, 1.5]$ \\ 
        && Gamma (log): $[2, 6]$ \\
        $\rho$ & Copula correlation & [-0.25, 0.25] \\
    \bottomrule
    \end{tabular} }
\end{table}

Across the Latin hypercube samples, we computed the mean, min, first quartile, median, and third quartile, and max for each GLM and each measure of effect. Table~\ref{tab:summary_sens} shows these summary statistics. Overall, the relative error in $\fsq$ varies across GLM types and effect measures, with the largest median relative errors observed under the logistic regression model: $-2.6\%$ when using $\phixz$ and $-3.6\%$ when using $\pspRsq$.

\begin{table}[H]
    \centering {
    \caption{Summary statistics of the relative error in $\fsq$ using $\phixz$ or $\pspRsq$ across sample parameters.}

    \medskip

    \label{tab:summary_sens}
    \begin{tabular}{lllc cccccc}
    \toprule
    \textbf{Distribution} & \textbf{Link} & \textbf{Effect} & \textbf{Mean} & \textbf{Min} & \textbf{Q1} & \textbf{Median} & \textbf{Q3} & \textbf{Max} \\
    \midrule
    Binomial & Logit & $\phixz$ & -2.9\%  & -17.6\%  & -4.8\%  & -2.6\%  & -0.5\%  & 8.8\%  \\
             &       & $\pspRsq$ & -4.2\%  & -19.3\%  & -6.6\%  & -3.6\%  & -1.3\%  & 5.4\%  \\
    Binomial & Identity & $\phixz$ & 3.4\%   & -8.1\%   & 0.4\%   & 2.3\%   & 5.4\%   & 41.3\%  \\
             &          & $\pspRsq$ & -       & -        & -       & -       & -       & -      \\
    Poisson  & Log      & $\phixz$ & -0.5\%  & -11.9\%  & -2.8\%  & -0.4\%  & 1.8\%   & 11.2\%  \\
             &          & $\pspRsq$ & -1.9\%  & -14.4\%  & -4.1\%  & -1.6\%  & 0.5\%   & 7.7\%  \\
    Gamma    & Log      & $\phixz$ & -       & -        & -       & -       & -       & -      \\
             &          & $\pspRsq$ & -0.4\%  & -8.6\%   & -2.0\%  & -0.4\%  & 1.1\%   & 6.6\%  \\
    \bottomrule 
    \multicolumn{9}{l}{\textbf{\textit{Note.}} The empty entries signify cases when the relative error equals zero.} \\
    \end{tabular} }
\end{table}

 To identify factors contributing to relative error, we computed partial rank correlation coefficients (PRCCs) between sampled parameters and each relative error. PRCCs measure the strength and direction of these relationships while accounting for the influence of other parameters. To accommodate nonlinear associations, we first rank-transformed the parameters and relative errors before computing partial correlations. Table~\ref{tab:prcc_results} presents these correlations. Our approach to parameter sensitivity analysis, which combines Latin Hypercube Sampling with PRCC, follows the methodology outlined in \cite{marino2008methodology}.

\begin{table}[H]
    \centering {
    \caption{Partial rank correlation coefficients (PRCCs) for different GLMs. PRCCs quantify the sensitivity of each model parameter to relative error ($\text{re}_\phi$ or $\text{re}_R$), while controlling for the effects of other parameters. Higher absolute values indicate greater sensitivity.}
    
    \medskip
    
    \label{tab:prcc_results} \small
    \begin{tabular}{lcccccccccc}  
        \toprule
        \textbf{Distribution} & \textbf{Link} & \textbf{Effect} &  
        $\boldsymbol{a_x}$ & $\boldsymbol{b_x}$ & $\boldsymbol{s_x}$ &  
        $\boldsymbol{a_z}$ & $\boldsymbol{b_z}$ & $\boldsymbol{s_z}$ &  
        $\boldsymbol{g^{-1}(\iota)}$ & $\boldsymbol{\rho}$ \\  
        \midrule
        Binomial  & Logit     &  $\phi$ & -0.90 &  0.89 & -0.69 &  0.04 & -0.07 & -0.34 & -0.15 & -0.08 \\
        Binomial  & Logit     &  $R^2$  & -0.90 &  0.89 & -0.87 &  0.01 & -0.05 & -0.01 &  0.36 &  0.10 \\
        Binomial  & Identity  &  $\phi$ &  0.87 & -0.84 &  0.47 &  0.01 & -0.08 &  0.44 & -0.66 & -0.04 \\
        Binomial  & Identity  &  $R^2$  &   -   &   -   &   -   &   -   &   -   &   -   &   -   &   -   \\
        Poisson   & Log       &  $\phi$ & -0.92 &  0.92 & -0.21 &  0.04 & -0.01 & -0.03 &  0.01 & -0.15 \\
        Poisson   & Log       &  $R^2$  & -0.92 &  0.92 & -0.70 &  0.03 &  0.01 & -0.03 &  0.01 & -0.16 \\
        Gamma     & Log       &  $\phi$ &   -   &   -   &   -   &   -   &   -   &   -   &   -   &   -   \\
        Gamma     & Log       &  $R^2$  & -0.93 &  0.93 & -0.43 & -0.01 &  0.05 &  0.02 & -0.00 & -0.05 \\
        \bottomrule
        \multicolumn{9}{l}{\textbf{\textit{Note.}} The empty entries signify cases when the relative error equals zero.} \\
    \end{tabular} }
\end{table}

Across GLMs, the shape parameters $a_x$ and $b_x$ of $\bbe'\bX$ show the largest PRCCs with the relative errors. These correlations often go in opposite directions: if increasing $a_x$ increases the error, then increasing $b_x$ decreases it, and vice versa. These shape parameters affect the skewness of $\bbe'\bX$. To correct for their impact on the error, we would need to get information about the skewness of $\bbe'\bX$ from practitioners. This may be difficult information to solicit. After the shape parameters $a_x$ and $b_x$, the next strongest correlations are with $s_x$, which determines the variance of $\bbe'\bX$.

\section{Case study results} \label{app:case_study}

\begin{table}[H]
\centering
\caption{Summary of participant characteristics and outcomes for case study on adults with major depression in the last year ($n=5185$).} \label{tab:case_study_summary}
{ \small 
\begin{tabular}{ll}
\toprule
\textbf{Variable} & \textbf{Summary} \\
\midrule
Female, n (\%) & 3476 (67\%) \\
Age group, n (\%) & \\
\quad 18-23 years old & 1753 (34\%) \\
\quad 24-34 years old & 1780 (34\%) \\
\quad 35 years old or older & 1652 (32\%) \\
Race/ethnicity, n (\%) & \\
\quad Non-Hispanic White & 3256 (63\%) \\
\quad Non-Hispanic Black/African American & 430 (8\%) \\
\quad Non-Hispanic Native American/Alaska Native & 67 (1\%) \\
\quad Non-Hispanic Native Hawaiian/Other Pacific Islander & 13 (0\%) \\
\quad Non-Hispanic Asian & 154 (3\%) \\
\quad Non-Hispanic more than one race & 358 (7\%) \\
\quad Hispanic & 907 (17\%) \\
Total family income, n (\%) & \\
\quad Less than \$20,000 & 1095 (21\%) \\
\quad \$20,000 - \$49,999 & 1576 (30\%) \\
\quad \$50,000 - \$74,999 & 865 (17\%) \\
\quad \$75,000 or more & 1649 (32\%) \\
Education level, n (\%) & \\
\quad Less than high school & 497 (10\%) \\
\quad High school graduate & 1362 (26\%) \\
\quad Some college/Associate degree & 1877 (36\%) \\
\quad College graduate & 1449 (28\%) \\
Mental health treatment - video or phone, n (\%) & 2263 (44\%) \\
Mental health treatment - medication, n (\%) & 2616 (50\%) \\
Mental health treatment - inpatient, n (\%) & 308 (6\%) \\
Mental health treatment - outpatient, n (\%) & 2538 (49\%) \\
Any mental health treatment, n (\%) & 3365 (65\%) \\
Total count of types of mental health treatment, mean (SD) & 1.49 (1.30) \\
SDS - home management, mean (SD) & 6.59 (6.07) \\
SDS - work/school, mean (SD) & 6.04 (7.77) \\
SDS - relationships, mean (SD) & 6.28 (5.93) \\
SDS - social life, mean (SD) & 6.88 (6.02) \\
SDS total score, mean (SD) & 24.20 (8.48) \\
\bottomrule \\
SDS = Sheehan Disability Scale.
\end{tabular} }
\end{table}

\begin{table}[H] \small
\centering {
\caption{Measures of effect size for receiving any mental health treatment in the last year from logistic regression models (Bernoulli distribution with logit link).}
\label{tab:models_1_3}
\begin{tabular}{lccc} 
\toprule
\textbf{Model} & \textbf{1} & \textbf{2} & \textbf{3} \\
\midrule
\textbf{Outcome} & Any treatment & Any treatment & Any treatment \\
\textbf{Family} & Binomial & Binomial & Binomial \\
\textbf{Link} & Logit & Logit & Logit \\
\textbf{Predictors} & Education & Education & Education, Sex:Education \\
\textbf{Adjustors} & None & Age, Sex & Age, Sex \\
\textbf{$\phixz$} & 0.51 & 0.38 & 0.40 \\
\textbf{Exp($\phixz$)} & 1.67 & 1.47 & 1.50 \\
\textbf{$\pspRsq$} & 0.015 & 0.008 & 0.009 \\
\textbf{$\fsq$} & 0.0149 & 0.0083 & 0.0090 \\
\textbf{$\fsqphi$} & 0.0149 & 0.0084 & 0.0093 \\
\textbf{$\fsqR$} & 0.0151 & 0.0083 & 0.0091 \\
\textbf{$\text{re}_\phi$} & 0.2\% & 0.9\% & 3.6\% \\
\textbf{$\text{re}_R$} & 1.7\% & -0.7\% & 0.7\% \\
$\bbe$ & -0.195, -0.620, -0.612 & -0.121, -0.475, -0.468 & 
\multirow{2}{4cm}{-0.078, -0.438, -0.449, {\color{white}-}0.193, {\color{white}-}0.192, {\color{white}-}0.074}
 \\ \\
\bottomrule
\end{tabular} }
\end{table}

\begin{table}[H] \small
\centering {
\caption{Measures of effect size for receiving any mental health treatment in the last year from GLMs with Bernoulli distribution and identity link (linear probability model).}
\label{tab:models_4_6}
\begin{tabular}{lccc} 
\toprule
\textbf{Model} & \textbf{4} & \textbf{5} & \textbf{6} \\
\midrule
\textbf{Outcome} & Any treatment & Any treatment & Any treatment \\
\textbf{Family} & Binomial & Binomial & Binomial \\
\textbf{Link} & Identity & Identity & Identity \\
\textbf{Predictors} & Education & Education & Education, Sex:Education \\
\textbf{Adjustors} & None & Age, Sex & Age, Sex \\
\textbf{$\phixz$} & 0.12 & 0.09 & 0.09 \\
\textbf{$\pspRsq$} & 0.015 & 0.008 & 0.009 \\
\textbf{$\fsq$} & 0.0150 & 0.0082 & 0.0088 \\
\textbf{$\fsqphi$} & 0.0150 & 0.0082 & 0.0086 \\
\textbf{$\fsqR$} & 0.0150 & 0.0082 & 0.0088 \\
\textbf{$\text{re}_\phi$} & 0.3\% & -0.6\% & -1.9\% \\
\textbf{$\text{re}_R$} & 0\% & 0\% & 0\% \\
$\bbe$ & -0.041, -0.141, -0.139
 & -0.025, -0.106, -0.105
 &\multirow{2}{4cm}{-0.013, -0.097, -0.101, {\color{white}-}0.036, {\color{white}-}0.029, {\color{white}-}0.002}\\ \\
\bottomrule
\end{tabular} }
\end{table}

\begin{table}[H] \small
\centering {
\caption{Measures of effect size for number of different types of treatment received in the last year (outpatient, inpatient, peer support, medication) from GLMs with Poisson distribution and log link.}
\label{tab:models_7_9}
\begin{tabular}{lccc} 
\toprule
\textbf{Model} & \textbf{7} & \textbf{8} & \textbf{9} \\
\midrule
\textbf{Outcome} & Treatment types & Treatment types & Treatment types \\
\textbf{Family} & Poisson & Poisson & Poisson \\
\textbf{Link} & Log & Log & Log \\
\textbf{Predictors} & Education & Education & Education, Sex:Education \\
\textbf{Adjustors} & None & Age, Sex & Age, Sex \\
\textbf{$\phixz$} & 0.22 & 0.17 & 0.17 \\
\textbf{Exp($\phixz$)} & 1.25 & 1.18 & 1.19 \\
\textbf{$\pspRsq$} & 0.018 & 0.010 & 0.011 \\
\textbf{$\fsq$} & 0.0172 & 0.0097 & 0.0107 \\
\textbf{$\fsqphi$} & 0.0181 & 0.0103 & 0.0114 \\
\textbf{$\fsqR$} & 0.0185 & 0.0103 & 0.0113 \\
\textbf{$\text{re}_\phi$} & 5.2\% & 5.8\% & 6.1\% \\
\textbf{$\text{re}_R$} & 7.0\% & 6.3\% & 5.3\% \\
$\bbe$ & -0.072, -0.271, -0.238
 & -0.042, -0.208, -0.175
 & \multirow{2}{4cm}{-0.012, -0.193, -0.177, {\color{white}-}0.086, {\color{white}-}0.044, -0.022}
 \\ \\
\bottomrule
\end{tabular} }
\end{table}

\begin{table}[H] \small
\centering {
\caption{Measures of effect size for functional impairment, as measured by the total score on the Sheehan Disability Scale (SDS), from GLM with Gamma distribution and log link.}
\label{tab:models_10_12}
\begin{tabular}{lccc} 
\toprule
\textbf{Model} & \textbf{10} & \textbf{11} & \textbf{12} \\
\midrule
\textbf{Outcome} & SDS Total$^a$ & SDS Total$^a$ & SDS Total$^a$ \\
\textbf{Family} & Gamma & Gamma & Gamma \\
\textbf{Link} & Log & Log & Log \\
\textbf{Predictors} & Education & Education & Education, Sex:Education \\
\textbf{Adjustors} & None & Income, Age, Sex & Income, Age, Sex \\
\textbf{$\phixz$} & 0.06 & 0.06 & 0.07 \\
\textbf{Exp($\phixz$)} & 1.07 & 1.07 & 1.07 \\
\textbf{$\pspRsq$} & 0.006 & 0.007 & 0.007 \\
\textbf{$\fsq$} & 0.0063 & 0.0064 & 0.0074 \\
\textbf{$\fsqphi$} & 0.0063 & 0.0064 & 0.0074 \\
\textbf{$\fsqR$} & 0.0065 & 0.0066 & 0.0075 \\
\textbf{$\text{re}_\phi$} & 0\% & 0\% & 0\% \\
\textbf{$\text{re}_R$} & 3.1\% & 2.5\% & 1.6\% \\
$\bbe$ & 0.070, 0.073, 0.070
 & 0.072, 0.079, 0.076
 & \multirow{2}{4cm}{{\color{white}-}0.077, {\color{white}-}0.076, {\color{white}-}0.067, {\color{white}-}0.016, -0.017, -0.045} \\ \\
\bottomrule \\
\multicolumn{2}{l}{$^a$ Shifted up by $0.5$ to avoid zeros.}
\end{tabular} }
\end{table}


\end{document}